\documentclass[aps,12pt,notitlepage,nofootinbib,eqsecnum]{revtex4-1}

\newcommand{\be}{\begin{eqnarray}}
\newcommand{\ee}{\end{eqnarray}}
\newcommand{\non}{\nonumber\\}

 \newcommand{\hatH}{{\hat{H}}}
 \newcommand{\hatO}{{\hat{O}}}
 \newcommand{\hatU}{{\hat{U}}}
 \newcommand{\calA}{{\cal A}}
 \newcommand{\calD}{{\cal D}}
 \newcommand{\calG}{{\cal G}}
 \newcommand{\calK}{{\cal K}}
 \newcommand{\calJ}{{\cal J}}
 \newcommand{\calL}{{\cal L}}
 \newcommand{\calM}{{\cal M}}
 \newcommand{\calO}{{\cal O}}
 \newcommand{\calP}{{\cal P}}
 \newcommand{\calW}{{\cal W}}
 \newcommand{\calH}{{\cal H}}
 \newcommand{\calZ}{{\cal Z}}

 \newcommand{\bra}[1]{\langle {#1} |}
 \newcommand{\ket}[1]{| {#1} \rangle}
 \newcommand{\ave}[1]{\langle {#1} \rangle}

\newcommand{\dotphi}{{\dot{\phi}}}
\newcommand{\hatrho}{{\hat{\rho}}}
\newcommand{\bfk}{{\bf k}}
\newcommand{\bfp}{{\bf p}}
\newcommand{\bfq}{{\bf q}}
\newcommand{\bfu}{{\bf u}}
\newcommand{\bfv}{{\bf v}}
\newcommand{\bfw}{{\bf w}}
\newcommand{\bfx}{{\bf x}}
\newcommand{\bfy}{{\bf y}}
\newcommand{\bfz}{{\bf z}}
\newcommand{\barJ}{{\bar{J}}}

\newcommand{\barphi}{{\bar{\phi}}}
\newcommand{\barchi}{{\bar{\chi}}}

\newcommand{\bfA}{{\bf A}}
\newcommand{\bfE}{{\bf E}}
\newcommand{\bfJ}{{\bf J}}

\newcommand{\bfkperp}{{\bf k}_\perp}
\newcommand{\bfpperp}{{\bf p}_\perp}
\newcommand{\bfqperp}{{\bf q}_\perp}

\newcommand{\bfuperp}{{\bf u}_\perp}
\newcommand{\bfvperp}{{\bf v}_\perp}

\newcommand{\bfxperp}{{\bf x}_\perp}
\newcommand{\bfyperp}{{\bf y}_\perp}
\newcommand{\bfzperp}{{\bf z}_\perp}

\newcommand{\deltaJ}{{\delta J}}

\newcommand{\ula}{{\underline{a}}}
\newcommand{\ulO}{{\underline{O}}}
\newcommand{\ulU}{{\underline{U}}}
\newcommand{\ulV}{{\underline{V}}}
\newcommand{\ulW}{{\underline{W}}}

\newcommand{\ul}[1]{\underline{#1}}

\newcommand{\tildecalA}{{\widetilde{\cal A}}}
\newcommand{\tildecalD}{{\widetilde{\cal D}}}
\newcommand{\tildecalJ}{{\widetilde{\cal J}}}
\newcommand{\tildecalG}{{\widetilde{\cal G}}}

\newcommand{\acutea}{{\acute{a}}}
\newcommand{\tildea}{{\tilde{a}}}
\newcommand{\bara}{{\bar{a}}}

\newcommand{\tildeA}{{\tilde{A}}}
\newcommand{\tildeS}{{\tilde{S}}}
\newcommand{\tildeG}{{\tilde{G}}}
\newcommand{\tildeJ}{{\tilde{J}}}

\newcommand{\tildeomega}{{\tilde{\omega}}}
\newcommand{\tildeeta}{{\tilde{\eta}}}
\newcommand{\tildealpha}{{\tilde{\alpha}}}
\newcommand{\tildelambda}{{\tilde{\lambda}}}
\newcommand{\tilderho}{{\tilde{\rho}}}
\newcommand{\tildezeta}{{\tilde{\zeta}}}
\newcommand{\tildesigma}{{\tilde{\sigma}}}
\newcommand{\tildenu}{{\tilde{\nu}}}
\newcommand{\tildexi}{{\tilde{\xi}}}

\newcommand{\tildeh}{{\tilde{h}}}
\newcommand{\tildeg}{{\tilde{g}}}

\newcommand{\mbf}[1]{{\mbox{\boldmath{$#1$}}}}
\newcommand{\tw}{\textwidth}

\usepackage{color}
\usepackage{graphicx}
\usepackage{hyperref}
\usepackage{amsmath}
\usepackage{slashed}
\usepackage[normalem]{ulem}

\begin{document}

\title{Color Glass Condensate in Schwinger-Keldysh QCD}
\author{Sangyong Jeon}
\affiliation
{Department of Physics, McGill University\\
3600 University Street\\
Montr\'eal QC, H3A-2T8\\
Canada}

\begin{abstract}
Within the Schwinger-Keldysh representation of many-body QCD, 
it is shown that the 
leading quantum corrections to the strong classical color field 
is ``classical'' in the sense that the fluctuation 
field still obeys the classical Jacobi-field equation, while the quantum effects
solely resides in the fluctuations of the initial field configurations.
Within this context,
a systematic derivation of the JIMWLK renormalization group equation 
is presented. 
A clear identification of
the correct form of gauge transformation rules and 
the correct form of the matter-field Lagrangian
in the
Schwinger-Keldysh QCD is also presented.

\end{abstract}

\maketitle

\section{Introduction}

The study of small $x$ gluons in a heavy nucleus in terms of classical gluon
fields was initiated by L.~McLerran, R.~Venugopalan, A.~Ayala and J.~Jalilian-Marian
in a series of seminal papers
\cite{McLerran:1993ni, 
McLerran:1993ka, 
McLerran:1994vd,
Ayala:1995kg, 
Ayala:1995hx}.
Since then the central idea called Color Glass Condensate (CGC) and its
generalization to nucleus-nucleus interaction, the Glasma, have
inspired much work among theoreticians and experimentalists alike.
Good reviews can be found in Refs.
\cite{Iancu:2002xk,
Kovner:2005pe,
JalilianMarian:2005jf,
Weigert:2005us,
McLerran:2010ub,
Gelis:2010nm,
Balitsky:2010jf,
Kovchegov:2012csa,
Iancu:2012xa,
Gelis:2012ri}.

As CGC posits an ensemble of strong color charges producing a strong gluon
field, its formulation properly belongs to the realm of many-body quantum field
theory. As discussed in the following,
the most natural language of many-body
quantum field theory is the Schwinger-Keldysh (SK) formalism 
\cite{Schwinger:1960qe,Keldysh:1964ud,Chou:1984es}
(also known as the in-in formalism or the closed-time-path (CTP) formalism).

Main purpose of the present paper is to re-derive the well known
JIMWLK
renormalization group\footnote{
JIMWLK stands for J.~Jalilian-Marian, E.~Iancu, L.~McLerran, H.~Weigert, 
A.~Leonidov and A.~Kovner.
}
equation 
\cite{JalilianMarian:1996xn,
JalilianMarian:1997jx,JalilianMarian:1997gr,JalilianMarian:1997dw,
JalilianMarian:1998cb,Iancu:2001ad,Iancu:2001md}
using the Schwinger-Keldysh formalism in a systematic way.  
The JIMWLK renormalization group equation is a non-linear generalization of
the BFKL (Balitsky-Fadin-Kuraev-Lipatov) renormalization group 
equation \cite{Kuraev:1977fs,Balitsky:1978ic}.  
The JIMWLK equation incorporates recombination processes 
leading to the gluon saturation \cite{Gribov:1984tu,Mueller:1985wy}
which becomes increasingly important as the density of small $x$ gluons
become higher in the ultrarelativistic limit.
Along the way, we identify where this study differs from previous
approaches and also
where NNLO contributions should appear.  This paper thus presents
the proof-of-principle calculations which sets up the stage for the
more elaborate NNLO calculations. 

In Ref.\cite{Iancu:2000hn}, Iancu, McLerran and Leonidov first introduced
the Schwinger-Keldysh formalism for CGC and JIMWLK.  
Subsequently, F.~Gelis, R.~Venugopalan and their collaborators 
developed the diagrammatic approach \cite{Gelis:2006yv,Gelis:2006cr}
which was later used in many applications
\cite{Gelis:2006ye,Gelis:2007pw,Gelis:2007hj,
Gelis:2008rw, Gelis:2008ad, Gelis:2008sz,
Dusling:2010rm, Dusling:2011rz, Dusling:2012yd}
such as particle productions and 
approach to thermalization, and factorizations.  
Yet, there are many benefits of using 
the Schwinger-Keldysh closed-time path integral explicitly
in contrast to the diagrammatic approach. 
The main benefit exploited in this paper is the clear and
clean separation of strong classical degree of freedom
and the quantum degree of freedom. This is well known in condensed
matter physics (for instance, see Refs.\cite{Kamenev1,Kamenev2}), and
similar conclusion was reached within the 2-PI effective action approach 
as well \cite{Berges:2007ym, Berges:2007re, Berges:2011sb, Berges:2012ev, 
Berges:2012cj}. However, this benefit appears to have not been 
widely appreciated in the CGC context.

One of the main conclusions of this paper is that the leading order
quantum corrections to the strong classical field
comes solely from the quantum fluctuations in the initial condition while
the field itself still obeys the classical field equation.
This makes it particularly simple to resum the leading log divergences
(the JIMWLK equation) and the secular divergences
\cite{Gelis:2006yv,Gelis:2006cr}
as those terms arise only from such quantum corrections. 
It also enables a clear derivation of the retarded, advanced and 
symmetric propagators in the classical background field
and where they should appear in any expression for a diagram.

The separation of 
the strong classical degrees of freedom and quantum corrections 
is actually rather easy to see within the Schwinger-Keldysh formalism,
especially in the $r$-$a$, or the Keldysh, representation. 
Consider for simplicity a scalar field theory.
In the usual closed-time-path formulation, 
the time contour starts from the initial time, goes forward 
to the final time and comes back to the initial time.
In the $r$-$a$ representation, the field on the forward time line
(call it $\phi_1$) and the backward time line (call it $\phi_2$) are
combined as the common part $\phi_r = (\phi_1 + \phi_2)/2$ and 
the difference part $\phi_a = \phi_1 - \phi_2$.
As will be discussed shortly in Section \ref{sec:SKScalar},
the generating functional in the $r$-$a$ representation 
is then given by \cite{Jeon:2004dh,Mueller:2002gd}
\be
\calZ[J_r, J_a]
& = & 
\int\calD\phi_r \calD\phi_a 
\rho_{\rm v}[\phi_r^i, \dot{\phi}_r^i]\,
\exp\left[i\int 
 \Big(
 \phi_a E[\phi_r, J_a]
 + J_r \phi_r
 -{1\over 24}\phi_a^3 V'''(\phi_r)
 \Big)
\right]
\label{eq:introcalZ}
\ee
where
\be
 E[\phi_r, J_a]
 =
 \left( -\partial^2\phi_r - m^2\phi_r 
 - V'(\phi_r) 
 + J_a
 \right)
\ee
is just the classical
equation of motion $\delta S_{\rm cl}[\phi_r, J_a]/\delta \phi_r$.
The functional
$ \rho_{\rm v}[\phi_r^i, \dot{\phi}_r^i] $
is the Wigner transform of the initial density matrix element 
$\rho_{\rm v}[\phi_1^i, \phi_2^i]$.
It can be
interpreted as the distribution of the initial values of the field and its
first time derivative.

Written this way, it is quite clear how the classical field emerges and where
the quantum corrections come from.
First, the integration over $\phi_a$ can be carried out explicitly
if the $\phi_a^3$ term in the Lagrangian is ignored.
This results in a delta-functional that
simply enforces the classical field equation $E[\phi_r, J_a] = 0$.
Therefore, the $\phi_a^3$ term
provides quantum corrections in the form of non-trivial quantum correlations.
Another more subtle 
source of quantum corrections is the initial density matrix. In vacuum, for
instance, the density operator is 
$\hat{\rho}_{\rm v} = | 0 \rangle\langle 0 |$
where $|0\rangle$ is the quantum vacuum state which contains 
zero-point oscillation of all momentum modes.
In contrast, classical vacuum corresponds to the truly empty space:
$\rho_{\rm cl.v}[\phi_r^i, \dot{\phi}_r^i] 
= \delta[\phi_r^i]\delta[\dot{\phi}_r^i]$.

Similarly,
the QCD Schwinger-Keldysh Lagrangian in the Keldysh representation is
\be
\calL 
& = &
\eta_\nu^a \left( [D_\mu, G^{\mu\nu}] 
- J^\nu\right)_a 
+{ig\over 4} [D_\mu, \eta_\nu]_a [\eta^\mu, \eta^\nu]_a
\label{eq:SKL}
\ee
where $A$ is the common field and $\eta$ is the difference field and 
$J^\nu_a$ is the \emph{physical} external colour current.
For the purpose of defining the generating functional, one may add
additional fictitious source terms to Eq.(\ref{eq:SKL}).
Here $D_\mu = \partial_\mu - igA_\mu$ is the covariant derivative containing
only the $A$ field and $G_{\mu\nu} = (i/g)[D_\mu,D_\nu]$ is the non-Abelian
field strength tensor.
Again the first term in Eq.(\ref{eq:SKL})
contains the classical Yang-Mills equation sourced by an external
colour current.
In fact without the triple $\eta$ term, integrating over $\eta$
will force $A$ to be a solution of the classical Yang-Mills equation.
Therefore, it is very natural to consider quantum fluctuations in
the classical background field by separating 
the field $A^\mu = \calA^\mu + a^\mu$ 
in terms of the classical field $\calA^\mu$ 
and the fluctuation field $a^\mu$.
As we have mentioned earlier,
the reason for the 
JIMWLK equation being essentially classical turns out to be
that the leading order quantum correction is contained
entirely in the fluctuations from the initial state while
the gluon field $A_\mu = \calA_\mu + a_\mu$
still obeys the classical Yang-Mills equation.
The main technical reason for this is that the difference-difference
propagator $\ave{\eta^\mu(x)\eta^\nu(y)}$ vanishes identically.
Consequently,
the triple $\eta$ terms contribute only at higher orders.
This is hard to see
without using the Schwinger-Keldysh formalism in the $r$-$a$ representation.

Another important aspect of the SK Lagrangian in Eq.(\ref{eq:SKL})
is how a physical source $J^\nu$ naturally couples to the classical field.  
The interaction Lagrangian is simply 
$\calL_{\rm SK\ int}= -\eta_\nu^a J_{a}^\nu$.
In previous single path formulations of CGC, the interaction Lagrangian
was postulated to
be either $\calL_W = {\rm Tr} J^+ W[A^-]$ 
or $\calL_{\ln W} = {\rm Tr}J^+ \ln W[A^-]$
where $W$ is a Wilson line in the light cone $x^+$ direction 
\cite{JalilianMarian:1997jx,JalilianMarian:2000ad,
Fukushima:2005kk,Fukushima:2006cj}.
These forms were postulated to satisfy two constraints.
First, the interaction term must be gauge invariant by itself.
Second, 
$\delta \calL/\delta A_\mu$ must generate a retarded current that satisfies
$[D_\mu, J^\mu] = 0$. The gauge invariant condition is satisfied by 
both $\calL_W$ an $\calL_{\ln W}$. 
However, neither $\calL_W$ nor $\calL_{\ln W}$
results in the exact retarded current without further approximations.

In contrast, there is no need for any postulation in the SK-QCD.
The interaction term that fulfills both constraints 
is just a part of the usual QCD Lagrangian.
As will be shown shortly, 
both $\eta_\nu$ and $J^\nu$
transforms covariantly under a gauge transformation. Hence, the interaction
term $\eta_\mu J^\mu$ is naturally gauge invariant.
In addition, since the physical current is given by $\delta \calL/\delta
\eta_\mu$, {\em not} $\delta\calL/\delta A_\mu$,
there is no need to guess an interaction term that will generate 
the retarded current. We can simply demand that the external current 
precess according to $[D_\mu, J^\mu] = 0$. This, of course, implies that
$J^\mu$ is a functional of $A_\mu$. But that does not interfere with the
$\delta/\delta \eta_\mu$ operation.  
The exact form of the external colour
source term is an ambiguity in the original formulation of CGC that 
did not really have a proper resolution until now. 
Our analysis in this paper provides a clear resolution.

Another benefit of re-deriving the JIMWLK equation in the SK-QCD is 
the clear identification of diagrams that contribute in the 
next to next to leading order (NNLO) corrections.
Since the leading order and the next-to-leading-order corrections 
all come from the classical part of Eq.(\ref{eq:SKL}),
the NNLO corrections must contain at least one triple $\eta$ vertex.
Since $\eta$ cannot connect to another $\eta$, the triple $\eta$ vertices
provide a non-trivial 3-point correlation in the fluctuation field $a^\mu$.
As we will shortly show, the correlator
$\ave{a^\mu(x)\eta^\nu(y)}$ is a retarded propagator.
Hence, $\eta^\nu(y)$ is must always be in the past of $a^\mu(x)$.
Therefore, it could be possible to package
the effect of a single triple $\eta$ vertex insertion 
as the 3-point correlation in the initial density matrix.
In this paper, we will concentrate on the NLO part while setting up
the stage for the NNLO corrections which will be the topic of future studies.

The rest of this paper is organized as follows.
In Section~\ref{sec:SKScalar}, a brief discussion of scalar theory
is presented as a simple illustration of the Schwinger-Keldysh formalism.
The main topic of Section~\ref{sec:SKQCD} is the gauge invariance
of the SK-QCD Lagrangian.
Section~\ref{sec:leading_order} contains power counting argument that
justifies the use of the classical Yang-Mills equation. It also contains
more detailed discussion on the external current Lagrangian.
A brief discussion on the classical solution found by McLerran and
Venugopalan is given in Section~\ref{sec:ClassicalSolution} mainly to 
fix notations.
In Section~\ref{sec:lightconevsaxial}, we discuss
the gauge transformation rules between the light-cone gauge fluctuation field
and the axial gauge fluctuation field. 
In Section~\ref{sec:qcd_props}, we carefully derive retarded and symmetric
propagators in the CGC background to prepare for the JIMWLK equation
derivation given in Section~\ref{sec:JIMWLK_derivation}.
Discussions and outlook is given in Section~\ref{sec:outlook}.
Appendices \ref{app:Scalar_props} and \ref{app:scalar_lightcone}
contain details of the scalar theory propagator derivations.
Appendix \ref{app:props_details} contains QCD propagator derivations.
Appendices~\ref{app:details_2pts} and
\ref{app:Lines1to6} have technical details of the JIMWLK
equation derivation.

\section{Schwinger-Keldysh Closed Time Path in Scalar Field Theory}
\label{sec:SKScalar}

In this section, we use the real scalar field theory to illustrate
why the Schwinger-Keldysh formalism is necessary and how the separation of
classical and quantum degrees of freedom can be naturally achieved.
The Lagrangian for the scalar field is 
\be
L = {1\over 2}\partial_\mu\phi\partial^\mu\phi 
- {m^2\over 2} \phi^2 - V(\phi)
\ee
The metric here is $g_{\mu\nu} = {\rm diag}(1,-1,-1,-1)$.
The need for the Schwinger-Keldysh formalism in addition to the usual
Feynman formalism is simple to state. 
In quantum mechanics and quantum field theory alike, 
we not only need to calculate the transition amplitude
\be
M_{fi}
= \bra{\phi_{f}}\hatU(t_f, t_i)\ket{\phi_{i}}
\label{eq:Mtransition}
\ee
but also the expectation value of an operator at times
$t \ge t_i$ for a given initial state
\be
\ave{\hatO(t)}_\phi &=& 
\bra{\phi_{i}} \hatO_H(t) \ket{\phi_{i}}
\non
& = &
\bra{\phi_{i}} \hatU(t_i, t) \hatO \hatU(t, t_i)\ket{\phi_{i}}
\label{eq:Oaverage}
\ee
or a given initial density operator
\be
\ave{\hatO(t)}_\phi &=& 
{\rm Tr}\left(
\hat{\rho}_{\rm init} \hatU(t_i, t) \hatO \hatU(t, t_i)\right)
\non
& = &
{\rm Tr}\left( \hat{\rho}(t) \hatO \right)
\label{eq:avehatO}
\ee
Here $\hatU(t, t_i) = e^{-i\hatH(t-t_i)}$ is the time evolution operator
with the Hamiltonian $\hatH$ and the subscript $H$ in
$\hatO_H(t)$ indicates that it is in the Heisenberg picture.
The time evolution of the density operator is given by
$\hat{\rho}(t) = \hatU(t, t_i)\hat{\rho}_{\rm init}\hatU(t_i, t)$.

By the usual procedure of inserting multiple resolutions of the identity,
the transition
amplitude can be expressed as a Feynman path integral
\be
\bra{\phi_{f}}\hatU(t_f, t_i)\ket{\phi_{i}}
=
\int_{\phi_i}^{\phi_f} \calD\phi\, e^{i\int_{t_i}^{t_f}dt\int d^3x\, L(\phi)}
\ee
In contrast, 
operator expectation values involve two transition amplitudes, one for each
time evolution operator in Eq.(\ref{eq:avehatO})\footnote{
Our convention here is that the functional integral at a fixed time is denoted
as $\int [d\phi]$ and the functional integral over both time and space is
denoted as $\int \calD\phi$.}
\be
\ave{\hatO(t)} 
& = &
\int[d\phi^i_1][d\phi^i_2][d\phi^f_1][d\phi^f_2]\,
\bra{\phi_1^i}\hat{\rho}_{\rm v}\ket{\phi^i_2}
\bra{\phi_2^i}\hatU(t_i, t)\ket{\phi_2^f}
\bra{\phi_2^f} \hatO\ket{\phi_1^f}
\bra{\phi_1^f} \hatU(t, t_i)\ket{\phi_1^i}
\label{eq:Oaverage3}
\ee
The generating functional therefore needs two path integrals, an ordinary
one for
$\bra{\phi_1^f} \hatU(t, t_i)\ket{\phi_1^i}$
and a Hermitian conjugate one for
$\bra{\phi_2^i}\hatU(t_i, t)\ket{\phi_2^f}$ both in the presence of sources
\be
\calZ[J_1, J_2]
& = &
\int [d\phi_f]\,
\bra{\phi_f} \hatU_{J_1}(t_f, t_i)\hatrho_{\rm v}
\hatU_{J_2}(t_i, t_f) \ket{\phi_f}
\non
& = &
\int [d\phi_f]\, [d\phi_1^i] \, [d\phi_2^i]\,
\bra{\phi_f} \hatU_{J_1}(t_f, t_i) \ket{\phi_1^i}
\bra{\phi_1^i} \hatrho_{\rm v}\ket{\phi_2^i}
\bra{\phi_2^i} \hatU_{J_2}(t_i, t_f) \ket{\phi_f}
\non
& = &
\int [d\phi_f][d\phi_1^i][d\phi_2^i]
\int_{\phi_i^1}^{\phi_f}\calD\phi_1 
\int_{\phi_i^2}^{\phi_f}\calD\phi_2\,
\rho_{\rm v}[\phi_1^i,\phi_2^i]\,
\non
& & {} \times
\exp\left(
i\int_{t_i}^{t_f} dt \int d^3x\, \left( L(\phi_1) +  J_1\phi_1 \right)
-
i\int_{t_i}^{t_f} dt \int d^3x\, \left( L(\phi_2) +  J_2\phi_2 \right)
\right)
\label{eq:ZJ1J2}
\ee
Here $\rho_{\rm v}[\phi_1^i, \phi_2^i] = 
\bra{\phi_1^i}\hat{\rho}_{\rm v}\ket{\phi_2^i}$ is the matrix element of the
initial density matrix at $t = t_i$. 
The final field configuration for $\phi_1$ and $\phi_2$ coincide,
$\phi_1^f = \phi_2^f = \phi_f$, because of the trace operation.

The separation of classical degree of freedom vs.~quantum one becomes
much clearer if
the following change of variables  (often referred to as the Keldysh
rotation) are made
\be
\phi_r &=& (\phi_1 + \phi_2)/2
\non
\phi_a & = & \phi_1 - \phi_2
\ee
with the corresponding change in the sources
\be
J_r &=& J_1 - J_2
\\
J_a &=& (J_1 + J_2)/2
\ee
The original field variables are 
$\phi_1 = \phi_r {+}\phi_a/2$
and $\phi_2 = \phi_r {-}\phi_a/2$.
Expressed this way, it is clear that
$\phi_r$ constitutes the common part of the 
two time evolutions and $\phi_a$ constitutes 
the difference part. Since classical evolution is deterministic,
it is very natural that the common field $\phi_r$ contains
the classical degree of freedom
and the difference field $\phi_a$ is purely quantum mechanical.

After making the substitutions and integrating by parts, 
we obtain a very suggestive end result (for details, see Ref.\cite{Jeon:2004dh})
\be
\calZ[J_r, J_a]
& = & 
\int\calD\phi_r \calD\phi_a 
\rho_{\rm v}[\phi_r^i, \dot{\phi}_r^i]\,
\exp\left(i\int_{t_i}^{t_f} dt\int d^3x\, \calL \right)
\label{eq:calZSK}
\ee
where
\be
\int_{t_i}^{t_f}dt\int d^3x\, \calL 
& = &
\int_{t_i}^{t_f}dt\int d^3x\, 
 \Big(
 \phi_a E[\phi_r, J_a]
 + J_r \phi_r
 -{1\over 24}\phi_a^3 V'''(\phi_r)
 \Big)
 \label{eq:calL}
\ee
is the Schwinger-Keldysh Lagrangian and
\be
 E[\phi_r, J_a]
 =
 -\partial^2\phi_r - m^2\phi_r 
 - V'(\phi_r) 
 + J_a
\ee
is the classical equation of motion.
The functional 
\be
\rho_{\rm v}[\phi_r^i, \dot{\phi}_r^i]
=
\int [d\phi_a^i] e^{i\int d^3x \dot{\phi}_r^i \phi_a^i}\, 
\rho_{\rm v}[\phi_r^i-\phi_a^i/2, \phi_r^i-\phi_a^i/2]
\label{eq:wigner_rho}
\ee
is the Wigner transform of the initial density matrix element
$\rho_{\rm v}[\phi_1^i, \phi_2^i]$ \cite{Mrowczynski:1994nf}.
The exponent $\int d^3x\,\dotphi_r^i\phi_a^i$ 
in Eq.(\ref{eq:wigner_rho})
is the surface term 
arising from integrating by parts in time. The surface term at the final time
$t_f$ is identically zero
since $\phi_a(t_f) = \phi_1(t_f)-\phi_2(t_f) = 0$.
Here
\be
\dot{\phi}_r^i = {\phi_r(t_i+\Delta t) - \phi_r(t_i)\over \Delta t}
\ee
needs to be defined in the context of the discretized-time functional integral.
Hence, $\rho_{\rm v}[\phi_r^i, \dot{\phi}_r^i]$ is essentially a
joint distribution of $\phi_r^i = \phi_r(t_i)$ 
and $\phi_r(t_i+\Delta t)$. Of course, the usual
Caveats for the Wigner functions such as the possibility of negative values
apply here as well.

The form of the generating functional (\ref{eq:calZSK})
is suggestive because this is almost in the form of a classical
initial value problem. 
In fact, if the term cubic in $\phi_a$
in Eq.(\ref{eq:calL}) is absent, integrating over
$\phi_a$ will just enforce the classical equation of motion with
the initial values of $\phi_r$ and $\dot{\phi}_r$ distributed according to
$\rho_{\rm v}[\phi_r^i, \dot{\phi}_r^i]$.
The $\phi_a^3$ term in $\calL$ prevents this interpretation.
Hence it provides quantum correlations.
This is, however, not the only place where quantum effects enter.
Consider the vacuum. 
In the classical field theory, vacuum really means an empty
space so that the initial values are strictly zero.
The quantum vacuum, however, contains the zero-point motion of all available
field modes.
For example, the perturbative vacuum in the Minkowski space corresponds to
\be
\rho_{\rm v}[\phi_r^i, {\pi}_r^i] 
=
\exp\left(-\int {d^3k\over (2\pi)^3 E_k}
\left( E_k^2 \phi_r^i(\bfk)\phi_r^i(-\bfk)
+ {\pi}_r^i(\bfk){\pi}_r^i(-\bfk)\right)\right)
\label{eq:rhovac}
\ee
where we set $\pi_r^i = \dot{\phi}_r^i$.
This introduces quantum fluctuations in the initial conditions even if
the subsequent evolution is strictly classical.
This form of the vacuum functional is essentially the Wigner transform of 
the product of the simple harmonic oscillator ground state wavefunctions
$\rho_0(x_1, x_2) = \ave{x_1|0}\ave{0|x_2}$. 
(At finite temperature, the integrand has an additional factor
$\tanh(E_k\beta/2)$. For more
details see Ref.\cite{Mrowczynski:1994nf,Jeon:2004dh}.) 

What we are interested in here is the case of strong classical fields.
To quantify what ``strong'' means,
let us specify the interaction potential to be
\be
V(\phi) ={g^2\over 4!}\phi^4
\ee
with $g \ll 1$.
A source is strong if it is $O(1/g)$. Consequently, the classical
field generated by it is also $O(1/g)$ and hence strong. 
What we would like to formulate now is the perturbation theory where 
the strong classical field acts as the un-perturbed solution.
When there is a physical external source
$\barJ$, it must be common to both $\phi_1$ and $\phi_2$.
That implies that $J_a = (J_1 + J_2)/2$ needs to be separated 
into the physical source $\barJ$ and the fictitious one for the generating
functional, $J_a$.
We also separate
\be
\phi_r \to \barphi + \phi_r
\ee
where $\barphi$ is a solution of the classical field equation 
\be
\partial^2\barphi + m^2\barphi + V'(\barphi) = \barJ
\label{eq:barphieq}
\ee
with the strong physical source term given by $\barJ$ and
the initial conditions given by
$\lim_{t\to t_i}\barphi(t) = 0$ and
$\lim_{t\to t_i}\partial_t{\barphi}(t) = 0$.
The fluctuation field $\phi_r$ satisfies the initial conditions
$\lim_{t\to t_i}\phi_r(t) = \phi_r^i$ and
$\lim_{t\to t_i}\dot{\phi}_r(t) = \dot{\phi}_r^i$.
These initial configurations are distributed
according to the Wigner functional $\rho_{\rm v}[\phi_r^i, \dotphi_r^i]$.

The Lagrangian becomes, after using the fact that $\barphi$
satisfies the classical field equation (\ref{eq:barphieq}),
\be
\calL & = & 
 \phi_a 
 \left( -\partial^2 - m^2 -V''(\barphi)\right)\phi_r
 + J_r\phi_r + J_a \phi_a 
 \non & & {}
 -{1\over 2}V'''(\barphi)\phi_a\phi_r^2
 -{1\over 6}V''''(\barphi)\phi_a\phi_r^3
 -{1\over 24}\phi_a^3 V'''(\barphi) 
 -{1\over 24}\phi_a^3 \phi_r V''''(\barphi)
\label{eq:calLra}
\ee
Since $\barphi = O(1/g)$,
we have $V''(\barphi) = O(1)$,
$V'''(\barphi) = O(g)$ and $V''''(\barphi) = O(g^2)$.
We can therefore carry out a formal perturbative expansion with
the cubic and quartic terms in Eq.(\ref{eq:calLra}).
Here $J_r$ and $J_a$ are the fictitious sources needed to define
the generating functional.

The terms linear in $\phi_a$ in Eq.(\ref{eq:calLra}) come from 
\be
\calL_{\rm cl} 
= \phi_a E[\barphi+\phi_r,\barJ] 
\ee
after using the fact that $E[\barphi, \barJ] = 0$.
Therefore if the $\phi_a^3$ terms in
Eq.(\ref{eq:calLra}) are once again ignored,
integrating over $\phi_a$ 
just enforces the classical field equation $E[\barphi+\phi_r, \barJ] = 0$
in addition to $E[\barphi, \barJ] = 0$.
In that case, $\phi_r$ is the difference between two classical solutions
and, by definition,
the equation it obeys is the full non-linear Jacobi-field equation
(c.f.~Refs.\cite{DeWittMorette:1976up,Bazanski:1976sa}).
For further analysis of the scalar theory including the derivation of
propagators, see Appendix \ref{app:Scalar_props}.

\section{Schwinger-Keldysh formulation of Many-body QCD}
\label{sec:SKQCD}

The closed-time-path version of the QCD Lagrangian is
\be
\calL_{SK}
=
-{1\over 2} G_{1,\mu\nu}^a G_{1,a}^{\mu\nu}
+{1\over 2} G_{2,\mu\nu}^a G_{2,a}^{\mu\nu}
- J_{a}^\mu A_{1,\mu}^a
+ J_{a}^\mu A_{2,\mu}^a
\ee
where $J^\mu$ is an external physical colour current. 
From now on, we will use the Greek letters 
for the space-time indices and Latin letters for the color indices.
The indices 1 and 2 indicates that the field lives on the forward time line
(1) or the backward time line (2).
We again perform the Keldysh rotation 
to form the common field $A_\mu$ and the difference field $\eta_\mu$:
\be
A_\mu^a & = & {1\over 2}\left(A_{1,\mu}^a + A_{2,\mu}^a \right)
\label{eq:QCD_rfield}
\\
\eta_\mu^a & = & A_{1,\mu}^a - A_{2,\mu}^a 
\ee
In terms of these fields and the physical source $J$,
the Lagrangian becomes
\be
\calL_{SK} 
& = &
\eta_\nu^a \left([D_\mu, G^{\mu\nu}] - J^\nu\right)_a 
+{ig\over 4} [D_\mu, \eta_\nu]^a[\eta^\mu, \eta^\nu]_a
\label{eq:LQCD}
\ee
Just as in the scalar field case,
the classical equation of motion,
$[D_\mu,G^{\mu\nu}] = J^\nu$, naturally emerges.
Unlike the case of single-path Lagrangian,
this term in the Schwinger-Keldysh Lagrangian
is actually fully gauge invariant even when $J^\mu$ is 
an external color current.
To show this, consider the gauge transformation properties of $A_\mu$ and
$\eta_\mu$.  Under the gauge transformation, the original fields transform
as
\be
A'_{1,2,\mu}
&=&
UA_{1,2,\mu} U^\dagger + i{1\over g}U\partial_\mu U^\dagger
\label{eq:gf_12}
\ee
Hence, the common field $A_\mu = (A_\mu^1 + A_\mu^2)/2$ transforms as expected, 
\be
A'_\mu 
& = &
U A_\mu U^\dagger + i{1\over g}U\partial_\mu U^\dagger
\label{eq:A_transform}
\ee
and $D_\mu=(\partial_\mu - igA_\mu) $ and 
$G_{\mu\nu} = (i/g)[D_\mu, D_\nu]$ also transforms covariantly.
The difference field $\eta_\mu = A_\mu^1 - A_\mu^2$ 
transforms covariantly as well without the derivative term since
it is canceled in the difference: 
\be
\eta'_\mu & = & U\eta_\mu U^\dagger 
\ee
The color current must transform covariantly 
\be
J'_\mu = U J_\mu U^\dagger
\label{eq:Jprimemu}
\ee
because the left hand side of the equation of motion 
$[D_\mu, G^{\mu\nu}] =J^\nu$ 
transforms covariantly.
Therefore, the classical
Yang-Mills equation term in the Lagrangian (\ref{eq:LQCD})
is fully gauge invariant including the coupling $\eta_\mu J^\mu$.
It is also trivial to see that 
$A_\mu J^\mu$ coupling would not be gauge invariant
even if $J^\mu$ is covariantly conserved, due to the derivative term 
in Eq.(\ref{eq:A_transform}). In QED, the derivative term would also
vanish f the charge conservation $\partial_\mu J^\mu = 0$ is imposed on the
current.
In non-Abelian Yang-Mills theory, however, it cannot vanish.

In previous studies of Color Glass Condensate where single path Lagrangian
was used, the coupling to the external current was postulated to be either
$ \calL_{W} = {\rm Tr} (J^+ W) $ or $\calL_{\ln W} = {\rm Tr}(J^+\ln W)$
\cite{JalilianMarian:1997jx,JalilianMarian:2000ad,
Fukushima:2005kk,Fukushima:2006cj}
where $J^+$ is the color charge density and $W$ is a Wilson line in 
the $x^+$ direction.
These forms were devised because they are gauge invariant while the
straight-forward $J^a_\mu A_a^\mu$ is not. The problem with the above forms,
however, was that the current 
derived from these interaction terms 
$J^\mu = {\delta \int \calL_{W,\ln W}\over \delta A_\mu}$
could not be a retarded solution of the covariant conservation law
$[D_\mu, J^\mu] = 0$.
Although using the closed-time-path with $\calL_{W}$ alleviates some of
this problem \cite{Iancu:2000hn}, it does not entirely solve it.

In our approach, having both $A$ and $\eta$ naturally
solves the problem that 
the external color current $J^\mu$ must also be a retarded solution of 
the covariant conservation equation.
The equation $[D_\mu, J^\mu] = 0$ implies that $J^\mu$ is a functional of
$A_\mu$.
If the interaction term was $\calL_{AJ} = -A_\mu J^\mu$,
this $A$ dependence of $J$ would prevent obtaining a proper classical
field equation because ${\delta \int \calL_{AJ}\over \delta A_\mu} \ne -J^\mu$.
However, as we have shown, the proper interaction term has the form
$\calL_{\eta J} = -\eta_\mu J^\mu$
and the fact that $J^\mu$ needs to be a functional of
$A_\mu$ has no relevance in deriving the classical equation of motion 
since it comes from varying $\eta_\mu$, not $A_\mu$.
The Wilson lines necessary for CGC calculations will reappear when $J^\mu$
is specified to be $J^\mu = \delta^{\mu+}\rho$ and demanded to be a retarded
solution of the covariant conservation equation.

In this way we have now resolved an old ambiguity in the
formulation of CGC, namely, the gauge invariant
coupling between the retarded current and the gauge field.
Neither ${\rm Tr}J^+ W$ nor ${\rm Tr}J^+\ln W$ is the right form.
The Schwinger-Keldysh formulation of QCD automatically yields the correct
form of the coupling between the external color current and the gluon field.
Although it can be argued that the 
leading order quantum correction can be obtained with 
$\calL_W$ and $\calL_{\ln W}$ \cite{JalilianMarian:2000ad}, 
for higher order correction it is 
essential that we have the correct form of the interaction term 
actually derived from QCD.
More discussion on this point is given in the next section.

We should note that invariance under the gauge transformation
Eq.(\ref{eq:gf_12}) is not the full symmetry of the Lagrangian $\calL_{SK}$.
If there is no physical source, $\calL_{SK}$ is invariant under
the gauge transformations of $A_1^\mu$ and $A_2^\mu$ separately.
However, with a physical external source term present, 
this can no longer be the case. The external source term in both branches
must behave the same way under gauge transformation.  Therefore only
the gauge invariance under this common transformation is of our concern here.

Another benefit of the above gauge transformation rules is 
the automatic appearance of the Gauss law constraint.
For a simple illustration, consider QED with the temporal gauge condition
$A^0 = 0$.
If we impose this condition on the single-path Lagrangian,
$\calL = -(1/4)F^2 - JA$, 
the source term $-J^\mu A_\mu$ becomes  $\bfJ{\cdot}\bfA$.
Therefore, the Euler-Lagrange equation from this Lagrangian no longer contains
the Gauss law $\nabla{\cdot}\bfE = J^0$. 
It then needs to be imposed as an extra constraint with $A^0$ reinstated
as the Lagrange multiplier.
In the Schwinger-Keldysh version of QED, this is not necessary.
The QED Lagrangian in the Schwinger-Keldysh formalism is
\be
\calL_{SK} = \eta_\nu ( \partial_\mu F^{\mu\nu} - J^\nu )
\ee
We can use the gauge transformation of 
$A'_\mu = A_\mu - \partial_\mu \Lambda$ to remove $A_0$. But since 
$\eta'_\mu = \eta_\mu$ does not change, $\eta_0$ 
cannot be set to zero at the same time.  
Hence, there is no need to impose the Gauss law as an extra constraint
with an extra Lagrange multiplier.
Euler-Lagrange equation with respect to $\eta_0$, or equivalently
carrying out $\eta_0$ integral in the path integral,
automatically enforces the Gauss law.

\section{Leading order quantum corrections in Schwinger-Keldysh QCD} 
\label{sec:leading_order}

In this section, we show that the leading order quantum correction to the
strong classical gluon field comes only from the initial state vacuum
fluctuations.
Let $\calA_\mu$ be the solution of the classical Yang-Mills equation
\footnote{
In this and the following sections, we largely (but not always)
follow notations in
Ref.\cite{Gelis:2008rw} for easier comparisons.
}
\be
[\calD_\mu, \calG^{\mu\nu}] - \calJ^\nu = 0
\label{eq:clYM}
\ee
where $\calD_\mu = \partial_\mu - ig\calA_\mu$ 
and $\calG_{\mu\nu} = (i/g)[\calD_\mu, \calD_\nu]$ with the strong 
external color current $\calJ^\nu = O(1/g)$.
The solution of Eq.(\ref{eq:clYM}) 
initially found by McLerran and Venugopalan
is by now well known. It is summarized
in Section~\ref{sec:ClassicalSolution}.

Letting $A_\mu = \calA_\mu + a_\mu$
where $a_\mu$ is the $O(1)$ fluctuation field,
the Lagrangian becomes
\be
\calL
& = &
\calL_{\rm cl}^{\calA}
+
\calL_{\rm cl}^a
+ \calL_{\rm quant}
\label{eq:FullL}
\ee
where
\be
\calL_{\rm cl}^\calA
= \eta_\nu \left( [\calD_\mu, \calG^{\mu\nu}] - \calJ^\nu \right)
\label{eq:calLcalA}
\ee
is proportional to the equation of motion (therefore vanishes).
The Jacobi-field part of the Lagrangian is
\be
\calL^a_{\rm cl}
&=&
\calL_{\rm cl}^{\calA+a}-\calL_{\rm cl}^\calA
\non
&=&
\eta_\nu [\calD_\mu, [\calD^\mu, a^\nu]]
-\eta_\nu [\calD^\nu, [\calD_\mu, a^\mu]]
+2ig\eta_\nu [\calG^{\mu\nu}, a_\mu]
\non & & {}
-2ig \eta_\nu [a_\mu, [\calD^\mu, a^\nu]]
+ig \eta_\nu [a^\nu, [\calD_\mu, a^\mu]]
\non & & {}
+ig \eta_\nu [a_\mu, [\calD^\nu, a^\mu]]
-g^2\eta_\nu [a_\mu, [a^\mu, a^\nu]]
\non
& & {}
- \eta_\nu \delta J^\nu
\label{eq:Lclassical}
\ee
and the purely quantum part is
\be
\calL_{\rm quant}
& = &
{ig\over 4}\left([\calD_\mu, \eta_\nu][\eta^\mu, \eta^\nu]\right)
+{g^2\over 4}\left( [a_\mu, \eta_\nu] [\eta^\mu, \eta^\nu]\right)
\non
& = &
{ig\over 4}\left([D_\mu, \eta_\nu][\eta^\mu, \eta^\nu]\right)
\label{eq:Lquant}
\ee
Both $\calL_{\rm cl}^a$ and $\calL_{\rm quant}$ are gauge invariant by themselves.
Here $\delta J^\mu = J^\mu - \calJ^\mu$ 
is the difference between $J$ that satisfies $[D_\mu,J^\mu] = 0$ (with
$A_\mu = \calA_\mu + a_\mu$) and 
$\calJ^\mu$ that satisfies $[\calD_\mu,\calJ^\mu] = 0$.
The first term $\calL_{\rm cl}^\calA$ in Eq.(\ref{eq:FullL})
vanishes when $\calA_\mu$ is the solution
of the equation of motion, $[\calD_\mu, \calG^{\mu\nu}] = \calJ^\nu$.
The first term and the second term together, 
$\calL_{\rm cl}^\calA + \calL_{\rm cl}^a$,
vanishes when $A_\mu = \calA_\mu + a_\mu$ is the solution of the
equation of motion, $[D_\mu, G^{\mu\nu}] = J^\nu$.
Consequently if the purely quantum part can be ignored,
$a_\mu$ is the difference between two classical solutions
and Eq.(\ref{eq:Lclassical}) imposes the Jacobi-field
equation on $a_\mu$ just as in the scalar case.

We can now carry out the power counting.
First, note that since we have a strong source
$\calJ_\mu = O(1/g)$, the classical Yang-Mills field is also strong,
$\calA_\mu = O(1/g)$, while the covariant derivative $\calD_\mu = O(1)$.
The interaction terms involving 3 fields 
in $\calL_{\rm cl}^a$ and $\calL_{\rm quant}$ are all $O(g)$ 
and the interaction terms involving 4 fields are all $O(g^2)$.
Hence perturbative expansion may be possible with these interaction terms,
provided $g \ll 1$ (For a purely classical Yang-Mills perturbation theory,
see Ref.\cite{Carta:2005fq}.).

\begin{figure}[t]
\centerline{
\includegraphics[width=8cm]{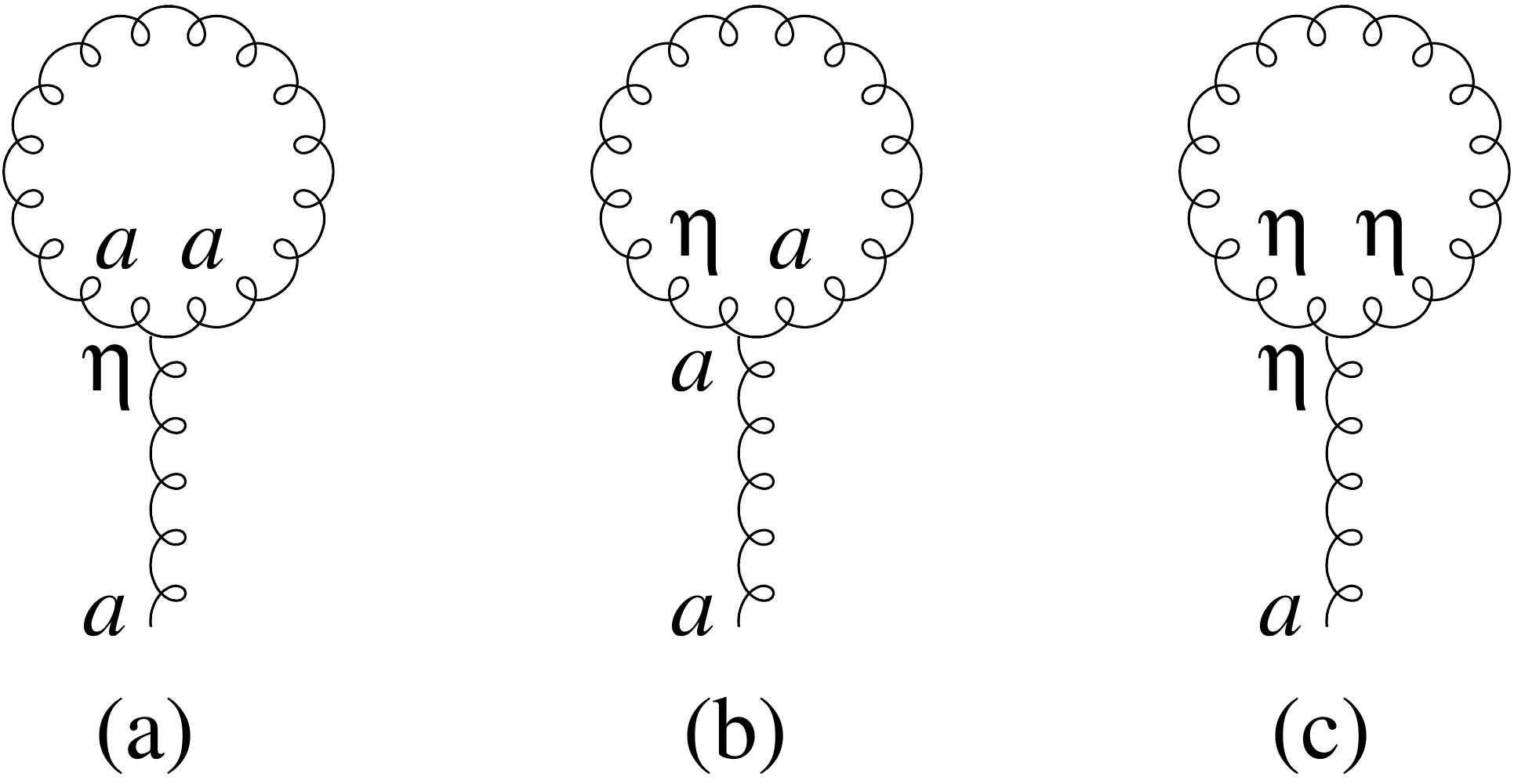}
}
\caption{One loop tadpole diagrams for $\ave{a}$.}
\label{fig:avea}.
\end{figure}

Consider first the quantum correction to the 1-point average $\ave{a(x)}$.
The first corrections to the 1-point average necessarily comes from
the terms involving 3 fields in Eq.(\ref{eq:Lclassical}) depicted in 
Fig.~\ref{fig:avea}.
Since the difference-difference propagator vanishes identically,
$\ave{\eta\eta} = 0$ (c.f.~Appendix \ref{app:scalar_full_props}),
the $g\calD\eta\eta\eta$ term in $\calL_{\rm quant}$
cannot result in leading order quantum corrections.
For the diagram (b), it vanishes because, as is shown
in Section~\ref{sec:qcd_props} and also in Appendix~\ref{app:Scalar_props},
the 2-point function $\ave{a^\mu(x)\eta^\nu(y)} = G_R^{\mu\nu}(x|y)$ 
is a retarded propagator. Hence, the diagram (b) contains a time-loop.
Therefore, it must vanish. Equivalently,
we need to define $\theta(0) = 0$ when the $\theta$-function is associated
with the retarded or the advanced propagator.

\begin{figure}[t]
\centerline{
\includegraphics[width=7cm]{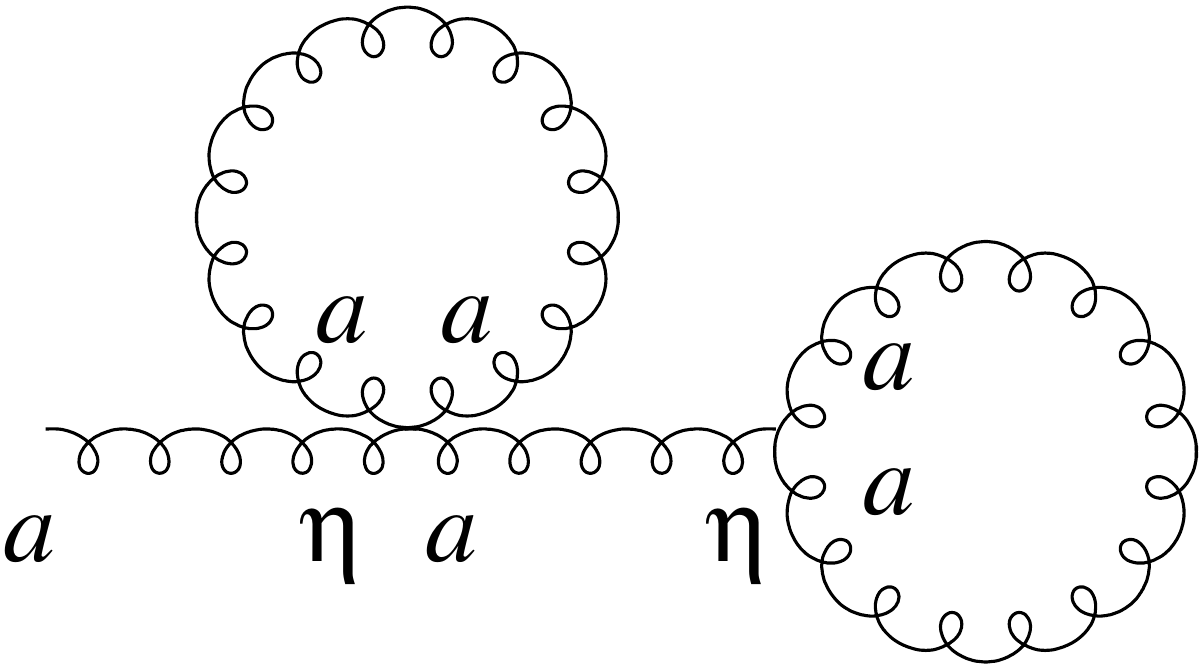}
}
\caption{A diagram with two tadpoles that can contribute at $O(1/g)$.}
\label{fig:avea2}.
\end{figure}

\begin{figure}[t]
\centerline{
\includegraphics[width=8cm]{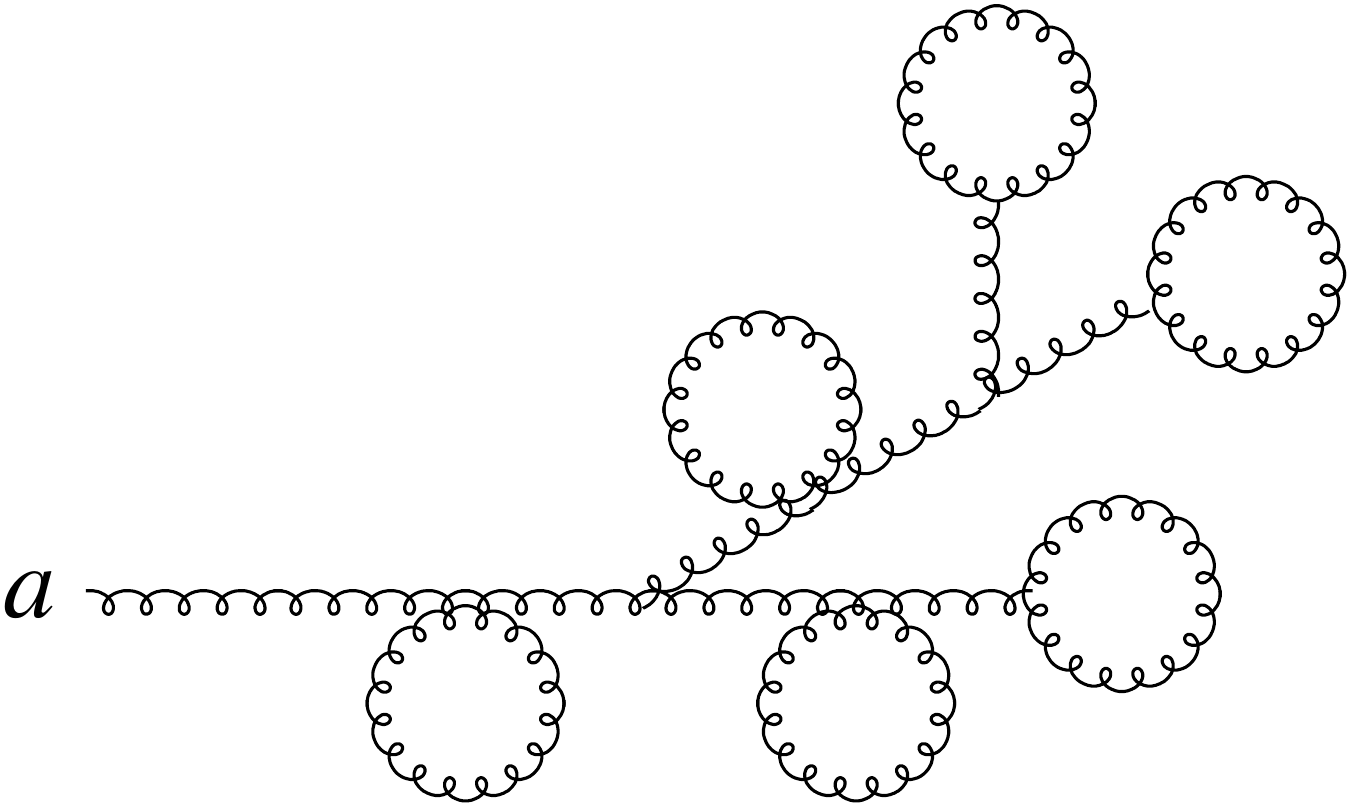}
}
\caption{A diagram with multiple tadpoles that can contribute at $O(1/g)$.}
\label{fig:avea3}.
\end{figure}

Disregarding $\delta J$ term for the moment
(since it does not change the power counting),
the first quantum correction to $\ave{a}$ comes solely from
the diagram (a) in Fig.~\ref{fig:avea}
\be
\ave{a_\lambda(x)}
& = &
-2ig\int d^4y\, G^R_{\lambda\nu}(x|y)
\ave{[a_\mu(y), [\calD^\mu(y), a^\nu(y)]]}_0
\non & & {}
+
ig\int d^4y\, G^R_{\lambda\nu}(x|y)
\ave{ [a^\nu(y), [\calD_\mu(y), a^\mu(y)]]}_0
\non & & {}
+ 
ig\int d^4y\, G^R_{\lambda\nu}(x|y)
\ave{ [a_\mu(y), [\calD^\nu(y), a^\mu(y)]]}_0
\label{eq:alambda1}
\ee
Here 
$\ave{a_\lambda(x) \eta_\nu(y)}_0 = G_{\lambda\nu}^R(x|y)$ is the retarded
propagator of the linear Jacobi field 
equation\footnote{
It is unfortunate that the conventional use of the letter $a$ for
the fluctuation field clashes with another conventional
use of the letter $a$ for the difference field in the $r$-$a$ formalism.
To avoid confusion, 
we will use $a$ solely for the fluctuation $r$-field 
(c.f.~Eq.(\ref{eq:QCD_rfield})) from now on,
and always use $\eta$ for the difference field.
}
\be
[\calD_\mu, [\calD^\mu, a^\nu]]
-  [\calD^\nu, [\calD_\mu, a^\mu]]
+2ig [\calG^{\mu\nu}, a_\mu]
=0
\ee
The 1-point function $\ave{a}$ is naively $O(g)$.  
However, as we will see shortly in Section~\ref{sec:JIMWLK_derivation},
the regularized tadpole can introduce a large small-$x$ log factor.
If this log enhancement is $O(1/g^2)$, then $\ave{a} = O(1/g)$,
making the NLO correction as big as the leading order.
In that case, not only the $g\eta \calD aa $ terms in Eq.(\ref{eq:Lclassical})
can contribute but so can the $g^2 \eta aaa$ terms. For instance, the diagram
shown in Fig.~\ref{fig:avea2} is also $O(1/g)$ because the explicit factor
of $g^2$ at the quartic vertex
is compensated by a large log from the tadpole.
In fact, any diagram that is composed of the combination of the cubic
vertex tadpoles and the quartic vertex tadpoles as shown in
Fig.\ref{fig:avea3}
can contribute at $O(1/g)$ when the tadpole provides a large log of the size
$1/g^2$.  

The power counting analysis for $\ave{a(x)}$ so far indicates that
when there is a logarithmic enhancement, naive perturbation theory
does not work. Insertions of any number of
the interaction terms in the Jacobi-field
Lagrangian (\ref{eq:Lclassical}) can contribute
at $O(1/g)$ and hence need to be resummed. 
As we have argued, this is equivalent to resummation of the leading
logs. This resummation can be fairly easily accomplished in the current
setting: We can simply start with the full Jacobi-field Lagrangian
$\calL_{\rm cl}^a$ in
(\ref{eq:Lclassical}) and do calculations {\em non-perturbatively}.
Ordinarily, such non-perturbative calculations would not be possible.
However, it is possible in this case because
$\calL_{\rm cl}^a$ is just another classical Lagrangian.
Integrating over $\eta_\mu$ 
enforces the full non-linear Jacobi field equation
for $a^\mu$ so that if $\calA^\mu$ is a solution of the classical field
equation with the source term $\calJ^\mu$, then
so is $A^\mu = \calA^\mu + a^\mu$ with the
source term $\calJ + \deltaJ$.
Quantum effect comes in via the tadpole 
$\lim_{y\to x}\ave{a^\mu(x)a^\nu(y)}$.
In Section~\ref{sec:qcd_props} as well as in Appendices
\ref{app:Scalar_props} and \ref{app:props_details}, it is shown
that the symmetric propagator, equivalently the $\ave{a^\mu(x)a^\nu(y)}$ 
correlator, is non-zero only if 
the initial vacuum contains the zero-point oscillation
of the momentum modes as in Eq.(\ref{eq:rhovac}) for the scalar field.
In this way, we have reduced the problem of finding the first order quantum
corrections, and also resumming all possible small-$x$ enhancements,
to a completely classical problem where quantum effect only resides 
in the fluctuations of initial conditions. Equivalently, in the fact
that $G_{S} \ne 0$.

The 2-point correlation function is given by
\be
\ave{A_\lambda(x)A_\sigma(y)}
& = &
\calA_\lambda(x) \calA_\sigma(y)
+\calA_\lambda(x) \ave{a_\sigma(y)}
+\calA_\sigma(y) \ave{a_\lambda(x)}
+\ave{a_\lambda(x)a_\sigma(y)}
\non & & {}
+\ave{a_\lambda(x)}\ave{a_\sigma(y)}
\label{eq:aveAA}
\ee
Since $\ave{a} = O(1/g)$ with the log enhancement,
one can see that all terms in Eq.(\ref{eq:aveAA}) can contribute
at $O(1/g)$ except for
the symmetric propagator $\ave{a_\lambda(x)a_\sigma(y)}$.
When we work out the derivation of the JIMWLK equation
in later sections (in particular Section~\ref{sec:2pt_corr}),
we will show that the leading order symmetric propagator is also
log-enhanced to become $O(1/g^2)$ when $x^- = y^-$. 
Therefore, the conclusion is the same as the 1-point average case: 
Solving classical Yang-Mills equation with quantum initial conditions
is enough for the leading-order plus leading-log calculations.
Again, triple $\eta$ vertices cannot
contribute since they do not result in non-vanishing tadpoles.

This pattern should hold for any $n$-point function.
For instance, consider the 3 point function $\ave{AAA}$.
The terms $\calA\calA\calA$, $\calA\calA\ave{a}$ and $\calA\ave{aa}$
are all $O(1/g^3)$. 
The 3-point correlator $\ave{aaa}$ is $O(g)$.
This is because $aaa$ can connect to either 
$g\eta\calD aa$ or $g\eta\calD\eta\eta$ vertex.
The resulting $\ave{aa}$ propagators do not become large because
they are not evaluated at equal times.
Hence, again all one needs are $\ave{a(x)}$ and $\ave{a(x)a(y)}$.
The triple $\eta$ vertices does not give rise to a leading order quantum
correction.

The estimate given above is for the case of a single nucleus
where the system is essentially static.
For the system created by two colliding heavy ions, 
the classical field $\calA_\mu$ evolves in time towards
the formation of QGP. In that case, 
the symmetric propagator $G_S(x|y) = \ave{a(x)a(y)}$
evaluated at $x=y$ can potentially exhibit
parametric resonances.
These can be again resummed by solving the classical Yang-Mills equation
\cite{Dusling:2010rm}. 
In this paper, we concentrate on the static single nucleus case and leave
the time-evolving background case for future studies.

In summary, in the presence of a strong source,
it is enough to consider just the classical part of the full 
Schwinger-Keldysh
Lagrangian:
\be
\calL_{\rm cl}
=
\eta_\nu^a \left( [D_\mu, G^{\mu\nu}]_a - J_{a}^\nu\right) 
\ee
together with initial state vacuum fluctuations 
for the leading order (LO) and the leading-log resummation
(following Ref.\cite{Gelis:2008rw} we call this ``NLO'').
The vacuum average of an operator $\calO$ is now
\be
\ave{\calO}_{\rm LO+NLO}
& = &
\int [da_i][d\pi_i] \rho_{\rm v}[a_i, \pi_i]\,
\int \calD A\, \calD \eta\, 
\exp\left(
i\int \calL_{\rm cl}
\right)
\calO[A]
\label{eq:calO}
\ee
where $a_i$ and $\pi_i$ are the initial values of $A$ and its momentum
and $\rho_{\rm v}[a_i, \pi_i]$ is the distribution of $a_i$ and $\pi_i$ due to
the vacuum fluctuations.
Since $\calL_{\rm cl}$ is linear in $\eta$, this integral can be carried out
yielding
\be
\ave{\calO}_{\rm LO+NLO}
& = &
\int [da_i][d\pi_i] \rho_{\rm v}[a_i, \pi_i]\,
\calO[A]
\label{eq:OLONLO1}
\ee
where $A$ is a classical solutions with the initial conditions given
by $a_i$ and $\pi_i$.
If we separate out a particular solution $\calA_\mu$ with null initial
conditions,
\be
A_\mu = \calA_\mu + a_\mu
\ee
then the average can be re-expressed as
\be
\ave{\calO}_{\rm LO+NLO}
& = &
\int [da_i][d\pi_i] \rho_{\rm v}[a_i, \pi_i]\,
\calO[\calA + a]
\label{eq:OLONLO2}
\ee
where $\calA_\mu$ depends on $\calJ^\mu$ via the equation of motion
$[\calD_\mu, \calG^{\mu\nu}] = \calJ^\nu$
and the fluctuation field $a_\mu$ satisfies
the non-linear Jacobi-field equation
(restoring $\delta J$)
\be
0
& = &
[\calD_\mu, [\calD^\mu, a^\nu]]
- [\calD^\nu, [\calD_\mu, a^\mu]]
+2ig [\calG^{\mu\nu}, a_\mu]
\non & & {}
-2ig [a_\mu, [\calD^\mu, a^\nu]]
+ig [a^\nu, [\calD_\mu, a^\mu]]
\non & & {}
+ig [a_\mu, [\calD^\nu, a^\mu]]
-g^2[a_\mu, [a^\mu, a^\nu]]
\non & & {}
- \delta J^\nu
\label{eq:calI}
\ee
This is of course, further subject to gauge fixing.

Let's now consider $\delta J$.
We have defined $\delta J$ to be the difference between the source terms
that satisfy $[D_\mu, J^\mu] = 0$ and $[\calD_\mu, \calJ^\mu] = 0$.
Explicitly,
\be
0
& = &
[D_\mu, J^\mu]
\non
& = &
[\calD_\mu - iga_\mu, \calJ^\mu + \delta J^\mu]
\non
& = &
[\calD_\mu, \delta J^\mu] -i[a_\mu, g\calJ^\mu] -ig[a_\mu, \delta J^\mu]
\ee
To be more concrete, we need to choose the form of $J^\mu$, $\calJ^\mu$ 
and also the
gauge condition for $A^\mu$. 
To anticipate next section's analysis, we choose
the current to be in the light-cone $x^+ = (t+z)/\sqrt{2}$ direction 
and independent of $x^+$,
$\calJ^\mu = \delta^{\mu+} \rho(x^-,\bfxperp)$, and the light-cone 
gauge condition to be $A^+ = 0$. 
Here $x^- = (t-z)/\sqrt{2}$.
As we will see in the next section,
this leads to $\calA^\pm = 0$ although $a^- = a_+\ne 0$.
Choosing an ansatz where only $\delta J^+$ is non-zero,
we get
\be
[\partial_+ - iga_+, \rho + \deltaJ^+] = 0
\ee
which is satisfied by
\be
J^+ = \rho + \deltaJ^+ = W(x) \rho(x^-,\bfxperp) W^\dagger(x)
\ee
where
\be
W(x) = \calP_{x^+} 
\exp\left(ig\int_{-\infty}^{x^+} dy^+\, a_+(x^-,y^+,\bfxperp)\right)
\ee
is a Wilson line in the $x^+$ direction
and $x^- = (t-z)/\sqrt{2}$.

It is instructive to compare our final expression
for the external source term in the light cone gauge
\be
S_{\eta J}
&= &
-\int d^4x\, \eta_+(x) W(x)\rho(x^-,\bfxperp)W^\dagger(x) 
\non
&= &
-\int d^4x\, \eta^a_+(x) \ulW_{ab}(x)\rho_b(x^-,\bfxperp)
\label{eq:ourS}
\ee
with that used previously in Ref.\cite{Iancu:2000hn}
\be
S_W 
& = &
-\int_C dz^+
\int dx^-\int d^2x_\perp
a^a_+(x^-, z^+, \bfxperp)
\ulW_{ab}(x^-,z^+,\bfxperp)
\rho_b(x^-,\bfxperp)
\label{eq:IancuSW}
\ee
where $C$ is the Schwinger-Keldysh contour in $x^+$. 
Here and after, underlines indicate
quantities in the color vector space.\footnote
{
Translations between the Lie-algebra 
representation and the color vector notation are as follows. 
Commutator between two Lie-algebra elements become
$[O, v]_a = \ulO_{ab}v_{b}$ where
$\ulO_{bc} = O_a (T^a)_{bc}$.
Here $(T^a)_{bc} = -if_{abc}$ is the $a$-th generator in the adjoint
representation.
In particular, 
$[\calD_\mu, v]_a = \ul{\calD}_{ab} v_b$
where
$\ul{\calD}_\mu
= \partial_\mu - ig\ul{\calA_\mu}$.
If $U = e^{i\theta_a t_a}$ is a group element,
$(U v U^\dagger)_a = \ulU_{ab}v_b$
where $\ulU = e^{i\theta_a T^a}= e^{i\ul{\theta}}$.
} 
Separating out the contours 1 and 2, this becomes
\be
S_W
& = &
-\int_{-\infty}^{\infty} dz^+
\int dx^-\int d^2x_\perp
a_{1,a}^-(x^-, z^+, \bfxperp)
\ulW_1^{ab}(x^-,z^+,\bfxperp)
\rho_b(x^-,\bfxperp)
\non & & {}
+
\int_{-\infty}^{\infty} dz^+
\int dx^-\int d^2x_\perp
a_{2,a}^-(x^-, z^+, \bfxperp)
\ulW_2^{ab}(x^-,z^-,\bfxperp)
\rho_b(x^-,\bfxperp)
\ee
where 
\be
\ulW_{1,2}(x)
= \exp\left(ig\int_{-\infty}^{x^+} dy^+ \ula^-_{1,2}(x^-,y^+,\bfxperp)\right)
\ee
This is almost, but not quite, the same as Eq.(\ref{eq:ourS}): 
The Wilson lines are in terms of $a^-_{1,2}$ instead of
$a^- = (a^-_1+a^-_2)/2$. 
Another difference is that in $S_W$, $a^-_{1,2}$ in front of the Wilson line
can be replaced by a derivative in $x^+$ so that upon integration in $x^+$
it can go back to the ${\rm Tr}J^+ W$ form,
while $S_{\eta J}$ does not permit this.

In Ref.\cite{Gelis:2008rw}, the authors advocated performing calculations in
the axial $A^- = 0$ gauge 
and gauge-rotating the results to the light-cone gauge afterwards
within the $A^\mu_{1,2}$ formalism (the authors call them $A^\mu_{\pm}$)
and took the time-evolution parameter to be $x^-$.
In some sense, this was done to avoid using $S_W$ explicitly.
Actually, the authors of Ref.\cite{Gelis:2008rw}
does not start their investigation at the
Lagrangian level, but at the classical field equation level.
Therefore, in effect they are essentially using Eq.(\ref{eq:ourS})
and their $a_\pm^\mu$ are effectively the common field $a^\mu$.

Now that we have established Eq.(\ref{eq:OLONLO2}), many further
investigations can be performed. Two main ones that are relevant to 
the study of heavy ion collisions are:
(i) The quantum evolution of gluon density in the initial state nuclei
{\em before} the collision and
(ii) the temporal evolution of gluon field {\em after} the collision. 
In the rest of this paper, we will deal with only the renormalization group
evolution of the nuclear gluon density before the collision.
The time evolution of the gluon field after the collision
will be discussed elsewhere.

For the study of the initial state of nuclei, the color current $\calJ^+$
represents the fast-moving hard partons while the classical field $\calA$
represents the abundant soft partons. 
Due to time dilation, the dynamics of fast
partons making up the color current is very slow and hence
can be regarded as static charge configuration.
For each collision, 
the distribution of the color charge density will be different. 
Therefore, the complete event
average of an observable requires averaging over the color charge
configurations
\be
\ave{\ave{\calO}}_{\rm LO+NLO}
& = &
\int \calD\rho\,W_{\rho}[\rho]
\int [da_i] \rho_{\rm v}[a_i]\,
\calO[\calA + a]
\label{eq:OLONLO2J}
\ee
where we switched to a more conventional notation for the color charge
density $\rho= \calJ^+$ and $W_{\rho}[\rho]$ is the distribution of the
color charges in a nucleus.
Here it is understood that $\calA$ satisfies the classical Yang-Mills
equation with the source $\rho(x^-,\bfxperp)$ 
for $x^- > 0$, but vanishes for $x^- < 0$.
The fluctuation field $a$
satisfies the Jacobi-field equation in the background of $\calA$
for all $x^-$.
The combined field $A = \calA + a$ also satisfies the classical Yang-Mills equation
with the boundary conditions at $x^- = x^-_{\rm init}$ 
specified by $a_i$. We do not need the initial condition for the
momentum field in the light-cone coordinate system because 
$\partial^2 = 2\partial_+ \partial_- - \nabla_\perp^2$ contains only one
$x^-$ derivative.

\section{Classical solution}
\label{sec:ClassicalSolution}

In the original McLerran Venugopalan (MV) model 
\cite{McLerran:1993ni,
McLerran:1993ka, 
McLerran:1994vd,
Ayala:1995kg,
Ayala:1995hx},
it was argued that classical Yang-Mills theory can
describe the density of small $x$ gluons in a heavy nucleus.
This was accomplished by considering the valance quarks to be static random
color sources that generate a strong gluonic Weizsacker-Williams field.
The formulation of the effective field theory of soft gluons 
by McLerran and Venugopalan is one of the major development
in many-body QCD in the recent years.
Subsequent developments in the
study of the initial states of the heavy ion collisions,
usually dubbed Color Glass Condensate (CGC),
owe much to the original McLerran-Venugopalan (MV) model.

The MV model calculates the gluon parton density in a heavy nucleus 
in terms of the classical Yang-Mills 
field~\cite{JalilianMarian:1996xn,Kovchegov:1996ty,Kovchegov:1997ke}.
However, 
as is well known, quantum fluctuations must play an important role 
in the scale dependence of a parton density function via renormalization
group equation.  The idea behind the JIMWLK equation is that this
renormalization group equation can be realized by studying
how the color charge density changes as the scale changes. 
To go further, we first need the classical solution.
In this section, we repeat the derivation of the MV solution mainly 
to fix the notations and sign conventions. Readers familiar with
the MV solution may skip this section.

To find a classical solution, we first need to fix the
coordinate system and the gauge. 
The light-cone coordinate system
\be
x^\pm = x_\mp =  {t \pm z\over \sqrt{2}}
\ee
provides a convenient description in our case because the particles
are moving at nearly the speed of light in ultrarelativistic heavy ion
accelerators such as RHIC or the LHC.
The metric for this part is off-diagonal and symmetric, with the non-zero
components given by $g^{\pm\mp}=g_{\pm\mp} = 1$.
The transverse part of the metric 
is diagonal with the non-zero components given by
$g^{11} = g^{22} = -1$. The d'Alembertian is 
\be
\partial_\mu \partial^\mu = 2\partial_+ \partial_- - \nabla_\perp^2
\ee
For our analysis, we choose the
direction of movement to be in the positive $z$ direction.
In other words, the nucleus occupies the $x^+$ axis.

When a particle is moving in the positive $z$ direction with the speed of
light, $x^+ = (t{+}z)/\sqrt{2}$ plays the role of physical time
since this is what determines the location of the particle at any given
moment.  
In our case, the system is static in the sense that it does not depend on
$x^+$. The evolution we are interested in is therefore
not the evolution in time, but the renormalization group flow.
Hence, it makes sense for us to consider 
$x^-$ as the ``evolution'' parameter since it is the conjugate of $p^+$.
This also makes a clean separation of the region 
behind the nucleus ($z < t$ or $x^- > 0$) which is already influenced by it,
and the region ahead of nucleus ($z > t$ or $x^- < 0$) which is not
yet influenced by it.  In the rest of the paper,
we will often refer to $x^-$ as the ``time''.

When the source current\footnote{
Both the color charge density and the vacuum density functional are denoted
by $\rho$ in this paper. The distinction should be clear from the context.
}
is given by
\be
\calJ^\mu_a(x) = \delta^{\mu +} \rho_a(x)
\ee
the light-cone gauge, $\calA^+ = 0$, is the most physical one.
However, this is not the most convenient gauge for explicit calculations
since the classical source term necessarily precess in this gauge.
Instead, it is more convenient to use the axial $\tildecalA^- = 0$ gauge
since  
the covariant conservation law then becomes independent of $\tildecalA_\mu$
\be
[\tildecalD_\mu, \tildecalJ^\mu]
&=& \partial_\mu \tildecalJ^\mu -ig[\tildecalA_\mu, \tildecalJ^\mu] 
\non
&=& \partial_+ \tilderho = 0
\label{eq:axialDJ}
\ee
which implies that the source $\tilderho$ is independent of $x^+$ and does
not precess.
We can always gauge-transform back to the light-cone gauge when necessary.
From now on, we will denote quantities in the axial
gauge with a tilde as in Eq.(\ref{eq:axialDJ}).
Quantities in the light-cone gauge will remain without any adornment.

The issue of whether $\tildecalA^- = \tildecalA_+ = 0$ 
gauge is allowed is a non-trivial one when the fields do not depend on
$x^+$.  To remove the $A_+$ component, we need to use the gauge transformation
$\tildecalA_+ = U\calA_+ U^\dagger + (i/g)U\partial_+ U^\dagger$.
However, if we want to keep the fields to be independent of $x^+$ 
even after the gauge transformation, $U$ cannot depend on $x^+$. 
This implies that we cannot normally
transform from the $\calA^+ = 0$ gauge
to the $\tildecalA^- = 0$ gauge without an explicit $x^+$ dependence.
The only way out of this is if in the $\calA^+ = 0$ gauge, $\calA^- = 0$ is
{\em an allowable solution} of the equations of motion. Then
$U\calA_+ U^\dagger + (i/g)U\partial_+ U^\dagger = 0$.
is trivial when $U^\dagger$ is independent of $x^+$.
Fortunately, with the current in the $x^+$ direction,
this is indeed the case for us. 

The classical Yang-Mills equation in the axial gauge is 
\be
[\tildecalD_\mu, \tildecalG^{\mu\nu}] = 
(i/g)[\tildecalD_\mu, [\tildecalD^\mu, \tildecalD^\nu]] = 
\delta^{\nu +}\tilderho(x^-, \bfx_\perp) 
\label{eq:DGrho}
\ee
Since $\tilderho$ is independent of $x^+$, we now 
look for a classical solution 
$\tildecalA^\mu(x^-,\bfxperp)$ which is also independent of $x^+$.
With this ansatz and our gauge choice $\tildecalA^- = \tildecalA_+ = 0$, 
we have $\tildecalG_{+\mu} = 0$.
The equations of motion then simplifies to
(with $i=1,2$)
\be
[\tildecalD_i, \tildecalG^{i+}] &=& \tilderho
\non {}
[\tildecalD_i, \tildecalG^{ij}] &=& 0
\ee
The second equation implies that the transverse components of the
classical solution is pure gauge which can be set to zero.
The only remaining equation is
\be
[\partial_i, \partial^i \tildecalA^+]
&= &
-\nabla_\perp^2 \tildecalA_-(x^-,\bfx_\perp)
\non
& = &
\tilderho(x^-, \bfx_\perp)
\ee
This has a simple solution
\be
\tildecalA^+(x^-,\bfx_\perp) = 
\tildecalA_-(x^-,\bfx_\perp) = 
\int_{\bfyperp} G_T(\bfxperp - \bfyperp)
\tilderho(x^-,\bfy_\perp)
\label{eq:Aplus}
\ee
where the transverse Green function satisfying 
$-\nabla_\perp^2 G_T = \delta$
is given by
\be
G_T(\bfxperp - \bfyperp)
& = &
-{1\over 2\pi} \ln\left({|\bfx_\perp-\bfy_\perp|\over R}\right)
\ee
where $R$ is a constant that sets the boundary condition in the transverse
plane.
For the null boundary condition in $x^-$,
$
\displaystyle
\lim_{x^-\to x^-_{\rm init}}\tildecalA^+(x^-,\bfxperp) = 0
$,
to be satisfied, it is enough to require 
$\displaystyle\lim_{x^-\to x^-_{\rm init}}\tilderho(x^-,\bfxperp) = 0$.
Here and after, we will often use the following
short-hands for coordinate space integrals
\be
\int_y \equiv \int_{x^-_{\rm init}}^\infty dy^-\int d^2y_\perp dy^+
\ \ \hbox{and}\ \ 
\int_{\bfy} \equiv \int d^2y_\perp dy^+
\ \ \hbox{and}\ \ 
\int_{\bfyperp} \equiv \int d^2y_\perp 
\ee
The initial time $x^-_{\rm init}$ is to be in the far past.
In fact, except when we solve initial value problems,
the limit $x^-_{\rm init}\to -\infty$ should be understood.
When the initial time is expressly needed, we will indicate it explicitly.

To go back to the light-cone gauge, we need to perform a gauge
transformation
\be
\calA_\mu = V \tildecalA_\mu V^\dagger + (i/g)V \partial_\mu V^\dagger
\ee
The light-cone gauge condition $\calA^+ = 0$ is solved by
a path-ordered exponential
\be
V^\dagger(x^-,\bfxperp)
& = &
\calP
\exp\left(
ig\int_{-\infty}^{x^-} dy^-\, \tildecalA_-(y^-,\bfxperp)
\right)
\label{eq:Vdaggersol}
\ee
The $\calA^-$ component also vanishes
because $\tildecalA^- = \tildecalA_+ = 0$
and $V$ is independent of $x^+$.
This is in accordance with our discussion above that $\calA^- = 0$
must be an allowed solution of the equations of motion.
The spatial components $\calA_i$ are the only non-zero components
and are given by
\be
\calA_i = (i/g)V\partial_i V^\dagger
\ee

For the derivation of the JIMWLK renormalization group equation,
we further specify the color density to be confined in a narrow strip
between $x^- = 0$ and $x^- = {\epsilon_Y}$.
The Wilson line is then given by
\be
V^\dagger(x^-,\bfxperp)
=
\left\{
\begin{array}{ll}
1  & \hbox{for $x^- \le 0$}
\\
\calP\exp\left(
ig\int_{0}^{x^-} dz^- \tildecalA^+(z^-,\bfxperp)
\right)
& \hbox{for $0 < x^- \le {\epsilon_Y}$}
\\
V^\dagger(\epsilon_Y,\bfxperp)
& \hbox{for $x^- > {\epsilon_Y}$}
\end{array}
\right.
\ee

We should note that the solution for $V^\dagger$, Eq.(\ref{eq:Vdaggersol}),
relies on the behavior of
a time-ordered exponential under a time derivative. Namely,
\be
\partial_- V^\dagger(x^-,\bfxperp) =
ig\tildecalA^+(x^-,\bfxperp)V^\dagger(x^-,\bfxperp)
\label{eq:partialminusVdagger}
\ee
If we have a true delta-function in $\tilderho$ as in
the original MV model, $V^\dagger$ cannot take the time-ordered
exponential form in Eq.(\ref{eq:Vdaggersol}).
This stems from the fact that $\theta(x^-)^n = \theta(x^-)$. Hence
$\exp(\theta(x^-)i\alpha(\bfxperp)) 
= \theta(x^-) \exp(i\alpha(\bfxperp)) + \theta(-x^-)$.
It is easy to verify that this form does not satisfy 
Eq.(\ref{eq:partialminusVdagger}). 
Therefore, $\tilderho(x^-,\bfxperp)$ should always
be regarded as having a non-zero support in $x^-$ if we want to go back
to the light-cone gauge.

\section{Transformation Between the Light-Cone Gauge and the Temporal Gauge
for the fluctuation fields}
\label{sec:lightconevsaxial}

For the classical solution we have considered in the last section, 
going from the axial to light-cone or vice-versa was accomplished by
applying gauge transformation with the Wilson line 
$V^\dagger(x^-,\bfxperp)$.
So far, however, we have not fixed any gauge for the fluctuation
field $a$.  

The classical Lagrangian
\be
\calL_{\rm cl} = \eta_\nu\Big (i/g) [D_\mu, [D^\mu, D^\nu]] - J^\nu \Big)
\label{eq:Lclassical2}
\ee
with $D_\mu = \partial_\mu -ig(\calA_\mu + a_\mu)$
is invariant under the gauge transformation
\be
\calA'_\mu + a_\mu'
= S\calA_\mu S^\dagger + Sa_\mu S^\dagger + (i/g)S\partial_\mu S^\dagger
\label{eq:gaugetr}
\ee
and $\eta_\mu' = S \eta_\mu S^\dagger$
where $S$ is an arbitrary gauge group element.
How the terms in the right hand side
of Eq.(\ref{eq:gaugetr})
is associated with the terms in the
left hand side of Eq.(\ref{eq:gaugetr}) is actually quite arbitrary.
One possibility is
\be
\calA_\mu' &=& S\calA_\mu S^\dagger
\non
a'_\mu & = & Sa_\mu S^\dagger + (i/g)S\partial_\mu S^\dagger
\label{eq:rule1}
\ee
Then we may choose a gauge
condition for $a$ by choosing a suitable $S$. However, in doing so,
we ruin the fact that the background field is a solution of the classical 
equation of motion. This is undesirable.
Another possibility is to set 
\be
\calA_\mu' &=& S\calA_\mu S^\dagger + (i/g)S\partial_\mu S^\dagger
\non
a'_\mu & = & Sa_\mu S^\dagger 
\label{eq:rule2}
\ee
In this way, we do maintain the fact that the background field $\calA'_\mu$
is the solution of the gauge-transformed classical equation of motion.
But we lose the possibility of eliminating a component of $a_\mu$ which is
not very convenient.
In order not to lose the property 
that $\calA$ is a classical solution and yet fix the gauge for the
fluctuation field, we need yet another grouping 
\cite{Iancu:2000hn}
\be
\calA'_\mu &=& \calA_\mu
\non
a'_\mu & = &
Sa_\mu S^\dagger + (i/g)S\partial_\mu S^\dagger
+ S\calA_\mu S^\dagger - \calA_\mu
\label{eq:rule3}
\ee
Recall that $\calA_\mu = O(1/g)$ and $a_\mu = O(1)$.
Therefore as long as $S = 1+ O(g)$, $a'_\mu$ remains to be $O(1)$.
This condition is equivalent to the residual gauge fixing condition
in Ref.~\cite{Gelis:2008rw}.

Transformation from the axial gauge to the light-cone gauge 
can be now done in two steps.
First, apply Eq.(\ref{eq:rule2}) to get the classical solution
in the light-cone gauge $\calA_\mu$ 
from the axial gauge solution $\tildecalA_\mu$
\be
\calA_\mu &=& V \tildecalA_\mu V^\dagger + (i/g) V\partial_\mu V^\dagger
\\
\acutea_\mu &=& V\tildea_\mu V^\dagger
\ee
where $\acutea_\mu$ is still in the axial gauge.
To eliminate $a^+$, we then apply Eq.(\ref{eq:rule3}).
\be
a^\mu 
=
S \acutea^\mu S^\dagger 
+ (i/g) S \partial^\mu S^\dagger 
+ S \calA^\mu S^\dagger
- \calA^\mu
\ee
The light cone gauge condition $a^+ = a_- = 0$ yields
\be
0 = 
S \acutea^+ S^\dagger 
+ (i/g) S \partial^+ S^\dagger 
\label{eq:acutea}
\ee
Since $\partial^+ = \partial_-$,
the solution is
\be
S^\dagger(x) =
\calP\exp\left(ig\int_{-\infty}^{x^-}dz^-\,\acutea^+(z^-,\bfx)\right)
\ee
where we defined $\bfx = (x^+,\bfxperp)$.
Other components are
\be
a_i &=&
S\acutea_i S^\dagger + (i/g)S\partial_i S^\dagger
+ S\calA_i S^\dagger - \calA_i
\non
a^- & = & (i/g)S\partial_+ S^\dagger
\label{eq:lightconea}
\ee
The last line follows because
$\acutea^\mu$ is in the axial gauge
and the light-cone gauge classical solution
has $\calA^- = 0$ as well as $\calA^+ = 0$
as shown in Section~\ref{sec:ClassicalSolution}.

We have thus shown that $U = SV$ transforms 
\be
U(\tildecalA_\mu + \tildea_\mu)U^\dagger
+
(i/g)U\partial_\mu U^\dagger
& = \calA_\mu + a_\mu
\ee
with $a_\mu$ given by Eq.(\ref{eq:lightconea}).

Using $\tildeS = V^\dagger S V$, a proper gauge transformation 
can be obtained between a light-cone gauge field
$\bara_\mu = V^\dagger a_\mu V$ and the axial gauge field
$\tildea_\mu$ as follows
\be
\bara_\mu &=&
\tildeS \tildea_\mu \tildeS^\dagger 
+ (i/g)\tildeS \partial_\mu \tildeS^\dagger
\ee
where we used the fact that $V$ does not depend on $x^+$.
Noting that
\be
&&
V^\dagger(x^-,\bfxperp) 
\acutea_-(z_1^-,\bfx) 
\acutea_-(z_2^-,\bfx) 
\cdots
\acutea_-(z_n^-,\bfx) 
V(x^-,\bfxperp)
\non
& = &
V^\dagger(x^-,\bfxperp) 
\acutea_-(z_1^-,\bfx) 
V(x^-,\bfxperp)
\cdots
V^\dagger(x^-,\bfxperp) 
\acutea_-(z_n^-,\bfx) 
V(x^-,\bfxperp)
\ee
we can re-express $\tildeS^\dagger$ as
\be
\tildeS^\dagger = 
\calP\exp\left(
ig\int_{-\infty}^{x^-}dz^-\,
V^\dagger(x^-, z^-;\bfxperp) \tildea^+(z^-,\bfxperp) V(x^-,z^-;\bfxperp)\right)
\ee
where
\be
V^\dagger(x^-,z^-;\bfxperp)
& = &
V^\dagger(x^-,\bfxperp) V(z^-,\bfxperp)
\non
& = &
\calP \exp\left(ig\int_{z^-}^{x^-} d{z'}^-\, 
\tildecalA_-({z'}^-,\bfxperp)\right)
\ee
is the incomplete Wilson line defined for $x^- > z^-$.

For our derivations in the later sections,
the leading order expression for $\tildeS^\dagger$ is sufficient
\be
\tildeS^\dagger
& \approx &
1 + ig \tildeomega(x)
\ee
where
\be
\tildeomega(x) = V^\dagger(x)\omega(x) V(x)
=
\int_{-\infty}^{x^-} dz^-\, V^\dagger(x^-, z^-;\bfxperp)
\tildea^+(z^-,\bfx) V(x^-,z^-;\bfxperp)
\label{eq:bdef}
\ee
with
\be
\omega(x) 
=
\int_{-\infty}^{x^-} dz^-\, 
\acutea^+(z^-,\bfx)
\label{eq:omega}
\ee
To this order, the light-cone fields are 
\be
a_i &=& V\bara_i V^\dagger 
=
V(\tildea_i - \partial_i \tildeomega)V^\dagger
+ O(\tildea^2)
\label{eq:baraeq1}
\\
a_+ &=& V\bara_+ V^\dagger 
= - V(\partial_+ \tildeomega)V^\dagger
+ O(\tildea^2)
\label{eq:baraeq2}
\ee
Or collectively,
\be
a_\mu = 
V\tildea_\mu V^\dagger - [\calD_\mu, \omega]
+ O(\tildea^2)
\label{eq:baraeqmu}
\ee
Ref.\cite{Iancu:2000hn} has equivalent transformation rules
for classical fields where 
$\omega(x) = -\int_{-\infty}^{x^+} dz^+\, a^-(x^-,z^+,\bfxperp)$
is used instead of Eq.(\ref{eq:omega}).  
This alternative solution for $S$ comes from solving  
Eq.(\ref{eq:lightconea}) instead of Eq.(\ref{eq:acutea}).
Transformation rules using Eq.(\ref{eq:omega}) was also identified 
in Ref.\cite{Gelis:2008rw} although both $a_1^\mu$ and $a_2^\mu$, 
that is both $a^\mu$ and $\eta^\mu$,
were transformed in Ref.\cite{Gelis:2008rw}. 
We have shown here that only the common field
$a^\mu$ transforms this way. The difference field just gets rotated.

For later convenience, we list here two mixed representation fields.
The following is the light-cone 
gauge field in the axial-gauge background field
\be
\bara_\mu = V^\dagger a_\mu V \approx \tildea_\mu - [\tildecalD_\mu, \tildeomega]
\ee
and the axial gauge field in the light-cone gauge background field is
\be
\acutea_\mu = V\tildea_\mu V^\dagger \approx a_\mu + [\calD_\mu, \omega]
\ee

\section{Propagators in the light-cone coordinates}
\label{sec:qcd_props}

Derivation of the JIMWLK equation requires 2-point correlation functions of
$a_\mu$, or the propagators, in the background of $\calA_\mu$.
Since the object to calculate is the gluon density, 
we only need the auto-correlator, or equivalently the symmetric propagator
$\ave{a_\mu(x)a_\nu(y)}$.
The symmetric propagators are in turn expressed in terms of the retarded
propagator.
In this section, we explicitly calculate the QCD
propagators in the axial gauge. Propagators in the light-cone
gauge can be obtained by gauge transforms. 
In Section \ref{subsec:formal}, we derive equations to solve
and develop formal relationships between
propagators.
In Section \ref{subsec:explicit_cal},
an explicit derivation of the propagators is given.

\subsection{Propagators in the axial gauge}
\label{subsec:formal}

We start with 
the bi-linear Lagrangian in the axial gauge with appropriate source
terms for the generating functional
\be
\calL_J
& = &
\tildeeta_+ [\tildecalD_\mu, [\tildecalD^\mu, \tildea^+]]
- \tildeeta_+  [\tildecalD^+, [\tildecalD_\mu, \tildea^\mu]]
-2ig \tildeeta_+ [\tildecalG^{+i}, \tildea_i]
\non & & {}
+
\tildeeta_i [\tildecalD_\mu, [\tildecalD^\mu, \tildea^i]]
- \tildeeta_i  [\tildecalD^i, [\tildecalD_\mu, \tildea^\mu]]
\non & & {}
- \tildeeta_-  [\tildecalD^-, [\tildecalD_\mu, \tildea^\mu]]
\non
& & {}
- \tildeJ_i \tildea^i
- \tildeJ_+ \tildea^+
- \tildeJ_\eta^i \tildeeta_i
- \tildeJ_\eta^+ \tildeeta_+
\ee
Recall that our analysis is in the context of a path integral.
Therefore, integrating out $\eta_\mu$ gives a $\delta$-functional that
enforces the $\mu$ component of the equations of motion.
The resulting linear Jacobi-field equations are
\be
\tildeJ_\eta^+
& = & [\tildecalD_\mu, [\tildecalD^\mu, \tildea^+]]
- [\tildecalD^+, [\tildecalD_\mu, \tildea^\mu]]
-2ig [\tildecalG^{+i}, \tildea_i]
\label{eq:etapluseq}
\\
\tildeJ_\eta^i
& = &
[\tildecalD_\mu, [\tildecalD^\mu, \tildea^i]]
- [\tildecalD^i, [\tildecalD_\mu, \tildea^\mu]]
\label{eq:etaieq}
\\
0
& = &
- \partial_+ [\tildecalD_\mu, \tildea^\mu]
\label{eq:etaminuseq}
\ee
Applying $\partial_+$ to Eq.(\ref{eq:etapluseq}) and
$\partial_i$ to Eq.(\ref{eq:etaieq}) and summing the results,
we get
\be
\partial_+ \tildeJ_\eta^+ 
+\partial_i \tildeJ_\eta^i 
=
[\tildecalD_-, \partial_+ [\tildecalD_\mu, \tildea^\mu]]
= 0
\ee
which is consistent with the covariant conservation law
$[\tildecalD_\mu, \tildeJ^\mu] = 0$.
The solution of Eq.(\ref{eq:etaminuseq}) is
\be
[\tildecalD_\mu, \tildea^\mu] = 
\partial_+ \tildea^+ 
+
\partial_i \tildea^i 
=
\tildeh(x^-,\bfxperp)
\label{eq:partialaeqh}
\ee
where $\tildeh$ is independent of $x^+$. Since the region $x^- < 0$ is
vacuum, it is consistent to set $\tildeh = 0$.

With the above constraint,
the transverse component of the equation of motion is simply two
independent scalar equations
\be
[\tildecalD_\mu, [\tildecalD^\mu, \tildea^i]] = \tildeJ_\eta^i 
\label{eq:DDai}
\ee
Applying
$\int_{x^-_{\rm init}}^\infty dy^-\int_\bfy \tildeG_R(x|y)$ 
to Eq.(\ref{eq:DDai}) from left 
and performing integrations by parts, we obtain the formal solution
\be
\tildea_a^i(x) = \tildealpha_a^i(x)
-
\int_{x^-_{\rm init}}^\infty dy^-
\int_\bfy\tildeG_R^{ab}(x|y)
\tildeJ_{\eta,b}^i(y)
\label{eq:tildeaisol}
\ee
where 
\be
\tildealpha_a^i(x)
& = &
2\int_{\bfy}\, 
(\partial_{y^+}\tildeG_{R}^{ab}(x|x_{\rm init}^-,\bfy))
\tildea_b^i(x_{\rm init}^-,\bfy)
\label{eq:alphaieq}
\ee
is the homogeneous solution coming from the surface term in carrying out the
integration by parts.
Here
$a$ and $b$ are color indices and
$\tildeG_R(x|y)$ is the retarded Green function
in the color vector space satisfying
\be
(\ul{\tildecalD}_\mu\ul{\tildecalD}^\mu)_x
\tildeG_R(x|y)
= -\delta(x-y)
\mbf{1}_{\rm adj.}
\label{eq:DDGRr}
\ee
and
\be
\tildeG_R(x|y)
(\ul{\tildecalD}_\mu\ul{\tildecalD}^\mu)^\dagger_y
= -\delta(x-y)
\mbf{1}_{\rm adj.}
\label{eq:DDGR}
\ee
where the underline in $\ul{\tildecalD}_\mu$ indicates that the 
covariant derivative is in the color vector notation,
and the derivatives on the right
should be interpreted in the sense of integration by parts.
The identity matrix $\mbf{1}_{\rm adj.}$ in the color space 
has the dimension $(N^2{-}1)\times (N^2{-}1)$.

Since $\tildeG_R(x|y)$ is the retarded propagator,
we can set
$\tildeG_R(x|y) = \theta(x^- - y^-)\tildeg_R(x|y)$.
Then $\tildeg_R(x|y)$ is 
required to satisfy (omitting $\mbf{1}_{\rm adj.}$ from now on)
\be
\partial_{y^+}\tildeg_R(x^-,\bfx|x^-,\bfy) = {1\over 2} \delta(\bfx-\bfy)
\label{eq:GR_delta_id_1}
\\
\partial_{x^+}\tildeg_R(x^-,\bfx|x^-,\bfy) = -{1\over 2} \delta(\bfx-\bfy)
\label{eq:GR_delta_id_2}
\ee

The $+$ component of the equation of motion is 
\be
[\tildecalD_\mu, [\tildecalD^\mu, \tildea^+]]
-2ig [\tildecalG^{+i}, \tildea_i]
= 
\tildeJ_\eta^+ 
\label{eq:DDtildeaplus}
\ee
Again by applying $\int_{x^-_{\rm init}}^\infty dy^- \int_\bfy \tildeG_R(x|y)$ 
from the left on
Eq.(\ref{eq:DDtildeaplus}) and performing integration by parts,
we obtain
the solution for $\tildea^+$ in the color vector space as
\be
\tildea^+(x)
& = &
\tildealpha^+(x)
-
\int_{x^-_{\rm init}}^\infty dy^- \int_{\bfy} 
\tildeG_R(x|y)
\tildeJ_\eta^+(y)
\non & & {}
+ 
2ig 
\int_{x^-_{\rm init}}^\infty dy^-
\int_{x^-_{\rm init}}^\infty dz^-
\int_{\bfy, \bfz} 
\tildeG_R(x|y) 
\ul{\tildecalG}^{+i}(y) \tildeG_R(y|z)
\tildeJ_i^\eta(z) 
\label{eq:tildeaplussol}
\non
\ee
where
\be
\tildealpha^+(x)
=
2\int_{\bfy}(\partial_{y^+}\tildeG_R(x|x^-_{\rm init},\bfy)
\tildea^+(x^-_{\rm init},\bfy)
- 2ig 
\int_{x^-_{\rm init}}^\infty dy^-
\tildeG_R(x|y) \ul{\tildecalG}^{+i}(y) \tildealpha_i(y)
\ee
is the homogeneous solution.
From 
Eq.(\ref{eq:tildeaisol}) and
Eq.(\ref{eq:tildeaplussol}),
the following retarded Green functions can be easily obtained 
\be
\tildeG_R{}_{ab}^{ij}(x|y) 
 =  
{\delta \tildea_a^i(x)\over \delta \tildeJ_\eta{}_j^b(y)}
=
\ave{\tildea_a^i(x)\tildeeta_b^j(y)}
= 
g^{ij} \tildeG_R^{ab}(x|y)
\ee
and
\be
\tildeG_R{}_{ab}^{+i}(x|y) 
&= & 
{\delta \tildea_a^+(x)\over \delta \tildeJ_\eta{}_i^b(y)}
=
\ave{\tildea_a^+(x)\tildeeta_b^i(y)}
\non
& = &
-
\int_{-\infty}^{x^+} dz^+ \partial^i_y \tildeG_R(x^-,z^+,\bfxperp|y)
+
2ig\int_z \tildeG_R(x|z)\ul{\tildecalG}^{+i}(z)\tildeG_R(z|y)
\ee
where we used $\partial_+\tildeJ_\eta^+ = -\partial_i\tildeJ_\eta^i$.
This can be also obtained using
\be
\tildea^+(x) = 
-\int_{-\infty}^{x^+} dz^+ \partial_i\tildea^i(x^-,z^+,\bfxperp)
\ee
and the following identity
\be
\partial_i^x \tildeG_R(x|y)
+ 
\partial_i^y \tildeG_R(x|y)
= 
-
2ig\int_{z}
\tildeG_R(x|z)\tildecalG^{+i}(z)\partial_+^z\tildeG_R(z|y)
\ee
The homogeneous parts of the two equations also satisfy
\be
\partial_+ \tildealpha^+ + \partial_i \tildealpha^i = 0
\label{eq:tildealpharel}
\ee
due to the same identity.
Other combination of indices 
for the retarded Green function
are a bit more complicated. Fortunately, we won't need them.
The symmetric propagators 
$\ave{\tildea^i(x)\tildea^j(y)}$,
$\ave{\tildea^i(x)\tildea^+(y)}$ and
$\ave{\tildea^+(x)\tildea^+(y)}$ are also all available through
the relationship Eq.(\ref{eq:tildealpharel}) once 
$\ave{\tildea^i(x)\tildea^j(y)}$ is known.

\subsection{Explicit Calculation of Propagators}
\label{subsec:explicit_cal}

To find the explicit solutions for $\tildeG_R$ and $\tildeG_S$,
we start with the following defining equations for
the Lorentz-scalar retarded Green function $\tildeG_R$ in the color vector
space
\be
\delta(x-y)
& = &
-\partial_x^2 \tildeG_R(x|y)
+ 2ig\ul{\tildecalA}_-(x^-,\bfxperp)(\partial_+^x \tildeG_R(x|y))
\label{eq:Gr_def}
\ee
and
\be
\delta(x-y)
=
-\partial_y^2 \tildeG_R(x|y)
- 2ig (\partial_{+}^y \tildeG_R(x|y)) \ul{\tildecalA}_-(y^-,\bfyperp)
\label{eq:Gl_def}
\ee
In our study of color field generated by a nucleus,
the classical field $\tildecalA_-^a$ satisfies
$\tildecalA^a_-(x^-,\bfxperp) 
= \int_{\bfyperp} G_T(\bfxperp-\bfyperp) \tilderho_a(x^-,\bfyperp)$.
Since the nucleus is moving along the $x^+$ axis,
$\tilderho(x^-,\bfyperp)$ exists
only within a thin strip $0 < x^- < \epsilon_Y$,
and so is $\tildecalA_-(x^-,\bfxperp)$.
When both $x^-$ and $y^-$ are outside this strip, 
there is no background field between the two times. Hence, we have, 
for $0 > x^- > y^-$ and $x^-> y^- > \epsilon_Y$,
\be
\tildeG_R(x|y) = G_R^0(x|y)
\ee
In general,
\be
\tildeG_R(x|y) 
& = &
G_R^0(x|y)
-2ig\int_u
G_R^0(x|u)
\ul{\tildecalA}_-(u^-,\bfuperp)
(\partial_{u_+} \tildeG_R(u|y))
\non
& = &
G_R^0(x|y)
+2ig\int_u 
(\partial_{u_+}\tildeG_R(x|u))
\ul{\tildecalA}_-(u^-,\bfuperp)G^0_R(u|y)
\label{eq:formal_GRl2}
\ee
The first line is for Eq.(\ref{eq:Gr_def}) where $\tildecalD^2_x$ is applied
to $\tildeG_R(x|y)$ from the left
and the second line is for Eq.(\ref{eq:Gl_def})
where $(\tildecalD^2_y)^\dagger$ 
is applied to $\tildeG_R(x|y)$ from the right.
The fact that these two expressions are equivalent 
can be shown by applying $\partial_x^2$ to the second line
(which is for Eq.(\ref{eq:Gl_def}))
and showing that it also satisfies Eq.(\ref{eq:Gr_def}).

The free-field propagator $G_R^0(x|y)$ is given by
\be
G_R^0(x|y)
& = &
\theta(x^- - y^-)
g_R^0(x|y)
\non
& = &
\theta(x^- - y^-)
\int_{\bfk}
{1\over 2ik^-}
e^{-i{\bfkperp^2\over 2k^-}(x^- - y^-)
- ik^- (x^+ - y^+) + i\bfk_\perp(\bfx_\perp-\bfy_\perp)}
\ee
one can easily check that 
the conditions in (\ref{eq:GR_delta_id_1})
and (\ref{eq:GR_delta_id_2})
are satisfied by
$g_R^0(x|y)$ by taking the derivatives and taking the limits.
From now on, we will frequently make use of
the following short-hands for momentum space integrals
\be
\int_{k} \equiv \int {d^4 k\over (2\pi)^4}
\ \ \hbox{and} \ \
\int_{\bfk} \equiv \int {d^2 k_\perp dk^- \over (2\pi)^3}
\ \ \hbox{also} \ \
\int_{\bfkperp} \equiv \int {d^2 k_\perp \over (2\pi)^2}
\ee
The momentum space representation of the free-field propagator is
\be
G_R^0(k)
& = &
{1\over 2k^+ k^- - \bfkperp^2 + ik^-\epsilon}
\label{eq:GRmom}
\ee
The advanced Green function is given by
\be
G_A^0(x|y) = G_R^0(y|x)
\ee
The following composition rule is often useful
\be
G_R^0(x|y) = 2 \int_{\bfz}
(\partial_{z^+}G_R^0(x|z^-,\bfz)) G_R^0(z^-,\bfz|y) 
\label{eq:GRrel}
\ee
where $z^-$ can be any point within the interval $x^- > z^- > y^-$.
However, this relationship needs to be used with a caution.
Since $\theta(x^- - y^-)$ does not behave the same way as \break
$\theta(x^--z^-)\theta(z^--y^-)$ under $\partial/\partial x^-$,
the relationship Eq.(\ref{eq:GRrel}) should be really interpreted
as the relationship between $g_R^0(x|y)$'s. Hence, this relationship
should be used only when there is a definite ordering of times.

Writing the second line of Eq.(\ref{eq:formal_GRl2})
as
\be
\tildeG_R(x|y) 
& = &
G_R^0(x|y)
+\int_u 
\tildeG_R(x|u)
\calM(u)
G_R^0(u|y)
\ee
where
\be
\calM(u)
=
-2ig \overleftarrow{\partial}_{u^+} \ul{\tildecalA}_-(u^-,\bfuperp)
\ee
is a matrix-valued operator with the right derivative
$\overleftarrow{\partial}$ defined as
\be
f(x|y)\overleftarrow{\partial_y} = -\partial_y f(x|y)
\ee
the series solution of the second line 
in Eq.(\ref{eq:formal_GRl2}) can be obtained as follows
\be
\tildeG_R(x|y)
& = &
G_R^0(x|y)
+ \int_{u_1} G_R^0(x|u_1) \calM(u_1) G_R^0(u_1|y)
\non & & {}
+ \int_{u_1,u_2}  
G_R^0(x|u_1) \calM(u_1) 
G_R^0(u_1|u_2) \calM(u_2) G_R^0(u_2|y) 
\non & & {}
+ \int_{u_1,u_2,u_3}
G_R^0(x|u_1) \calM(u_1) 
G_R^0(u_1|u_2) \calM(u_2) 
G_R^0(u_2|u_3) \calM(u_3) 
G_R^0(u_3|y) 
\non & & {}
+ \cdots
\ee
To find an approximation solution, notice that $u^-_i$ are all confined
between $0$ and $\epsilon_Y$ because of
$\ul{\tildecalA}_-(u_i^-,\bfu_{\perp\,i})$.
We know that
\be
\lim_{u^-\to 0^+} 2\partial_{v^+}G_R^0(u^-,\bfu|0,\bfv) = \delta(\bfu-\bfv)
\ee
where $u^-\to 0^+$ means that the limit should be approached from above.
Hence, for small $u^-$ and $v^-$, we may approximate
\be
2\partial_{v^+}G_R^0(u^-,\bfu|v^-,\bfv) \approx
\theta(u^- - v^-)\delta(\bfu-\bfv)
\ee
This has the effect of replacing
\be
\int_{\bfu_{i+1}}\,
G_R^0(u_i|u_{i+1})\calM(u_{i+1})
& \to &
ig
\theta(u_i^- - u_{i+1}^-)
\ul{\tildecalA}_-(u^-_{i+1},\bfuperp{}_1)
\ee
so that
\be
\tildeG_R(x|y)
& \approx &
G_R^0(x|y)
+ 
2ig\int_{u_1}
(\partial_{u_1^+}G_R^0(x|u_1))
\ul{\tildecalA}_-(u_1^-,\bfu_{\perp 1}) G_R^0(u_1|y)
\non & & {}
+ 
2(ig)^2\int_{u_1}\int du_2^-\, 
(\partial_{u_1^+}G_R^0(x|u_1))
\ul{\tildecalA}_-(u_1^-,\bfu_{\perp 1}) 
V^\dagger(u_1^-, u_2^-; \bfu_{\perp 1}) 
\non & & {} \qquad\qquad\qquad\qquad\times
\theta(u^-_1 - u^-_2)
\ul{\tildecalA}_-(u^-_2, \bfu_{\perp 1})
G_R^0(u_2^-,\bfu_1|y) 
\non
\label{eq:resummedGR}
\ee
where we resummed time-ordered integrals to
\be
\ulV^\dagger(u_1^-, u_2^-;\bfuperp{}_1)
=
\calP\exp\left(
ig\int_{u_2^-}^{u_1^-} dz^-\, \ul{\tildecalA}_-(z^-,\bfuperp{}_1)
\right)
\ee
Using $\partial_{u^-}\theta(u^- - v^-) = \delta(u^- - v^-)$ and
\be
\partial_{u^-} \ul{V}^\dagger(u^-,v^-;\bfu)
&=& ig\ul{\tildecalA}_-(u^-,\bfu) \ul{V}^\dagger(u^-,v^-;\bfu)
\\
\partial_{v^-} \ul{V}^\dagger(u^-,v^-;\bfu)
&= &
-ig\ul{V}^\dagger(u^-,v^-;\bfu) \ul{\tildecalA}_-(v^-,\bfu)
\ee
this becomes
\be
\tildeG_R(x|y)
& \approx &
G_R^0(x|y)
\non & & {}
-
2
\int_{\bfu_1}\int_0^{\epsilon_Y} du_1^-\, 
(\partial_{u_1^+} G_R^0(x|u_1))
\non & & {} \qquad\times
\partial_{u_1^-}
\int_0^{\epsilon_Y} du_2^-\,
\theta(u_1^- - u_2^-)
\left(\partial_{u_2^-} \ulV^\dagger(u_1^-, u_2^-; \bfuperp{}_1) \right)
G_R^0(u_2^-,\bfu_1|y)
\ee
What we are mainly interested in is the case when none of $x^-$ or $y^-$ are
in the $(0,\epsilon_Y)$ interval. In that case,
integrating by parts yields
\be
\tildeG_R(x|y)
& \approx &
G_R^0(x|y)
\non & & {}
+
2
\int_{\bfu_1}
\big(\partial_{u_1^+} G_R^0(x|\epsilon_Y,\bfu_1)\big)
\big(
\ulV^\dagger(\epsilon_Y,0; \bfuperp{}_1)
G_R^0(0,\bfu_1|y)
-
G_R^0(\epsilon_Y,\bfu_1|y)
\big)
\label{eq:tildeGRfinal}
\ee
where we ignored $O(\epsilon_Y)$ terms. For details, see Appendix
\ref{app:props_details}.
When $x^- > y^- > \epsilon_Y$, or $ y^- < x^- < 0$,
Eq.(\ref{eq:tildeGRfinal}) gives
\be
\tildeG_R(x|y)
=
\tildeG_R^0(x|y)
\label{eq:GR_outside}
\ee
as it should.
When $x^- > \epsilon_Y$ and $y^- < 0$, we have
\be
\tildeG_R(x|y)
& \approx &
2
\int_{\bfu}\,
(\partial_{u^+}G_R^0(x|\epsilon_Y,\bfu))
\ulV^\dagger(\epsilon_Y,0; \bfuperp) G_R^0(0,\bfu|y)
\label{eq:tildeGRneeded}
\ee
upon using the composition rule Eq.(\ref{eq:GRrel}).
We will frequently need
\be
\lim_{x^-\to \epsilon_Y+0^+}\tildeG_R(x|y)
\approx \ulV^\dagger(\epsilon_Y,0;\bfxperp)G_R^0(0,\bfu_1|y)
\label{eq:tildeGRepsilonY}
\ee
which comes from Eq.(\ref{eq:tildeGRneeded}) and 
Eq.(\ref{eq:GR_delta_id_2}).
Expressions for $\tildeG_R(x|y)$ when either one or both  $x^-$ and $y^-$
are in $(0,\epsilon_Y)$ can be
found in Appendix \ref{app:props_details}.

For the light-cone gauge, 
the retarded propagator is obtained as
\be
G_R(x|y) = \ulV(x)\tildeG_R(x|y)\ulV^\dagger(y)
\ee
which comes from the following relationship
\be
-\delta(x-y)
& = &
\ulV(x) \tildecalD_\mu \tildecalD^\mu \tildeG_R(x|y) \ulV^\dagger(y) 
\non
& = &
\calD_\mu\calD^\mu G_R(x|y)
\ee

Using Eq.(\ref{eq:tildeaisol}), the symmetric propagator
is given by
\be
\ave{\tildea_a^i(x)\tildea_b^j(y)}
& = &
\ave{\tildealpha_a^i(x)\tildealpha_b^j(y)}
\ee
with $\ave{\tildealpha_a^i(x)} = 0$.
Using Eq.(\ref{eq:alphaieq}),  
\be
\ave{\tildea^i_a(x) \tildea^j_b(y)}
& = &
4\int_{\bfu,\bfv}\, 
(\partial_{u^+}\tildeG_{R}^{ac}(x|x_{\rm init}^-,\bfu))
\ave{
\tildea_c^i(x_{\rm init}^-,\bfu)
\tildea_d^i(x_{\rm init}^-,\bfv)
}
(\partial_{v^+}\tildeG_{A}^{db}(x_{\rm init}^-,\bfv|y))
\ee
If we have the perturbative vacuum at $x^-_{\rm init}$,
then 
$ \ave{ \tildea_c^i(x_{\rm init}^-,\bfu) \tildea_d^i(x_{\rm init}^-,\bfv) } $
is the free-field propagator at equal times,
\be
\ave{\tildea^i_a(x^-_{\rm init},\bfx) \tildea^j_b(x^-_{\rm init},\bfy)}
&=&
\delta_{ab} \delta^{ij} G_S^0(0,\bfx-\bfy)
\non
&=&
-\delta_{ab} g^{ij} G_S^0(0,\bfx-\bfy)
\ee
The free field propagator in the momentum space
is just the on-shell $\delta$-function:
\be
G_S^0(x)
& = &
\int_k
e^{-ik^+ x^- - ik^- x^+ + i\bfkperp{\cdot}\bfxperp}\,
\pi\delta(2k^+ k^- - \bfkperp^2)
\non
& = &
\int_{\bfk}
{1\over 4|k^-|}
e^{-i(\bfkperp^2/2k^-) x^- - ik^- x^+ + i\bfkperp{\cdot}\bfxperp}\,
\ee
Hence,
and for $x^-, y^- > x^-_{\rm init}$
\be
\ave{\tildea_a^i(x)\tildea_b^j(y)}
= 
\ave{\tildealpha_a^i(x)\tildealpha_b^j(y)}
= 
-g^{ij} \tildeG_S^{ab}(x|y)
\ee
where
\be
\tildeG_S^{ab}(x|y)
& = &
4
\int_{\bfu,\bfv}
\left[\partial_{u^+}\tildeG_R(x|x^-_{\rm init},\bfu) \right]_{ac}
G_S^0(0,\bfu-\bfv)
\left[\partial_{v^+}\tildeG_A(x^-_{\rm init},\bfv|y) \right]_{cb}
\label{eq:aveaiaj}
\ee
with the understanding that $x^-_{\rm init}$ is in the far-past.
All other symmetric propagators can be derived from this by using 
\be
\tildea^+(x) = -\int_{-\infty}^{x^+} dx'^+\,
\partial_i \tildea^i(x^-,x'^+,\bfxperp)
\ee

In the next sections, we will need $\tildeG_S(x|y)$ 
for 5 different cases: 
\begin{enumerate}
 \item $x^- < 0, y^-< 0$\mbox{}\\
 In this case $\tildeG_R = G_R^0$ and $\tildeG_A = G_A^0$
 in Eq.(\ref{eq:aveaiaj}).
 Then using the relationship Eq.(\ref{eq:aveaiaj}) for the free-field,
 \be
 \tildeG^{ab}_S(x|y) = \delta_{ab} G_S^0(x|y)
 \label{eq:GS_both_neg}
 \ee
 
 \item $x^- = \epsilon_Y, y^- = \epsilon_Y$\mbox{}\\
 Using Eq.(\ref{eq:tildeGRepsilonY}), and the fact that
 (c.f.~Eq.(\ref{eq:GS0h}))
 \be
 G_S^0(0,\bfxperp-\bfyperp)
 = B(x^+ - y^+)\delta(\bfxperp - \bfyperp)
 \ee
 we get
 \be
 \tildeG^{ab}_S(\epsilon_Y,\bfx|\epsilon_Y,\bfy)
 = \delta_{ab} B(x^+ - y^+)\delta(\bfxperp-\bfyperp)
 \label{eq:GS_both_eps}
 \ee
 where we defined a distribution
 \be
 B(x^+) = \int {dk^-\over 8\pi |k^-|} e^{-ik^- x^+}
 \ee

 \item $x^- = \epsilon_Y, y^- < 0$\mbox{}\\
 Using Eq.(\ref{eq:tildeGRepsilonY}) and 
 the relationship Eq.(\ref{eq:aveaiaj}) for the free-field,
 we get
 \be
 \tildeG^{ab}_S(\epsilon_Y, \bfx|y) 
 = \ulV^\dagger_{ab}(\epsilon_Y,\bfxperp)G_S^0(0,\bfx|y)
 \label{eq:GS_y_neg}
 \ee
 
 \item $x^- < 0, y^- = \epsilon_Y$\mbox{}\\
 This is just the Hermitian conjugate of the previous case,
 \be
 \tildeG^{ab}_S(x|\epsilon_Y, \bfy) 
 = G_S^0(x|0,\bfy) \ulV_{ab}(\epsilon_Y,\bfyperp)
 \label{eq:GS_x_neg}
 \ee

 \item $x^- > \epsilon_Y, y^- > \epsilon_Y$\mbox{}\\
 This is also just the free-field propagator
 \be
 \tildeG^{ab}_S(x|y) = \delta_{ab}G_S^0(x|y)
 \label{eq:GS_both_pos}
 \ee
 Details are in Appendix \ref{app:props_details}.
\end{enumerate}

\section{Deriving the JIMWLK Equation}
\label{sec:JIMWLK_derivation}

The original derivation of the JIMWLK equation was carried out in
Refs.\cite{Ayala:1995hx,
JalilianMarian:1996xn,
JalilianMarian:1997jx,
JalilianMarian:1997gr,
JalilianMarian:1997dw}
using the single-path formulation.
An alternate derivation of the JIMWLK equation using the Schwinger-Keldysh
formalism with the 1-2 representation was carried out in 
Refs.~\cite{Iancu:2000hn,Gelis:2008rw}.
In this section, we use the $r$-$a$ formalism developed so far
to derive the JIMWLK equation.

The JIMWLK equation is a renormalization group equation for 
the color density distribution $W_{\rho}[\rho]$ 
(c.f.~Eq.(\ref{eq:OLONLO2J})) but including the vacuum
fluctuations of $a$. The idea is as follows.
Quantum fluctuations with high $p^+$ components have correspondingly large
Lorentz $\gamma$ factors. Therefore, their interaction time scale is 
correspondingly longer
than that of soft fluctuations with low $p^+$ components.
The hard modes then play the role of 
the ``static'' color source for the soft components. This additional source
(call it $\delta\rho$) can be added to
$\rho$ which was originally composed of only valence quarks. The soft field
can be added to the classical field to form a new classical field sourced by 
both the original source $\rho$ and the additional source $\delta\rho$.
As we move the cut-off scale (call it $\Lambda$)
between the hard and the soft modes,
the modified color-density distribution $W_\Lambda[\rho]$ is found to
satisfy a renormalization group equation. This is the JIMWLK equation.

To derive JIMWLK equation, we first need to choose a coordinate system
that is convenient for high energy collisions. 
The light-cone coordinate system
$
x^\pm = x_\mp =  {(x^0 \pm x^3)/\sqrt{2}}
$
provides a convenient description.
For our analysis, we choose the
direction of the movement to be in the positive $z$ direction as before.
Equivalently, 
the source location in the space-time is in a small neighborhood of $x^- = 0$.
As explained earlier, we cannot use our analysis so far for a source that is
strictly confined at $x^- = 0$. We need to
consider the source to be spread out in a thin strip between $x^- =0$ and
$x^- = \epsilon_Y$.
For the classical solution in Eq.(\ref{eq:OLONLO2}), 
the problem then becomes
solving the source-less classical Yang-Mills equation 
in the region $x^- > \epsilon_Y$ with the boundary
condition given at $x^- = \epsilon_Y$. 
Correlation functions in the region $x^- > \epsilon_Y$ are then functionals
of the field values 
$\calA_\mu(\epsilon_Y,\bfxperp)$, 
$\ave{a_\mu(\epsilon_Y,\bfx)}$
and the equal time correlators
$\ave{a_\mu(\epsilon_Y,\bfx) a_\nu(\epsilon_Y,\bfy)}$.
These quantities are exactly the ingredients of the JIMWLK equation.

In the presence of a strong source, 
an observable up to NLO is given by
(c.f.~Eq.(\ref{eq:OLONLO2J}))
\be
\ave{\ave{\calO}}_{\rm LO+NLO}
& = &
\int \calD\rho\,W_{\rho}[\rho]
\int [da_i] \rho_{\rm v}[a_i]\,
\calO[A[\rho, a_i]]
\label{eq:OLONLO3J}
\ee
where $A[\rho,a_i]$ is the classical solution with the strong
color source $\rho$ and the initial condition $a_i$.
Again, only the initial field value
in the initial condition needed because 
the d'Alembertian is only linear in $\partial_-$.
The main idea behind JIMWLK equation is to replace the initial field
value fluctuations with equivalent source fluctuations.
Namely, we would like to identify Eq.(\ref{eq:OLONLO3J}) with
\be
\ave{\calO}_{Y,\rm LO+NLO}
& = &
\int \calD\rho\,W_{\rho}[\rho]
\int \calD\lambda\, Y[\lambda;\rho]\,
\calO[A[\rho + \lambda]]
\label{eq:OLONLO4J}
\ee
where $Y[\lambda;\rho]$ is the distribution of the additional source 
$\lambda$ in the background of $\rho$. 
The requirement is then that the two expressions
Eq.(\ref{eq:OLONLO3J}) and Eq.(\ref{eq:OLONLO4J})
match in the $x^- > \epsilon_Y$ region.
Since there is no source in this region, 
it is enough to match the initial values 
of $\ave{a^\mu(x)}$ and $\ave{a^\mu(x)a^\nu(y)}$ at $x^-, y^- = \epsilon_Y$
for the NLO calculations.

\subsection{2-point correlator}
\label{sec:2pt_corr}

In the physical light-cone gauge, 
the linear Jacobi-field equations in the region where
$x^- > \epsilon_Y$ are given by
\be
[\calD_\mu, [\calD^\mu, a^-]]
& = &
\partial_+ [\calD_\mu, a^\mu]
\label{eq:linjacobiminus}
\\ {}
[\calD_\mu, [\calD^\mu, a^i]]
& = &
[\calD^i, [\calD_\mu, a^\mu]]
\label{eq:linjacobispace}
\\ {}
0
& = &
\partial_- [\calD_\mu, a^\mu]
\label{eq:partialminusDa}
\ee
Eq.(\ref{eq:partialminusDa}) implies that the covariant divergence
$\sigma_1(\bfx) \equiv [\calD_\mu, a^\mu]$
is independent of $x^-$. 
Its values are fixed at $x^- = \epsilon_Y$.
Then the right hand sides of Eqs.(\ref{eq:linjacobiminus}) and
(\ref{eq:linjacobispace}) can be regarded as static sources
whose behaviors are fully determined by $\sigma_1(\bfx)$
at $x^- = \epsilon_Y$.
It is then natural to expect $\sigma_1(\bfx)$
to play the role of the additional source $\lambda$ as we will see below.

For $x^-, y^- > \epsilon_Y$,
the 2-point function matching requirement between 
Eq.(\ref{eq:OLONLO3J}) and Eq.(\ref{eq:OLONLO4J}) is
\be
\lefteqn{
\int [da_i] \rho_{\rm v}[a_i]\,
a_a^\mu[\tilderho, a_i](x)
a_b^\nu[\tilderho, a_i](y)
} && 
\non
& = &
\int \calD\tildelambda\, Y[\tildelambda;\tilderho]\,
\non & & {}
\times
\left(
\int_{\epsilon_Y}^{\infty} dz^- \int_{\bfz} \tildelambda_e(z) 
\left({\delta A_a^\mu[J^+](x)\over \delta \tildeJ_e^+(z)}
\right)
\int_{\epsilon_Y}^{\infty} dz'^-\int_{\bfz'} \tildelambda_f(z') 
\left({\delta A_b^\nu[J^+](y)\over \delta \tildeJ_f^+(z')}
\right)
+ O(\tildelambda^3)
\right)
_{A\to\calA\atop \tildeJ^+\to \tilderho}
\label{eq:matching_cond2}
\non
\ee
where we have indicated that even though the gauge fields are in the light-cone gauge,
we would like the sources to be in the axial gauge because 
it is simpler to use them as the functional integration variable as they 
do not precess.
It should be understood that the matching here is between two connected 
correlation functions.
The purpose of this matching condition is to fix the noise correlation function 
$\ave{\tildelambda(x)\tildelambda(y)}$ in terms of the quantum correlation function
$\ave{a^\mu(x)a^\nu(y)}$.

Due to causality, the classical solution $A^\mu(x)$ cannot depend on the
behavior of the source $\tildeJ^+(z)$ in the future. 
Therefore, we naturally expect 
\be
\left({\delta A_a^\mu[J^+](x)\over \delta \tildeJ_e^+(z)}\right)
\propto \theta(x^- - z^-)
\ee
The $z^-$ integral is then restricted to be in $\epsilon_Y \le z^- < x^-$.
This presents a problem. In the limit $x^- \to \epsilon_Y$ from above,
the left hand side of Eq.(\ref{eq:matching_cond2}) does exist while the
right hand side vanishes {\em unless} we set
\be
\tildelambda(z) 
= \delta(z^- - \epsilon_Y^+)\tildezeta(\bfz)
\ee
where we have defined $\epsilon_Y^+$
to be a value infinitesimally above $\epsilon_Y$ so that 
$\int_{\epsilon_Y}^\infty dz^- \delta(z^- - \epsilon_Y^+) = 1$.

To calculate
\be
{\delta A_a^\mu(x)\over \delta\tildeJ^+_e(z)}
& = &
\int_u 
{\delta J^+_b(u)\over \delta\tildeJ^+_e(z)} 
{\delta A_a^\mu(x)\over \delta J^+_b(u)}
\ee
we need a general classical solution 
$A^\mu$,
not just $\calA^\mu$. The limit $A_\mu \to \calA_\mu$ should be
taken only at the end.
The relationship between the axial gauge source
$\tildeJ^+$ and the light cone gauge source $J^+$ is 
\be
J^+ = U\tildeJ^+ U^\dagger 
\label{eq:utilderhou}
\ee
In the color vector notation, this is
\be
J^+_a = \ulU_{ab} \tildeJ^+_b
\label{eq:utilderhou_adj}
\ee
As long as we assume that the $x^+$ dependence of $J^+$ is adiabatically
slow, the gauge group element is still given by the Wilson line
\be
\ulU^\dagger(x) =
\calP\,\exp\left(
ig\int_{-\infty}^{x^-} dy^-\,\ul{\tildeA}_a^+(y^-,\bfx) T^a
\right)
\ee
where $T^a$ is the $SU(3)$ generator in the adjoint representation
and $\tildeA^+$ satisfies
\be
\tildeA_a^+(x^-,\bfx) = \int_{\bfyperp} 
G_T(\bfxperp-\bfyperp) \tildeJ_a^+(x^-,x^+,\bfyperp)
\ee
Using Eq.(\ref{eq:utilderhou_adj}), we get
\be
{\delta A_a^\mu(x)\over \delta\tildeJ^+_e(z)}
& = &
\int_u \,
{\delta J_b^+(u)\over \delta\tildeJ^+_e(z)} 
{\delta A_a^\mu(x)\over \delta J^+_b(u)}
\non
& = &
\int_{u}
{\delta \ulU_{bc}(u) \over \delta\tildeJ^+_e(z)} 
\tildeJ^+_c(u)
{\delta A_a^\mu(x)\over \delta J^+_b(u)}
+
\ulU_{be}(z)
{\delta A_a^\mu(x)\over \delta J^+_b(z)}
\label{eq:deltaAdeltaJ}
\ee
where again we let the functions vary in $x^+$ adiabatically slowly.
In the first term of Eq.(\ref{eq:deltaAdeltaJ}),
the factor $\delta\ulU_{bc}(u)/\delta\tildeJ_e^+(z)$ is non-vanishing
only when 
$u^-$ is larger than $z^-$
since $\ulU$ is path-ordered. 
On the other hand the factor $\tildeJ_c^+(u)$
is non-zero only when 
$u^-$ is smaller than $\epsilon_Y$
in the limit $\tildeJ^+\to\tilderho$.
Hence in the limit $z^- \to \epsilon_Y^+$, 
these two conditions cannot be fulfilled simultaneously.
Consequently,
\be
\lim_{z^- \to \epsilon_Y^+\atop \tildeJ^+ \to \tilderho}
{\delta A_a^\mu(x)\over \delta\tildeJ^+_e(z)}
=
\ulV_{be}(z^-,\bfzperp)
\calK^\mu_{a,b}(x|z) 
\ee
with $\ulV^\dagger(z^-,\bfzperp)$ defined in Eq.(\ref{eq:Vdaggersol}).
The response function $\calK^\mu_{a,b}(x|z)$ is defined by
\be
\calK^\mu_{a,b}(x|z) 
=
\lim_{z^-\to \epsilon_Y^+\atop \tildeJ^+ \to \tilderho}
{\delta A_a^\mu(x)\over \delta J^+_b(z)}
\ee
To find the the equation satisfied by
the response functions,
we start from the full equation of motion:
\be
(i/g)[D_\mu, [D^\mu, D^\nu]] = \delta^{\nu + }J^+
\ee
Taking the derivative with respect to $J_e^+$ and letting 
$A_\mu \to \calA_i$,
we get
\be
-\delta(z-x)\delta_{eg}
& = &
\partial_- \ul{\calD}^{gb}_\mu \calK_{b,e}^\mu(x|z)
+2ig \ul{\calG}_{ga}^{+i}(x) \calK_i^{a,e}(x|z)
\label{eq:calKeq1}
\\
{}
\ul{\calD}^{ga}_\mu \ul{\calD}_{ab}^\mu \calK_{b,e}^i(x|z)
& = &
\ul{\calD}_{ga}^i\ul{\calD}^{ab}_\mu \calK_{b,e}^\mu(x|z)
+2ig \ul{\calG}_{ga}^{i+}(x) \calK_{a,e}^-(x|z)
\label{eq:calKeq2}
\\
{}
\ul{\calD}^{ga}_\mu\ul{\calD}_{ab}^\mu \calK_{b,e}^-(x|z)
& = &
\partial_+\ul{\calD}^{gb}_\mu\calK_{b,e}^\mu(x|z)
\label{eq:calKeq3}
\ee
Integrating over $x^-$, Eq.(\ref{eq:calKeq1}) becomes
\be
\ul{\calD}^{gb}_\mu \calK_{b,e}^\mu(x|z)
& = &
-\theta(x^- - z^-)\delta(\bfx-\bfz)\delta_{ge}
- 2ig
\int^{x^-}_{-\infty} d{x'}^-\,
\ul{\calG}_{ga}^{+i}({x'}^-,\bfxperp) \calK_i^{a,e}({x'}^-,\bfx|z)
\label{eq:calDcalK}
\ee
Due to the causality, it is clear that
\be
\calK^i_{a,e}(x|z) = 
\left( {\delta A_a^i(x)\over \delta J_e^+(z)} \right)_{\calA,\rho}
\propto \theta(x^- - z^-)
\ee
In the limit $z^- \to \epsilon_Y^+$,
the second term 
in the right hand side of Eq.(\ref{eq:calDcalK}) vanishes
because $\calG^{+i}({x'}^-,\bfxperp)$
is non-zero only for $0 < {x'}^- < \epsilon_Y$
while
$\calK^e_i({x'}^-,\bfx|\epsilon_Y,\bfz)$ is non-zero only when ${x'}^- >\epsilon_Y$.
Hence, the the second term in Eq.(\ref{eq:calDcalK}) does not contribute.

Since Eq.(\ref{eq:calDcalK}) is simple, it suggests that
instead of using $\ave{a^\mu(x)a^\nu(y)}$ for matching,
we should use
$\ave{[\calD_\mu, a^\mu(x)]_a [\calD_\nu, a^\nu(y)]_b} = 
\ave{\sigma_1^a(x)\sigma_1^b(y)}$.
Using Eq.(\ref{eq:calDcalK} then yields a simpler matching condition
\be
\ave{\sigma_1^a(\bfx)\sigma_1^b(\bfy)}_{\rm v}
&=&
\int_{z,z'}
\ulV_{ce}(z) \calD^{ad}_\mu\calK_{d,c}^\mu(x|z) 
\ulV_{dh}(z') \calD^{bg}_\nu \calK_{g,d}^\nu(y|z')
\ave{\tildelambda_e(z)\tildelambda_h(z')}_Y
\non
& = &
\ulV_{ac}(\epsilon_Y,\bfxperp) \ulV_{bd}(\epsilon_Y,\bfy)
\ave{ \tildezeta_c(\bfx)\tildezeta_d(\bfy) }_{Y}
\label{eq:matching_cond_mm}
\ee
where
we used the fact that $\sigma_1$ does not depend on $x^-$ when $x^- > \epsilon_Y$.
The subscripts ${\rm v}$ and $Y$ indicates that the average is with respect to 
the vacuum density matrix $\rho_{\rm v}[a_\mu]$ for $\sigma_1$
and with respect to $Y[\tildelambda, \tilderho]$ for $\tildezeta$.
From now on, these subscripts will be omitted unless possible confusion can arise.
Note that all reference to the time variables disappeared as they should.
Defining
$\tildesigma_1^a \equiv \ulV_{ab}^\dagger \sigma_1^b$, 
we finally get
\be
\ave{\tildesigma^a_1(\bfx)\, \tildesigma^b_1(\bfy)}
& = &
\ave{ \tildezeta_a(\bfx)\tildezeta_b(\bfy) }
\ee

Using the relationship between $a_\mu$ and $\tildea_\mu$ and $\tildeomega$,
we have
\be
\tildesigma^a_1 
\approx 
-[\tildecalD^\mu, [\tildecalD_\mu, \tildeomega]]_a
=
\partial_i\left(2\tildea_a^i - \partial^i\tildeomega_a\right)
\label{eq:sigma1_in_chi}
\ee
Evaluating the quantum correlation function
\be
\bar\chi^{ab}(\bfx|\bfy) = \ave{\tildesigma_1^a(\bfx)\tildesigma_1^b(\bfy)}
\label{eq:barchi}
\ee
then requires evaluating the correlation functions of
$\tildea^j(\epsilon_Y,\bfx)$ and $\tildeomega(\epsilon_Y,\bfx)$.
Since $\barchi$ is fully determined at $x^- = \epsilon_Y$,
we must set $x^- = y^- = \epsilon_Y$ in the following calculations.
In the color vector space, we have
\be
\tildeomega_a(\epsilon_Y, \bfx) 
& = &
-
\ulV_{ab}^\dagger(\epsilon_Y,\bfxperp)
\int_{-\infty}^{\epsilon_Y}dz^-
\int_{-\infty}^{x^+} dz^+\,
\ulV_{bc}(z^-,\bfxperp)
\partial_i \tildea_c^i(z^-,z^+,\bfxperp) 
\non
& = &
-
\ulV_{ab}^\dagger(\epsilon_Y,\bfxperp)
\int_{-\infty}^{0}dz^-
\int_{-\infty}^{x^+} dz^+\,
\partial_i \tildea_b^i(z^-,z^+,\bfxperp) 
+
O(\epsilon_Y)
\label{eq:bapprox}
\ee
where in the second line we used the fact that
the contribution from
the strip $0 < z^- < \epsilon_Y$ is $O(\epsilon_Y)$ 
since the integrand is regular.
Hence all needed combinations of $\tildea^j$ and $\tildeomega$ 
can be calculated from 
$\ave{\tildea^j(x)\tildea^k(y)}$ with both $x^-, y^-$ not in 
the interval $(0,\epsilon_Y)$.
The relevant symmetric propagators were given in 
Eqs.(\ref{eq:GS_both_neg} -- \ref{eq:GS_both_pos}).
From Eq.(\ref{eq:GS_both_eps}),
we get the transverse correlation function
as
\be
\ave{\tildea_a^j(\epsilon_Y,\bfx)\tildea_b^k(\epsilon_Y,\bfy)}
& \approx &
g^{jk}
\ulV_{ac}^\dagger(\epsilon_Y,\bfxperp)
G_{S}^0(0,\bfx|0,\bfy)
\ulV_{cb}(\epsilon_Y,\bfyperp)
\non
& = &
-g^{jk}\delta_{ab}
\delta(\bfxperp-\bfyperp)B(x^+ - y^+)
\ee
where 
\be
B(x^+) = 
\int {dk^-\over 8\pi|k^-|} e^{- ik^- x^+}
\label{eq:B_def}
\ee
is a divergent integral which becomes a large small-$x$ logarithm
when regularized.

Using Eq.(\ref{eq:bapprox}) and Eq.(\ref{eq:GS_x_neg}), we next get
\be
{\ave{\tildeomega_a(\epsilon_Y,\bfx)
\tildea_b^j(\epsilon_Y,\bfy)}}
& = &
-
\ulV_{ac}^\dagger(\epsilon_Y,\bfxperp)
\int_{-\infty}^{0}dz^-
\int_{-\infty}^{x^+} dz^+\,
\ave{ \partial_i \tildea_c^i(z^-,z^+,\bfxperp) 
\tildea_b^j(\epsilon_Y,\bfy)}
\non
& = &
-2B(x^+ - y^+)\ulV_{ac}^\dagger(\epsilon_Y,\bfxperp)
\ulV_{cb}(\epsilon_Y,\bfyperp)
\partial^j_x 
G_T(\bfxperp - \bfyperp)
\ee
where $G_T(\bfxperp)$ is the Green function of the 2-D Laplacian.
This comes from
\be
\int_{-\infty}^0 dx^-\int_{-\infty}^{x^+} dx'^+\,
G_S^0(x^-, x'^+, \bfxperp) 
& = &
-2B(x^+)G_T(\bfxperp)
\ee
where
\be
G_T(\bfxperp) & = &
\int {d^2k_\perp\over(2\pi)^2}\, 
{e^{i\bfkperp{\cdot}\bfxperp}\over \bfkperp^2}
\ee
satisfies $ -\nabla_\perp^2 G_T(\bfxperp) = \delta(\bfxperp) $.
Similarly,
\be
{\ave{\tildea_a^i(\epsilon_Y,\bfx) \tildeomega_b(\epsilon_Y,\bfy)}}
& = &
-
\ulV_{bc}^\dagger(\epsilon_Y,\bfyperp)
\int_{-\infty}^{0}dz^-
\int_{-\infty}^{y^+} dz^+\,
\ave{\tildea_a^i(\epsilon_Y,\bfx)
\partial_j \tildea_c^j(z^-,z^+,\bfyperp)} 
\non
& = &
-
2B(x^+ - y^+)
\ulV^\dagger_{bc}(\epsilon_Y,\bfyperp)
\ulV_{ac}^\dagger(\epsilon_Y,\bfxperp)
\partial^i_y G_T(\bfxperp-\bfyperp)
\ee
For $\ave{\tildeomega\tildeomega}$, we get
\be
{\ave{\tildeomega_a(\epsilon_Y,\bfx)\tildeomega_b(\epsilon_Y,\bfy)}}
& \approx &
\ulV_{ac}^\dagger(\epsilon_Y,\bfxperp)
\ulV_{bd}^\dagger(\epsilon_Y,\bfyperp)
\non & & {}
\int_{-\infty}^{0}dz^-
\int_{-\infty}^{x^+} dz^+\,
\int_{-\infty}^{0}dz'^-
\int_{-\infty}^{y^+} dz'^+\,
\ave{
\partial_i \tildea_c^i(z^-,z^+,\bfxperp) 
\partial_j \tildea_d^j(z'^-,z'^+,\bfyperp) 
}
\non
& = &
B(x^+ - y^+)
\ulV_{ac}^\dagger(\epsilon_Y,\bfxperp)
\ulV_{bc}^\dagger(\epsilon_Y,\bfyperp)
G_T(\bfxperp-\bfyperp)
\ee
using Eq.(\ref{eq:GS_both_neg}).

The correlator we want is 
\be
\barchi_{ab}(\bfx|\bfy)
& =  &
\ave{\tildesigma_1^a(\bfx)\tildesigma_1^b(\bfy)}
 = 
\ave{\tildezeta^a(\bfx)\tildezeta^b(\bfy)}
\non
& = & 
\partial_{i}^x\partial_{j}^y 
\chi_{ab}^{ij}(\epsilon_Y,\bfx|\epsilon_Y,\bfy)
\ee
where 
\be
\chi^{ij}(x|y)
= 
\left(
4\ave{\tildea_a^i(x)\tildea_b^j(y)}
-2\ave{\tildea_a^i(x) \partial^j \tildeomega_b(y)}
-2\ave{\partial^i \tildeomega_a(x) \tildea_b^j (y)}
+\ave{\partial^i \tildeomega_a(x) \partial^j \tildeomega_b(y)}
\right)
\non
\ee
Anticipating what enters the JIMWLK equation,
we may define
\be
\eta(\bfxperp|\bfyperp) 
&=& 
{1\over 4B(x^+ - y^+)}
\int_{\bfuperp,\bfvperp}
G_T(\bfxperp-\bfuperp)\,
\barchi(\epsilon_Y,x^+,\bfuperp|\epsilon_Y,y^+,\bfvperp)
G_T(\bfvperp-\bfyperp)
\label{eq:eta_result_0}
\ee
Few integrations by parts give
\be
\lefteqn{\eta(\bfxperp|\bfyperp)}
&&
\non
& = &
-
\int d^2u_\perp\,
\left(
1 
-
\ulV^\dagger(\epsilon_Y,\bfuperp)
\ulV(\epsilon_Y,\bfyperp)
-
\ulV^\dagger(\epsilon_Y,\bfxperp) 
\ulV(\epsilon_Y,\bfuperp)
+
\ulV^\dagger(\epsilon_Y,\bfxperp)
\ulV(\epsilon_Y,\bfyperp)
\right)
\non & & {}
\qquad\qquad
\partial_x^i G_T(\bfxperp-\bfuperp)
\partial_i^y G_T(\bfuperp-\bfyperp)
\label{eq:eta_result}
\ee
which is the standard result 
\cite{Ferreiro:2001qy}.
Some details of the 2-point function calculation
can be found in Appendix~\ref{app:details_2pts}.

\subsection{1-point average}

Applying the same logic as in the last section,
the 1-point average matching requirement is
\be
\lefteqn{
\ave{\sigma_1^a(x)}}
&&
\non
& = &
\calD_\mu^{ab}
\int \calD\tildelambda\, Y[\tildelambda;\tilderho]\,
\left(
\int_z \tildelambda_e(z) 
\left(
{\delta A_b^\mu(x)\over \delta \tildeJ_e^+(z)}
\right)
+
{1\over 2}
\int_{z, z'}
\tildelambda_e(z)\tildelambda_r(z')
\left(
{\delta^2 A^\mu(x) \over \delta\tildeJ_e^+(z) \delta\tildeJ_r^+(z')}
\right)
\right)_{A\to\calA,\tildeJ^+\to\rho\atop z^-,z'^- \to \epsilon_Y^+}
\non
& & {}
+O(\lambda^3)
\ee
For the 2-point function case, $\sigma_1$ was expanded up to the linear order
$\sigma_1 = -[\calD^\mu,[\calD_\mu,\omega]] + O(a^2)$.
This was adequate because what we needed was the correlator.
For the 1-point average, this is no longer the case. Since
the tadpole-like contributions are going to be the dominant contribution
to the 1-point average,
it is necessary to
expand $\sigma_1$ up to the quadratic order in $\tildea^\mu$. 
The purpose of this matching condition is to fix $\ave{\tildelambda}$ in terms of
the quantum average $\ave{\sigma_1(x)}$ and the quantum correlator
$\ave{\sigma_1(x)\sigma_1(y)}$, or equivalently
the noise correlator $\ave{\tildelambda(x)\tildelambda(y)}$.

By differentiating Eq.(\ref{eq:deltaAdeltaJ}) once more
and disregarding terms that vanish 
in the limits $z^-, z'^- \to \epsilon_Y^+$ and $\tildeJ^+ \to \tilderho$,
we get
\be
{{\delta^2 A_a^\mu(x)\over
\delta \tildeJ_e^+(z) \delta \tildeJ_r^+(z') }}
& = &
{{\delta\ulU_{br}(z')\over\delta \tildeJ_e^+(z)}}
{{\delta A^\mu_a(x)\over\delta J_b^+(z')}}
+
{\delta \ulU_{be}(z)
\over \delta \tildeJ_r^+(z')}
{\delta A_a^\mu(x)\over \delta J^+_b(z)}
+
\ulU_{be}(z)
\ulU_{gr}(z')
{\delta^2 A_a^\mu(x)\over \delta J^+_g(z') \delta J^+_b(z)}
\non
& & {}
+ \hbox{(terms that vanish in the limits)}
\label{eq:delta2AdeltaJdeltaJ}
\ee
Recall that $\delta A/\delta J$ is basically a propagator because
$J$ sources the classical field $A$.
The second derivative $\delta^2 A/\delta J \delta J$ then involves 
two propagators because this is the kernel for 2 sources.
As such, when Eq.(\ref{eq:delta2AdeltaJdeltaJ}) 
is closed with another propagator
$\ave{\tildezeta(z)\tildezeta(z')}$,  
the first two terms will generate UV divergences, but the second derivative
term will not generate a UV divergence because the resulting
one-loop will involve at least 3 propagators. 
Since we are interested in deriving a renormalization group equation,
we can safely ignore the second derivative term.
It is, however, a possible source of NNLO contribution..

Using $\tildecalA_- = \int G_T\tilderho$ and assuming adiabatically
slow change in $x^+$, we get
\be
\lefteqn{
\left.
{1\over 2}\int_{z,z'}
\tildelambda_e(z)\tildelambda_r(z')
{{\delta^2 A_a^\mu(x)\over
\delta \tildeJ_e^+(z) \delta \tildeJ_r^+(z') }}
\right|_{\calA,\tilderho,z^-,z'^-}
} && 
\non
& \approx &
-
\theta(0)
\int_{\bfz,\bfz'}
\tildezeta_e(\bfz)\tildezeta_r(\bfz')
\ulV_{bg}(\epsilon_Y,\bfz') ig(T^e)_{gr}
\delta(z'^+ - z^+)G_T(\bfzperp' - \bfzperp)
\calK^\mu_{a,b}(x|\epsilon_Y,\bfz')
\ee
In this context,
we need to define $\theta(0) = 1/2$ \cite{Iancu:2000hn}.

The matching condition now becomes for $x^- > \epsilon_Y$
\be
\ave{\sigma_1^a(x)}
& = &
-\ulV_{ab}(\epsilon_Y, \bfxperp)\ave{\tildezeta_b(\bfx)}
\non & & {}
+
{1\over 2}
\ulV_{ab}(\epsilon_Y, \bfxperp)
f_{ebr}
\int_{\bfzperp} 
\ave{\tildezeta_{e}(x^+, \bfzperp)\tildezeta_{r}(x^+,\bfxperp)}
G_T(\bfxperp - \bfzperp)
\label{eq:matching_cond_zeta}
\ee
using $(T^e)_{br} = -if_{ebr}$.
Equivalently, in terms of axial gauge quantities
the one-point noise average is 
\be
\ave{\tildezeta_a(\bfx)}
& = &
-\ave{\tildesigma_1^a(x)}
-
{1\over 2}
f_{abc}
\int_{\bfzperp} 
\ave{\tildezeta_{b}(x^+, \bfzperp)\tildezeta_{c}(x^+,\bfxperp)}
G_T(\bfxperp - \bfzperp)
\label{eq:matching_cond_temp}
\ee
The two-point noise correlator
on the right hand side of Eq.(\ref{eq:matching_cond_temp}) 
is already calculated in the previous section.

In order to calculate $\ave{\tildesigma_1}$,
we need the full Jacobi-field equations in the light-cone gauge in 
the $x^- > \epsilon_Y$ region 
\be
0 & = &
[\calD_\mu, [\calD^\mu, a^\nu]]
-[\calD^\nu, [\calD_\mu, a^\mu]]
\non & & {}
-2ig [a_\mu, [\calD^\mu, a^\nu]]
+ig [a^\nu, [\calD_\mu, a^\mu]]
+ig [a_\mu, [\calD^\nu, a^\mu]]
- g^2 [a_\mu, [a^\mu, a^\nu]]
\ee
The $\nu = +$ part of the equation  is
\be
\partial_-\sigma_1
& = &
ig [a_i, \partial_- a^i]
\label{eq:jacobiplusfull}
\ee
There is no $O(a^3)$ term.
Note that $\sigma_1$ is no longer independent of $x^-$.
This actually presents another problem. 
If we go back to the matching condition
Eq.(\ref{eq:matching_cond_zeta}),
it is clear that the right hand side does not
depend on $x^-$ while 
Eq.(\ref{eq:jacobiplusfull}) indicates that
$\sigma_1$ is no longer independent of $x^-$.
Therefore the matching condition makes sense only 
in the limit $x^-\to \epsilon_Y$:
\be
\ave{\tildezeta_a(\bfx)}
& = &
-\ave{\tildesigma_1^a(\epsilon_Y, \bfx)}
- {1\over 2} 
gf_{abc}
\int_{\bfzperp} 
\ave{ \tildezeta_{b}(x^+,\bfzperp) \tildezeta_{c}(x^+, \bfxperp) }
G_T(\bfzperp - \bfxperp)
\label{eq:matching_cond_temp2}
\ee

In the $x^- > \epsilon_Y$ region, the solution of 
Eq.(\ref{eq:jacobiplusfull}) is
\be
\ave{ \sigma_1(x) }
& = &
ig 
\int_{\epsilon_Y}^{x^-} dy^-\, 
\ave{
[a_i(y^-,\bfx), \partial_- a^i(y^-,\bfx)]
}
+ \ave{\sigma_1(\epsilon_Y, \bfx)}
\label{eq:sigma_sol}
\ee
Unfortunately, the constant of integration 
$\ave{\sigma_1(\epsilon_Y, \bfx)}$ is actually what we want.
The analysis so far cannot determine this boundary value.
It turned out that to 
derive the JIMWLK equation, we must use
\be
\ave{\sigma_1(\epsilon_Y, \bfx)}
& = &
ig 
\int_{\epsilon_Y}^{\infty} dy^-\, 
\ave{ [a_i(y^-,\bfx), \partial_- a^i(y^-,\bfx)]}
\label{eq:sigmaepsY}
\ee
This is also consistent with previous calculations,
for instance, in Ref.\cite{Ferreiro:2001qy}.
Eq.(\ref{eq:sigmaepsY})
is the weakest part of arguments laid out in this paper.
If we put Eq.(\ref{eq:sigmaepsY}) back into Eq.(\ref{eq:sigma_sol}),
then it would imply that
\be
\lim_{x^-\to \infty} \ave{\sigma_1(x)}
= 2\ave{\sigma_1(\epsilon_Y,\bfx)}
\ee
This does not contradict anything we have argued so far. However,
there does not seem to be good reason for this relationship, either.
Nevertheless, to derive the JIMWLK equation we need to assume that
Eq.(\ref{eq:sigmaepsY}) is valid.

Since the right hand side of Eq.(\ref{eq:sigmaepsY})
is already quadratic, we can now use
\be
a_i \approx \acutea_i - [\calD_i, \omega]
\ee
to get
\be
\ave{ \sigma_1(\epsilon_Y,\bfx) }
& = &
ig 
\int_{\epsilon_Y}^{\infty} dy^-\, 
\ave{
[\acutea_i(y^-,x^+,\bfxperp), \partial_- \acutea^i(y^-,x^+,\bfxperp)]
}
\non & & {}
+ig
\int_{\epsilon_Y}^{\infty} dy^-\, 
\ave{
[[\calD_i,\omega(y^-,x^+,\bfxperp)], 
[\calD^i, \acutea^+(y^-,x^+,\bfxperp)]]
}
\non & & {}
-2ig
\int_{\epsilon_Y}^{\infty} dy^-\, 
\ave{
[\acutea_i(y^-,x^+,\bfxperp), 
[\calD_i, \acutea^+(y^-,x^+,\bfxperp)]]
}
\non & & {}
+ig
\ave{
[[\calD_i, \omega(\epsilon_Y,x^+,\bfxperp)], \acutea^i(\epsilon_Y,x^+,\bfxperp)]
}
\label{eq:xi}
\ee
where we used one integration by parts to get the last two terms.
We also used $\partial_-\omega = \acutea^+$
and the fact that when $y^- > \epsilon_Y$, $\tildecalA_i$ is
independent of $y^-$ so that $[\partial_-, \calD_i] = 0$.
Among these terms, only the second line is finite.
All others are either zero or $O(\epsilon_Y)$.
For instance, the first term vanishes because the bracket is anti-symmetric
in the color index while the symmetric propagator in the $y^- > \epsilon_Y$
region is symmetric in the color index (c.f.~Eq.(\ref{eq:GS_both_pos})).
For details including the evaluation of the second line, see Appendix
\ref{app:Lines1to6}.
The end result is
\be
\lefteqn{\ave{\sigma_1^a(\epsilon_Y,\bfxperp)}} &&
\non
& = &
2B(0)g 
\ulV_{ab}(\epsilon_Y,\bfxperp)
f_{bcd}
\non & & {}
\partial_i^x 
\left(
\ulV_{de}^\dagger(\epsilon_Y, \bfxperp)
\partial_x^i
\int_{\bfuperp} \,
(\partial_l^x G_T(\bfxperp-\bfuperp))
\ulV_{ec}(\epsilon_Y, \bfuperp) 
(\partial^l_x G_T(\bfuperp-\bfxperp))
\right)
\ee

To get $\ave{\tildezeta}$, we need 
in addition
\be
R & = &
{1\over 2} 
gf_{abc}
\int_{\bfzperp} 
\ave{ \tildezeta_{b}(x^+,\bfzperp) \tildezeta_{c}(x^+, \bfxperp) }
G_T(\bfzperp - \bfxperp)
\ee
Using the expression for $\barchi$ we obtained in the 
previous section to evaluate $R$, we finally get
\be
{\ave{\tildezeta_a(\bfx)}/(4B(0))} 
& = &
\partial^i\partial_i \nu_a(\bfxperp)
-
{1\over 2}
gf_{abc}
(\partial_j^x \partial_x^j \ulV_{cd}^\dagger(\epsilon_Y,\bfxperp))
\ulV_{db}(\epsilon_Y,\bfxperp)
G_T(0)
\label{eq:zeta_result}
\ee
where
\be
\nu_a(\bfxperp)
& = &
-
{g\over 2}
f_{abc}
\ulV_{cd}^\dagger(\epsilon_Y, \bfxperp)
\int_{\bfuperp} \,
(\partial_l^x G_T(\bfxperp-\bfuperp))
\ulV_{db}(\epsilon_Y, \bfuperp) 
(\partial^l_x G_T(\bfuperp-\bfxperp))
\label{eq:nu_result}
\ee
Details can be found in Appendix \ref{app:Lines1to6}.
Again, we recover the known results from Ref.\cite{Ferreiro:2001qy} 
except for the presence of the second term in Eq.(\ref{eq:zeta_result}).
This second term is needed to cancel the UV divergence in the first term.
In Ref.\cite{Ferreiro:2001qy}, this divergent term was mis-identified as cancelling 
the tadpoles coming from other parts of the calculations (which we do not
have). This is because the authors showed 
that $ \nu_a(\bfxperp) $
is UV finite.
But that does not guarantee that $\nabla_\perp^2 \nu(\bfx)$ is UV finite.
In fact, we have just shown that it is not finite and needs 
the naturally arising counter term.

\subsection{JIMWLK Equation}
\label{sec:JIMWLK_eq}

Armed with $\barchi$ and $\eta$, we can now formulate the renormalization
group equation for the color density distribution.
We start with Eq.(\ref{eq:OLONLO4J}).
Since the $x^-$ dependence of 
$\lambda = \delta(x^- - \epsilon_Y)\tildezeta(\bfx)$ is not allowed to vary,
Eq.(\ref{eq:OLONLO4J}) should really be written as
\be
\ave{\calO}_{Y,\rm LO+NLO}
& = &
\int \calD\tilderho\,W_{\tilderho}[\tilderho]
\int \calD\tildezeta\, Y[\tildezeta;\tilderho]\,
\calO[A[\tilderho + \tildelambda]]
\ee
What we would like to do is to define 
$\tilderho' = \tilderho + \tildelambda$ and
rewrite the above as
\be
\ave{\calO}_{Y,\rm LO+NLO}
& = &
\int \calD\tilderho'\, \calW[\tilderho'] \calO[A[\tilderho']]
\ee
with 
\be
\calW[\tilderho']
= \int \calD\tildezeta\, W_\tilderho[\tilderho'-\tildelambda]
Y[\tildezeta;\tilderho'-\tildelambda]
\ee
and expand in $\tildelambda$.
Strictly speaking, however, this operation is not permitted.
The classical source 
$\tilderho(x^-, \bfxperp)$ 
by design does {\em not} depend
on $x^+$. If it does, the classical solution we found in
Section\ref{sec:ClassicalSolution} is not the right solution.
In the time domain,
$\tilderho(x^-,\bfxperp)$ is defined only within $0 \le x^- \le \epsilon_Y$.
On the other hand, $\tildelambda = \delta(x^- - \epsilon_Y)\tildezeta(\bfx)$
does depend on $x^+$ and has a point support in $x^-$.
In other words, $\tilderho$ and $\tildelambda$
(even $\tildezeta$) are not defined over the
same space and cannot be combined as 
$\tilderho'(x^-,\bfx) = 
\tilderho(x^-,\bfxperp) +\delta(x^- - \epsilon_Y)\tildezeta(x^+,\bfxperp)$
to form a new functional integration variable.
What is permissible is to let $\tilderho(x^-,\bfxperp)$ slowly vary in $x^+$
and use the $x^-$ integrated sources as the proper functional integration
variables. That is, let 
\be
\tilderho(x) = \tildesigma(\bfx)\delta_{\epsilon_Y}(x^-)
\ee
where $\delta_{\epsilon_Y}(x^-)$ is a smeared $\delta$-function with the width of 
$\epsilon_Y$. 
Then combine $\tilderho$ and $\tildelambda$ as
\be
\tildesigma'(\bfx) = \int dx^- (\tilderho(x) + \tildelambda(x))
= \tildesigma(\bfx) + \tildezeta(\bfx)
\ee

Expansion of $W_\tilderho$ and $Y$ can now be done consistently 
(dropping primes)
\be
\lefteqn{
\calW[\tildesigma] - W_\tilderho[\tildesigma]
} && 
\non
& \approx &
\int \calD\tildezeta\,
\left(
-
\int_{\bfu}
\tildezeta(\bfu)
{\delta \over \delta\tildesigma(\bfu)}
W_\tilderho[\tildesigma] Y[\tildezeta;\tildesigma]
+
{1\over 2}
\tildezeta(\bfu)\tildezeta(\bfv)
\int_{\bfu,\bfv}
{\delta^2 \over \delta\tildesigma(\bfu)\tildesigma(\bfv)}
W_\tilderho[\tildesigma] Y[\tildezeta;\tildesigma]
\right)
\non
& = &
-
\int_{\bfu}
{\delta \over \delta\tildesigma_a(\bfu)}
\ave{\tildezeta_a(\bfu)}
W_\tilderho[\tildesigma]
+
{1\over 2}
\int_{\bfu,\bfv}
{\delta^2 \over \delta\tildesigma_a(\bfu)\tildesigma_b(\bfv)}
\ave{ \tildezeta_a(\bfu)\tildezeta_b(\bfv)}
W_\tilderho[\tildesigma]
\label{eq:DeltaW}
\ee
again restricting $\tildesigma(\bfx)$ to vary adiabatically slowly in $x^+$.
The results Eqs.(\ref{eq:eta_result_0}) and (\ref{eq:zeta_result})
indicate that both 
$\ave{\tildezeta(\bfx)\tildezeta(\bfy)}$ 
and $\ave{\tildezeta(\bfx)}$ contain the divergent function
\be
B(x^+ - y^+)
& = &
\int {dk^- \over 8\pi |k^-|} e^{-ik^-(x^+ - y^+)}
\non
& = &
\int_{0}^{\infty} {dk^- \over 4\pi k^-} \cos(k^-(x^+ - y^+)) 
\ee
which needs an infrared cut-off.
Integrating between two cut-offs, $k^-_2 < k^- < k^-_1$, we get
\be
B(x^+ - y^+) \approx \ln(k^-_1/k^-_2)/4\pi = (Y_2 - Y_1)/4\pi
\ee
when $k^-_{1,2}|x^+ - y^+| \ll 1$ and 
$Y$ is the rapidity defined by $k^\pm = k_T e^{\pm Y}/\sqrt{2}$.

Factoring this out, the rest of $\ave{\tildezeta}$ and $\ave{\tildezeta\tildezeta}$
are functions of transverse coordinates only.
Hence, we can safely revert $\tildesigma(\bfz)\to \tildesigma(\bfzperp)$, too.
Eq.(\ref{eq:DeltaW}) then becomes in the $\Delta Y = \to 0$ limit,
\be
\pi {\partial W\over \partial Y}
& = &
-
\int_{\bfuperp}
{\delta \over \delta\alpha_a(\bfuperp)}
\nu_a(\bfuperp)
W[\tildesigma]
+
{1\over 2}
\int_{\bfuperp,\bfvperp}
{\delta^2 \over \delta\alpha_a(\bfuperp)\alpha_b(\bfvperp)}
\eta_{ab}(\bfuperp|\bfvperp)
W[\tildesigma]
\label{eq:JIMWLK}
\ee
where $\alpha(\bfzperp)$ is defined by 
\be
-\nabla^2_\perp \alpha(\bfxperp)
=
\tildesigma(\bfxperp)
\ee
To go from Eq.(\ref{eq:DeltaW}) to Eq.(\ref{eq:JIMWLK}), 
the counter term in $\ave{\tildezeta}$ is dropped because 
what is required in Eq.(\ref{eq:JIMWLK}) is now that $\nu$ be finite
instead of $\nabla_\perp^2 \nu$.

The functions $\nu$ and $\eta$ are related:
Taking a functional derivative of $\eta(\bfxperp|\bfyperp)$ in
Eq.(\ref{eq:eta_result}) and using the fact that 
$(T^a)_{bc} = -if_{abc}$ is totally anti-symmetric,
we can show (see Appendix \ref{app:Lines1to6})
\be
\nu^a(\bfuperp)
& = &
{1\over 2}\int_{\bfvperp} 
{\delta 
\eta^{ab}(\bfuperp|\bfvperp) 
\over \delta\alpha^b(\bfvperp)}
\ee
This allows the above to be written as an evolution equation
\be
{\partial W\over \partial Y}
& = &
\calH W
\ee
with the ``Hamiltonian'' given by
\be
\calH
& = &
{1\over 2\pi}
\int_{\bfuperp,\bfvperp}
{\delta\over \delta\alpha_a(\bfuperp)}
\eta^{ab}(\bfu_\perp|\bfv_\perp)
{\delta\over \delta\alpha_b(\bfvperp)}
\ee
This concludes our derivation of the JIMWLK equation.

\section{Discussions and Outlook}
\label{sec:outlook}

In this paper, we have derived the JIMWLK evolution 
equation starting from the Schwinger-Keldysh (SK) formulation of many-body QCD. 
Previously, Refs.\cite{Iancu:2000hn}
and \cite{Gelis:2008rw} introduced the SK formalism to the derivation of the 
JIMWLK equation.
The main difference here is the usage of the Keldysh rotation, or the $r$--$a$
formalism.  The main benefits of doing so includes: clear identification of
of classical and is quantum degrees of freedom, proper gauge transformation
rules, and proper identification of the interaction Lagrangian between
the external color current and the gluon field.

In the single-path formalism used in the original derivation, 
the gluon field is usually broken into 3 pieces
\be
A_\mu = \calA_\mu + \delta A_\mu + a_\mu
\ee
where $\calA_\mu$ is the classical field, $\delta A_\mu$ is the 
soft component of the fluctuation field to be identified as additional 
classical degree of freedom, and $a_\mu$ is regarded as the quantum degree
of freedom. This division was necessary because in the single path formalism,
the classical Yang-Mills equation arises in the context of saddle point
approximation. 
In contrast, we have 
\be
A_\mu = \calA_\mu + a_\mu
\ee
where the fluctuation field $a_\mu$ is explicitly identified as the classical
Jacobi-field and the source of quantum fluctuation is identified as residing only
in the distribution of the initial configurations of $a_\mu$.

In previous literature, gauge transformation rules were incompletely understood.
Consequently, the gauge invariant coupling between the external source and the
gluon field was also incomplete. By correctly identifying the gauge transformation
rules, we have identified that the correct coupling takes the following form
\be
S_{\rm int} = \int_x \eta^a_\mu(x)J_a^\mu(x)
\ee
The separation of $\eta$ and $A$ makes it completely natural
that the current is exactly given by
\be
{\delta S_{\rm int}\over \delta\eta_\mu^a(x)} = J_a^\mu(x)
\ee
which precesses according to
the covariant conservation equation $ [D_\mu, J^\mu] = 0 $
with the covariant derivative $D_\mu = \partial_\mu - igA_\mu$.
In previous literature, retarded current could arise
only within the context of saddle point approximation.

Despite these differences, the leading order JIMWLK equation 
is correctly derived in previous studies as well as here.
This is a testament to the 
correctness of the JIMWLK equation and the physical picture it represents.

In retrospect, it is very natural that the $r$-$a$ formalism
provides the cleanest separation of the classical and the quantum degrees of freedom.
First of all, classical mechanics is the mechanics of the expectation values
of the observables. The Schwinger-Keldysh formalism is for the
expectation values.
Second, since the classical mechanics is deterministic, it must correspond to
the common part of the bra and ket state time evolution in 
$\bra{\rm init}e^{i\hatH t}\hatO e^{-i\hatH t}\ket{\rm init}$. 
The Keldysh rotation accomplishes exactly that.

\begin{figure}[th]
\centerline{\includegraphics[width=0.9\tw]{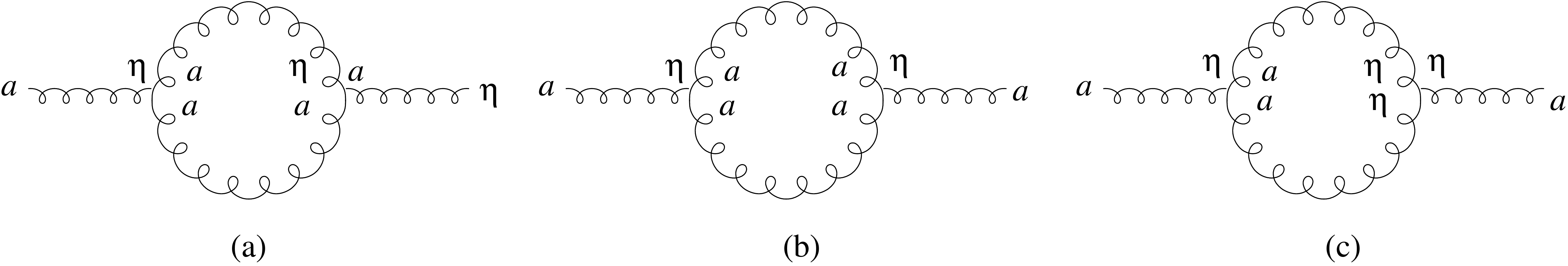}}
\caption{Gluon self-energy diagrams that contribute at NNLO.
Diagram (a) is for the retarded propagator and (b), (c) are for the symmetric
propagator.  These diagrams are all $O(g^2)$.
Distinguishable mirror images of these diagrams also contribute at NNLO.}
\label{fig:gluon_one_loops}
\end{figure}

\begin{figure}[th]
\centerline{\includegraphics[width=0.6\tw]{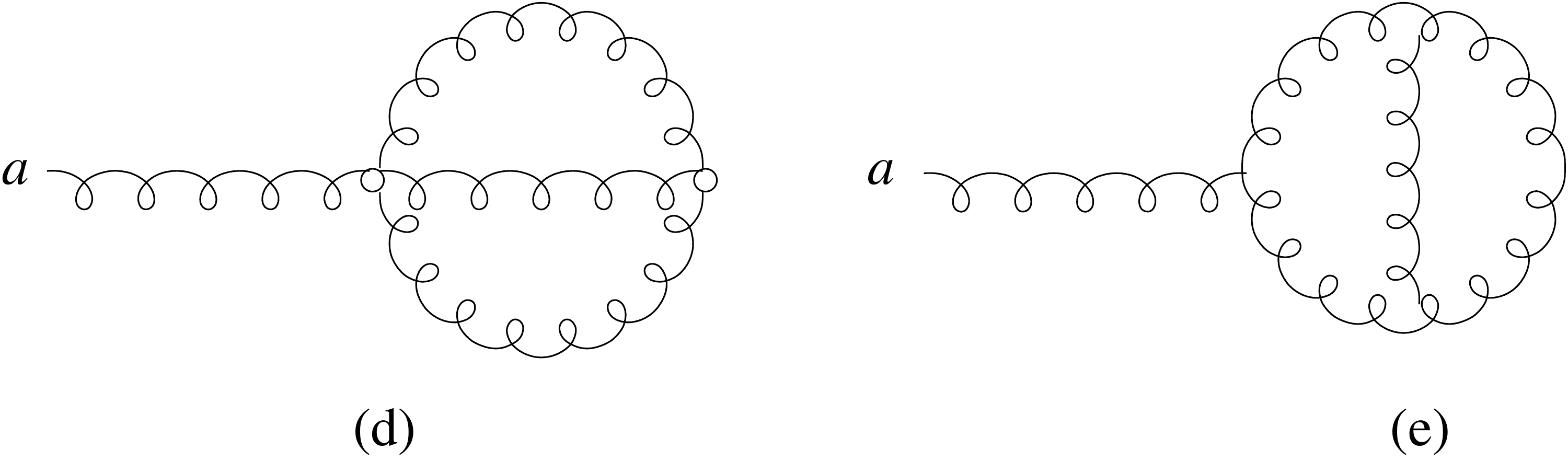}}
\caption{Gluon tadpole diagrams that contribute at NNLO to $\ave{a}$.
These diagrams are all $O(g^3)$. One cubic vertex can be a triple $\eta$
vertex.
Self-energy insertions to one-loops are not shown
since they can be regarded as a part of resummed propagators.
Distinguishable mirror images of these diagrams also contribute at NNLO.
}
\label{fig:gluon_tadpoles}
\end{figure}

Looking ahead, it isn't clear whether classical analysis can be still
used for the NNLO calculations. Certainly, the triple $\eta$ vertex
(which has the form $(ig/4)[\calD_\mu,\eta_\nu]_a[\eta^\mu,\eta^\nu]_a$) 
does contribute. 
In Fig.\ref{fig:gluon_one_loops}, we show all diagrams that contribute
to the 2-point functions at NNLO.
It is possible to draw diagrams that contain the retarded (advanced)
self-energy (diagram (a)
in Fig.\ref{fig:gluon_one_loops}) for the symmetric propagator.
However, as discussed in Appendix \ref{app:scalar_full_props}, they should be thought of as a part of the
resummed retarded (advanced) propagator.
The NNLO tadpole diagrams contributing to the 1-point average is shown in
Fig.\ref{fig:gluon_tadpoles}.

Diagrams involving the triple $\eta$ vertex
such as (c) are not contained
in the classical field program laid out in this paper.
On the surface, therefore, it looks as though the classical field approach will not
work at NNLO. However, notice that in these diagrams, no symmetric propagator
appears.  Equivalently, vacuum zero-point motions are not contained in these 
diagrams. 
In these diagrams, time flows from
$\eta$ to $a$ since $\ave{a(x)\eta(y)} = G_R(x|y)$ is the retarded propagator.
Hence, we can interpret the role of the triple $\eta$ vertex 
as a
quantum mechanism that produces three virtual particles out of either vacuum or
out of the strong background field in the causally connected past. 
In other words,
the role of the triple $\eta$ vertex here is to provide a
non-trivial 3-point correlations among $a$'s. 
It is then conceivable that these may be treated as a non-Gaussian part of the
vacuum density functional $\rho_{\rm v}$ or the source distribution 
$W[\tilderho]$, yet keeping the dynamics part to be still classical.
Whether this really is the case is yet to be seen. Work in this direction
is underway.

\section*{Acknowledgements}

This work is supported in part by the Natural Sciences and
Engineering Research Council of Canada.
Discussions with R.~Venugopalan, F.~Gelis, J.~Jalilian-Marian,
E.~Iancu, K.~Dusling, G.D.~Moore, and C.~Gale are gratefully acknowledged. 
I am especially
grateful to R.~Venugopalan for careful reading of 
earlier versions of the manuscript and numerous helpful suggestions,
and A.~Clerk for pointing out the condensed matter connections.
I also thank the 
hospitality of the Brookhaven National Laboratory where
part of this work was carried out.

\bibliography{big_list}{}

\begin{thebibliography}{10}

\bibitem{McLerran:1993ni}
L.~D. McLerran and R.~Venugopalan,
\newblock Phys.Rev. {\bf D49}, 2233 (1994).

\bibitem{McLerran:1993ka}
L.~D. McLerran and R.~Venugopalan,
\newblock Phys.Rev. {\bf D49}, 3352 (1994).

\bibitem{McLerran:1994vd}
L.~D. McLerran and R.~Venugopalan,
\newblock Phys.Rev. {\bf D50}, 2225 (1994).

\bibitem{Ayala:1995kg}
A.~Ayala, J.~Jalilian-Marian, L.~D. McLerran, and R.~Venugopalan,
\newblock Phys.Rev. {\bf D52}, 2935 (1995).

\bibitem{Ayala:1995hx}
A.~Ayala, J.~Jalilian-Marian, L.~D. McLerran, and R.~Venugopalan,
\newblock Phys.Rev. {\bf D53}, 458 (1996).

\bibitem{Iancu:2002xk}
E.~Iancu, A.~Leonidov, and L.~McLerran,
\newblock {\em {The Color glass condensate: An Introduction}}, 2002,
  hep-ph/0202270.

\bibitem{Kovner:2005pe}
A.~Kovner,
\newblock Acta Phys.Polon. {\bf B36}, 3551 (2005).

\bibitem{JalilianMarian:2005jf}
J.~Jalilian-Marian and Y.~V. Kovchegov,
\newblock Prog.Part.Nucl.Phys. {\bf 56}, 104 (2006).

\bibitem{Weigert:2005us}
H.~Weigert,
\newblock Prog.Part.Nucl.Phys. {\bf 55}, 461 (2005).

\bibitem{McLerran:2010ub}
L.~McLerran,
\newblock Acta Phys.Polon. {\bf B41}, 2799 (2010).

\bibitem{Gelis:2010nm}
F.~Gelis, E.~Iancu, J.~Jalilian-Marian, and R.~Venugopalan,
\newblock Ann.Rev.Nucl.Part.Sci. {\bf 60}, 463 (2010).

\bibitem{Balitsky:2010jf}
I.~Balitsky,
\newblock {\em {High-energy amplitudes in the next-to-leading order}}, 2010,
  1004.0057.

\bibitem{Kovchegov:2012csa}
Y.~V. Kovchegov,
\newblock AIP Conf.Proc. {\bf 1520}, 3 (2012).

\bibitem{Iancu:2012xa}
E.~Iancu,
\newblock {\em {QCD in heavy ion collisions}}, 2012, 1205.0579.

\bibitem{Gelis:2012ri}
F.~Gelis,
\newblock {\em {Color Glass Condensate and Glasma}}, 2012, 1211.3327.

\bibitem{Schwinger:1960qe}
J.~S. Schwinger,
\newblock J.Math.Phys. {\bf 2}, 407 (1961).

\bibitem{Keldysh:1964ud}
L.~Keldysh,
\newblock Zh.Eksp.Teor.Fiz. {\bf 47}, 1515 (1964).

\bibitem{Chou:1984es}
K.-c. Chou, Z.-b. Su, B.-l. Hao, and L.~Yu,
\newblock Phys.Rept. {\bf 118}, 1 (1985).

\bibitem{JalilianMarian:1996xn}
J.~Jalilian-Marian, A.~Kovner, L.~D. McLerran, and H.~Weigert,
\newblock Phys.Rev. {\bf D55}, 5414 (1997).

\bibitem{JalilianMarian:1997jx}
J.~Jalilian-Marian, A.~Kovner, A.~Leonidov, and H.~Weigert,
\newblock Nucl.Phys. {\bf B504}, 415 (1997).

\bibitem{JalilianMarian:1997gr}
J.~Jalilian-Marian, A.~Kovner, A.~Leonidov, and H.~Weigert,
\newblock Phys.Rev. {\bf D59}, 014014 (1998).

\bibitem{JalilianMarian:1997dw}
J.~Jalilian-Marian, A.~Kovner, and H.~Weigert,
\newblock Phys.Rev. {\bf D59}, 014015 (1998).

\bibitem{JalilianMarian:1998cb}
J.~Jalilian-Marian, A.~Kovner, A.~Leonidov, and H.~Weigert,
\newblock Phys.Rev. {\bf D59}, 034007 (1999).

\bibitem{Iancu:2001ad}
E.~Iancu, A.~Leonidov, and L.~D. McLerran,
\newblock Phys.Lett. {\bf B510}, 133 (2001).

\bibitem{Iancu:2001md}
E.~Iancu and L.~D. McLerran,
\newblock Phys.Lett. {\bf B510}, 145 (2001).

\bibitem{Kuraev:1977fs}
E.~Kuraev, L.~Lipatov, and V.~S. Fadin,
\newblock Sov.Phys.JETP {\bf 45}, 199 (1977).

\bibitem{Balitsky:1978ic}
I.~Balitsky and L.~Lipatov,
\newblock Sov.J.Nucl.Phys. {\bf 28}, 822 (1978).

\bibitem{Gribov:1984tu}
L.~Gribov, E.~Levin, and M.~Ryskin,
\newblock Phys.Rept. {\bf 100}, 1 (1983).

\bibitem{Mueller:1985wy}
A.~H. Mueller and J.-w. Qiu,
\newblock Nucl.Phys. {\bf B268}, 427 (1986).

\bibitem{Iancu:2000hn}
E.~Iancu, A.~Leonidov, and L.~D. McLerran,
\newblock Nucl.Phys. {\bf A692}, 583 (2001).

\bibitem{Gelis:2006yv}
F.~Gelis and R.~Venugopalan,
\newblock Nucl.Phys. {\bf A776}, 135 (2006).

\bibitem{Gelis:2006cr}
F.~Gelis and R.~Venugopalan,
\newblock Nucl.Phys. {\bf A779}, 177 (2006).

\bibitem{Gelis:2006ye}
F.~Gelis and R.~Venugopalan,
\newblock Nucl.Phys. {\bf A782}, 297 (2007).

\bibitem{Gelis:2007pw}
F.~Gelis, S.~Jeon, and R.~Venugopalan,
\newblock Nucl.Phys. {\bf A817}, 61 (2009).

\bibitem{Gelis:2007hj}
F.~Gelis and R.~Venugopalan,
\newblock Nucl.Phys. {\bf A783}, 149 (2007).

\bibitem{Gelis:2008rw}
F.~Gelis, T.~Lappi, and R.~Venugopalan,
\newblock Phys.Rev. {\bf D78}, 054019 (2008).

\bibitem{Gelis:2008ad}
F.~Gelis, T.~Lappi, and R.~Venugopalan,
\newblock Phys.Rev. {\bf D78}, 054020 (2008).

\bibitem{Gelis:2008sz}
F.~Gelis, T.~Lappi, and R.~Venugopalan,
\newblock Phys.Rev. {\bf D79}, 094017 (2009).

\bibitem{Dusling:2010rm}
K.~Dusling, T.~Epelbaum, F.~Gelis, and R.~Venugopalan,
\newblock Nucl.Phys. {\bf A850}, 69 (2011).

\bibitem{Dusling:2011rz}
K.~Dusling, F.~Gelis, and R.~Venugopalan,
\newblock Nucl.Phys. {\bf A872}, 161 (2011).

\bibitem{Dusling:2012yd}
K.~Dusling, T.~Epelbaum, F.~Gelis, and R.~Venugopalan,
\newblock {\em {Quantum chaos in the perfect fluid: spectrum of initial
  fluctuations in the little bang}}, 2012, arXiv/1210.6053.

\bibitem{Kamenev1}
A.~Kamenev and A.~Levchenko,
\newblock Advances in Physics {\bf 58}, 197 (2009).

\bibitem{Kamenev2}
A.~Kamenev,
\newblock {\em {Many-body theory of non-equilibrium systems}}, 2004,
  arXiv/cond-mat/0412296.

\bibitem{Berges:2007ym}
J.~Berges and T.~Gasenzer,
\newblock Phys.Rev. {\bf A76}, 033604 (2007).

\bibitem{Berges:2007re}
J.~Berges, S.~Scheffler, and D.~Sexty,
\newblock Phys.Rev. {\bf D77}, 034504 (2008).

\bibitem{Berges:2011sb}
J.~Berges, S.~Scheffler, S.~Schlichting, and D.~Sexty,
\newblock Phys.Rev. {\bf D85}, 034507 (2012).

\bibitem{Berges:2012ev}
J.~Berges, S.~Schlichting, and D.~Sexty,
\newblock Phys.Rev. {\bf D86}, 074006 (2012).

\bibitem{Berges:2012cj}
J.~Berges and S.~Schlichting,
\newblock Phys.Rev. {\bf D87}, 014026 (2013).

\bibitem{Jeon:2004dh}
S.~Jeon,
\newblock Phys.Rev. {\bf C72}, 014907 (2005).

\bibitem{Mueller:2002gd}
A.~Mueller and D.~Son,
\newblock Phys.Lett. {\bf B582}, 279 (2004).

\bibitem{JalilianMarian:2000ad}
J.~Jalilian-Marian, S.~Jeon, and R.~Venugopalan,
\newblock Phys.Rev. {\bf D63}, 036004 (2001).

\bibitem{Fukushima:2005kk}
K.~Fukushima,
\newblock Nucl.Phys. {\bf A770}, 71 (2006).

\bibitem{Fukushima:2006cj}
K.~Fukushima,
\newblock Nucl.Phys. {\bf A775}, 69 (2006).

\bibitem{Mrowczynski:1994nf}
S.~Mrowczynski and B.~Muller,
\newblock Phys.Rev. {\bf D50}, 7542 (1994).

\bibitem{DeWittMorette:1976up}
C.~DeWitt-Morette,
\newblock Annals Phys. {\bf 97}, 367 (1976).

\bibitem{Bazanski:1976sa}
S.~Bazanski,
\newblock Acta Phys.Polon. {\bf B7}, 305 (1976).

\bibitem{Carta:2005fq}
P.~Carta, E.~Gozzi, and D.~Mauro,
\newblock Annalen Phys. {\bf 15}, 177 (2006).

\bibitem{Kovchegov:1996ty}
Y.~V. Kovchegov,
\newblock Phys.Rev. {\bf D54}, 5463 (1996).

\bibitem{Kovchegov:1997ke}
Y.~V. Kovchegov and D.~H. Rischke,
\newblock Phys.Rev. {\bf C56}, 1084 (1997).

\bibitem{Ferreiro:2001qy}
E.~Ferreiro, E.~Iancu, A.~Leonidov, and L.~McLerran,
\newblock Nucl.Phys. {\bf A703}, 489 (2002).

\bibitem{Greiner:1998vd}
C.~Greiner and S.~Leupold,
\newblock Annals Phys. {\bf 270}, 328 (1998).

\bibitem{Greiner:1998ri}
C.~Greiner and S.~Leupold,
\newblock Eur.Phys.J. {\bf C8}, 517 (1999).

\bibitem{Heinzl:2007ca}
T.~Heinzl and A.~Ilderton,
\newblock J.Phys. {\bf A40}, 9097 (2007).

\end{thebibliography}
\bibliographystyle{mybst}

\appendix

\section{Scalar propagators in the background field -- Minkowski space}
\label{app:Scalar_props}

In this section, we work out in some detail propagators in a background field
that arise from the generating functional
\be
\calZ[J_r, J_a]
& = & 
\int\calD\phi_r \calD\phi_a 
\rho_{\rm v}[\phi_r^i, \dot{\phi}_r^i]\,
\exp\left(i\int_{t_{i}}^{\infty} dt\int d^3x\, \calL_J \right)
\label{eq:calZapp}
\ee
where
\be
\int_{t_{i}}^{\infty}dt\int d^3x\, \calL_J 
& = &
\int_{t_{i}}^{\infty}dt\int d^3x\, 
 \Big(
 \phi_a E[\phi_r, J_a]
 + J_r \phi_r
 \Big)
\ee
is the linear Jacobi field Lagrangian with the linear Jacobi-field equation
\be
 E[\phi_r, J_a]
 =
 -\partial^2\phi_r - m^2\phi_r 
 - V''(\barphi) \phi_r
 + J_a
\ee
enforced by the integration over $\phi_a$.

\subsection{Retarded propagator}

In the free-field limit, the retarded propagator for scalar field 
satisfies
\be
(\partial^2 + m^2)G_R^0(x|y) = -\delta(x-y)
\ee
The solution is
\be
G_R^0(t-t',\bfk) = -\theta(t-t') {\sin(E_k (t-t'))\over E_k}
\ee
where we have Fourier-transformed in space.

When there is a strong background field, the retarded propagator
is required to satisfy
\be
(\partial_x^2 + m^2 + V''(\barphi(x)))G_R(x|y) = -\delta(x-y)
\ee
This is solved by 
\be
G_R(x|y) = G_R^0(x|y) + \int_{u,v} G_R^0(x|u)V''(\barphi(u))G_R(u|y)
\ee
By applying $\int_y G_R(x|y)$ to
\be
(\partial_x^2 + m^2 + V''(\barphi))_y \phi_r(y) = J_a(y)
\ee
and integrating by parts twice, 
the formal solution of the Jacobi-field equation is obtained as follows
\be
\phi_r(x) = \phi_h(x) 
-\int_{t_i}^{\infty}dy^0 \int_{\bfy} G_R(x|y) J_a(y)
\label{eq:phir_sol}
\ee
where the homogeneous solution $\phi_h(x)$
comes from the boundary terms at $y^0 = t_i$
\be
\phi_h(x) = 
\int d^3y\,
\left[
\partial_0^yG_R(x|y^0,\bfy) 
\right]_{y^0=t_i}
\phi_r(t_{i}, \bfy)
-
\int d^3y\,
G_R(x|t_{i},\bfy) \dot{\phi}_r(t_{i}, \bfy)
\ee
Consistency demands
\be
\lim_{x^0\to t_{i}} 
\lim_{y^0\to t_{i}} 
\partial_0^yG_R(x|y) 
= \delta(\bfx-\bfy)
\ee
and
\be
\lim_{y^0\to t_{i}} 
\lim_{x^0\to t_{i}} 
\partial_0^xG_R(x|y) 
= -\delta(\bfx-\bfy)
\ee
These conditions are consistent with the operator definition of the retarded
propagator
\be
G_R(t,t') = -i\theta(t-t')
\ave{[\phi(t,\bfx), \phi(t',\bfx')]}
\ee
Setting $g(t,\bfx|t',\bfy) = -i\ave{[\phi(t,\bfx), \phi(t',\bfy)]}$,
we have 
\be
\partial_t g(t|t') 
&=& 
-i\ave{[\pi(t,\bfx), \phi(t',\bfy)]}
\ee
where $\dot{\phi} = \pi$ is the momentum field.
In the equal time limit, $\partial_tg$ just becomes equal time commutator
relationship between the field and its canonical momentum
\be
\lim_{t\to t'}
\partial_t g(t,\bfx|t',\bfy) 
& = &
-i\ave{[\pi(t,\bfx), \phi(t',\bfy)]}
\non 
& = & -\delta(\bfx-\bfy)
\ee
Likewise, 
$ \lim_{t\to t'} \partial_{t'} g(t,\bfx|t',\bfy) =  \delta(\bfx-\bfy) $.
It is also clear that 
$G_R(t,\bfx|t,\bfy) = 0$ due to causality.

Putting the solution Eq.(\ref{eq:phir_sol}) back into the generating functional
Eq.(\ref{eq:calZapp}) then yields
\be
\int_x \calL_J 
= 
\int_x \phi_h(x) J_r(x)
-
\int_{x,y} J_r(x) G_R(x|y) J_a(y)
\ee
Differentiating $\calZ$ with respect to $iJ_r(x)$ and $iJ_a(y)$ 
results in $G_R(x|y)$.

\subsection{The symmetric propagator}

To obtain the $rr$ correlator, the homogeneous solution obtained above
needs to be averaged over the vacuum fluctuations
\be
\rho_{\rm v}[\phi^i_r, \pi^i_r] = 
\exp\left(-\int_\bfk 
\left(
{1\over E_k} \pi^i_r(\bfk)\pi^i_r(-\bfk) + E_k\phi^i_r(\bfk)\phi^i_r(-\bfk) \right)
\right)
\ee
Fourier-transforming in space, we have
\be
\lefteqn{
\int[d\phi^i_r][d\pi^i_r] 
\rho_{\rm v}[\phi_r^i,\pi_r^i] e^{i\int \phi_h J_r}
} & &
\non
& = &
\int[d\phi^i_r][d\pi^i_r] 
\exp\Bigg(
-
i\int dt\, J_r(t,-\bfp)\,
\partial_{t'} g(t,\bfp|t_{i},-\bfk)\,\phi^i_r(\bfk) 
+
i\int dt\, J_r(t,-\bfp)\,
g(t,\bfp|t_{i},-\bfk)\,\pi_r^{i}(\bfk)
\non & & {}\qquad\qquad
-
{1\over E_k} \pi^i_r(\bfk)\pi^i_r(-\bfk) 
-
E_k\phi^i_r(\bfk)\phi^i_r(-\bfk) 
\Bigg)
\ee
where integrations over $\bfk$ and $\bfp$ should be understood.
Completing square for $\phi_r^i$ and integrating over the resulting
Gaussian yields
\be
S_\phi
& = &
-{1\over 4E_k}
\int dt dt'\,
J_r(t,-\bfp)
\partial_{t'}g(t,\bfp|t_{i}, -\bfk)
\partial_{t'}g(t,\bfq|t_{i}, -\bfk) 
J_r(t',-\bfq)
\ee
in the exponent.
Doing the same for $\pi_r^i$ yields
\be
S_\pi
& = &
-{E_k\over 4}
\int dt dt'\,
J_r(t,-\bfq)\,
g(t,\bfq|t_{i},-\bfk)
g(t,\bfp|t_{i},-\bfk) 
J_r(t',-\bfp)
\ee
again in the exponent.
Differentiating with respect to $iJ_r$ twice, we finally get
\be
\lefteqn{
G_{S}(t, \bfp| t', \bfq)
}&&
\non
&=&
\lim_{u\to t_{i}} \lim_{v\to t_{i}}
\int {d^3k\over (2\pi)^3 2E_k}
\Bigg[
\partial_{u}g(t,\bfp|u, -\bfk)
\partial_{v}g(t',\bfq|v, -\bfk) 
+
E_k^2
g(t,\bfp|u,-\bfk)
g(t',\bfq|v,-\bfk) 
\Bigg]
\non
\label{eq:appGSinit}
\ee
For the free field case,
\be
G_{S}^0(t-s,\bfk)
& = &
{1\over 2E_k}
\left(
\cos(E_k(t-t_i))\cos(E_k(s-t_i))
+
\sin(E_k(t-t_i))\sin(E_k(s-t_i))
\right)
\non
& = &
{\cos(E_k(t-s))\over 2 E_k}
\ee
or in the full momentum space
\be
G_S^0(k) = \pi\delta(k^2 - m^2)
\ee

Using the identity
\be
\int_{t_i}^{\infty} dt'\, 
\partial_{t'}( e^{\pm iE_kt'} (\partial_{t'} \mp iE_k)g(t|t'))
& = &
\int_{t_i}^{\infty} dt'\, 
e^{\pm iE_kt'} (\partial_{t'}^2 + E_k^2)g(t|t'))
\ee
the symmetric propagator can be re-expressed as
\be
\lefteqn{G_{S}(t,\bfp|s,\bfq)}
&& 
\non
& = &
\int {d^4 k\over (2\pi)^4}\,\pi\delta(k_0^2 - E_k^2)\,
\non
& & {} \times
\left[
\int_{t_i}^\infty dt'\,
e^{-ik^0t'}(\partial_{t'}^2 + E_k^2)g(t,\bfp|t',-\bfk)
\int_{t_i}^\infty ds'\,
e^{ik^0s'}(\partial_{s'}^2 + E_k^2)g(s,\bfq|s',-\bfk)
\right]
\ee
Symbolically, this may be written as
\be
G_{S}(x|y) = \int_{u,v}
\left( (G_R^0)_u^{-1} G_R(x|u) \right) 
G_{S}^0(u|v) 
\left( (G_A^0)_v^{-1}G_A(v|y) \right)
\label{eq:appGS}
\ee
The same results can be also obtained by diagramatical analysis
\cite{Gelis:2006yv,Gelis:2006cr,Dusling:2010rm}. 
It should be emphasized here that the appearance of $G_{R,A}$ as a propagator
has nothing to do with quantum mechanics as they come from the classical
solution Eq.(\ref{eq:phir_sol}).
We should also note that if we had a classical vacuum where the initial
state density is given by
$\rho_{\rm cl, v}[\phi_r^i, {\pi}_r^i]
=
\delta[\phi_r^i]\delta[{\pi}_r^i]$,
then the symmetric propagator vanishes, $G_S = 0$. 
Non-zero $G_{S}$ is obtained only if we have a non-trivial
$\rho_{\rm v}$.

At finite temperature, the equilibrium symmetric propagator is given by
\be
G^0_S(p) = (n(E_p) + 1/2)(2\pi)\delta(p^2 - m^2) 
\ee
where $n(p^0) = 1/(e^{\beta p^0} - 1)$ is the Bose-Einstein distribution
function. 
This shows that calculating $G_S$ is equivalent to calculating
the phase space density. For the definition of the phase space density
in non-equilibrium systems in this context,
see Refs.\cite{Greiner:1998vd,Greiner:1998ri}.

\subsection{Full Propagators}
\label{app:scalar_full_props}

In this section, we work out relationship between full propagators, 
not just the Jacobi-field Green functions.
The full propagators are defined by
\be
iG_R(x|y) &=& G_{ra}(x|y) = \ave{\phi_r(x)\phi_a(y)}.
\\
iG_A(x|y) &=& G_{ar}(x|y) = \ave{\phi_a(x)\phi_r(y)}.
\\
G_S(x|y) &=& G_{rr}(x|y) = \ave{\phi_r(x)\phi_r(y)}.
\ee
The interaction terms are of the form 
$V'''(\barphi)\phi_a\phi_r^2$, $V''''(\barphi)\phi_a\phi_r^3$
and $ V''''(\barphi) \phi_a^3\phi_r $.
There are three different kinds of 
self-energies in this theory:
\be
\Sigma_R(x|y) & = & \Sigma_{ar}(x|y)
\\
\Sigma_A(x|y) & = & \Sigma_{ra}(x|y)
\\
\Sigma_S(x|y) & = & \Sigma_{aa}(x|y)
\ee
where, for instance,
the subscripts $ra$ indicates
that the available field to connect from left is 
the type $r$ and the one on the right is type $a$.

In the perturbative expansion, the leading order correlator
$\ave{\phi_a(x)\phi_a(y)}_0 = G^0_{aa}(x|y)$ does not exist.
If there exists $\ave{\phi_a(x)\phi_a(y)}$ beyond the leading order,
it must come from self-energy insertions. 
Since this self-energy must connect two $\phi_a$'s,
the perturbative expansion must contain at least one $\Sigma_{rr}(u|v)$.
In the Schwinger-Keldysh formalism,
the interaction terms must contain an odd number of $\phi_a$'s.
Therefore, if we want to form $\Sigma_{rr}(u|v)$ by connecting the
interaction terms, the connected part of this
self-energy must include at least one $\ave{\phi_a(u')\phi_a(v')}$.
Otherwise, a time-loop forms. Therefore, the 
existence of $\Sigma_{rr}$ depends on the existence of $G_{aa}$.
But then, the existence of $G_{aa}$ depends on the existence of
$\Sigma_{rr}$. This circular argument can be resolved only if both vanishes.

In perturbative expansion of the retarded propagator $G_{ra}$, 
insertions of $\Sigma_A$ or $\Sigma_S$ cannot appear because 
they always lead to $\ave{\phi_a(x)\phi_a(y)} = 0$.
Therefore
\be
G_R(x|y) = G_R^0(x|y) + \int_{u,v} G_R^0(x|u)\Sigma_R(u|v)G_R(v|y)
\ee
By applying $\partial^2 + m^2$ it is easy to check that
this satisfies
\be
(\partial^2 + m^2)G_R(x|y) + \int_u \Sigma_R(x|u)G_R(u|y) = -\delta(x|y)
\label{eq:app_GRdef}
\ee
Below, we will need the alternative form
\be
G_R(x|y) = G_R^0(x|y) + \int_{u,v} G_R(x|u)\Sigma_R(u|v)G_R^0(v|y)
\label{eq:alt_GR}
\ee
For $G_A$, the same logic applies to yield
\be
G_A(x|y) = G_A^0(x|y) + \int_{u,v} G_A^0(x|u)\Sigma_R(u|v)G_A(v|y)
\label{eq:app_GA}
\ee
and
\be
G_A(x|y) = G_A^0(x|y) + \int_{u,v} G_A(x|u)\Sigma_R(u|v)G_A^0(v|y)
\label{eq:alt_GA}
\ee
For the symmetric propagator, all three types of self-energies can appear
\be
G_S(x|y)
& = & G_S^0(x|y)
\non
& & {}
+ \int_{u,v} G_R(x|u)\Sigma_R(u|v)G_S^0(v|y)
+ \int_{u,v} G_S^0(x|u)\Sigma_A(u|v)G_A(v|y)
\non
& & {}
+ \int_{u,v} G_R(x|u)\Sigma_S(u|v)G_A(v|y)
+ \int_{u,v,w,z} G_R(x|u)\Sigma_R(u|v)G_S^0(v|w)\Sigma_A(w|z)G_A(z|y)
\non
\ee
The leading order symmetric propagator $G_S^0$ can appear only once
in any $G_{r\cdots}^0 \Sigma G^0 \cdots \Sigma G^0_{\cdots r}$ chain. 
This is because $G_S^0 = G_{rr}^0$ can either start or end a chain,
or it must appear accompanied by $\Sigma_R$ and
$\Sigma_A$ in the following way
\be
\Sigma_R(u)G_S^0(u|v)\Sigma_A(v)
& = &
\Sigma_{ar}(u)G_{rr}^0(u|v)\Sigma_{ra}(v)
\ee
since $\Sigma_{rr}$ does not exist.
In this combination,
both free-ends are $\phi_a$. Hence more than one insertion of this
combination inevitably leads to $G_{aa}(u'|v') = 0$ somewhere in the chain.
The symmetric self-energy $\Sigma_S = \Sigma_{aa}$ can appear in the chain
also only once for exactly the same reason.

From Eq.(\ref{eq:alt_GR}), we get
\be
G_R(x|u)\Sigma_R(u|y) = -\delta(x-y) -(\partial^2 + m^2)_yG_R(x|y)
\ee
and also
\be
\Sigma_A(x|v)G_A(v|y) = -\delta(x-y) -(\partial^2 + m^2)_xG_A(x|y)
\ee
The symmetric propagator then simplifies to
\be
G_S(x|y) 
& = &
\int_u G_R(x|u)\Sigma_S(u)G_A(u|y)
\non & & {}
+ 
\int_{u,v}
\left(
(\partial^2 + m^2)_u G_R(x|u)
\right)
G_S^0(u|v)
\left(
(\partial^2 + m^2)_v G_A(v|y)
\right)
\label{eq:appGSfull}
\ee
Compared to Eq.(\ref{eq:appGS}), there is an additional term 
$\int_u G_R(x|u)\Sigma_S(u)G_A(u|y)$. Since $\Sigma_S$ contains more powers
of coupling constants, this is explicitly higher order.
If the time integrations in Eq.(\ref{eq:appGSfull}) extends from $-\infty$ to
$\infty$, then the last term can be made to
vanish provided that all functions vanish
at $t = \pm\infty$.  
For the finite initial time, the second term is the same as
Eq.(\ref{eq:appGSinit}).

\section{Scalar Field Light-Cone Propagators}
\label{app:scalar_lightcone}

One important difference in using the light-cone coordinate system 
is the fact that the d'Alembertian 
$\partial^2 = 2\partial_+ \partial_- - \nabla_\perp^2$ is
only linear in the time derivative $\partial_-$.
Therefore, the vacuum functional can depend only on the initial value of the
field, but not on the time-derivative of the field.

The retarded Green function $G_R(x|y)$ vanishes when $x^- < y^-$.
Hence it can be written as
\be
G_R(x|y) = \theta(x^- - y^-)g_R(x|y)
\label{eq:GRdef}
\ee
where $g_R(x|y)$ is a homogeneous solution satisfying 
\be
(2\partial_+ \partial_- - \nabla_\perp^2 + m^2 + V''(\barphi))_x g_R(x|y) = 0
\\
(2\partial_+ \partial_- - \nabla_\perp^2 + m^2 + V''(\barphi))_y g_R(x|y) = 0
\ee
Applying
$\partial_+ \partial_-$ to Eq.(\ref{eq:GRdef}) yields
\be
\partial_{x^+}\partial_{x^-} G_R(x|y)
= 
\delta(x^- - y^-)\partial_{x^+}g_R(x|y)
+
\theta(x^- - y^-)\partial_{x^+} \partial_{x^-} g_R(x|y)
\ee
Since we require $(\partial^2 + m^2 + V'')_xG_R(x|y) = -\delta(x-y)$,
the homogeneous solution must satisfy
\be
\partial_{x^+} g_R(x^-, \bfx| x^-, \bfy) = -{1\over 2}\delta(\bfx-\bfy)
\ee
with $\bfx \equiv (x^+, \bfx_\perp)$.
We also require $g_R$ to be symmetric in the sense that
\be
\partial_{y^+}g_R(x^-,\bfx|x^-,\bfy) = {1\over 2}\delta(\bfx-\bfy)
\label{eq:yplusg}
\ee
so that $(\partial^2 + m^2 + V'')_yG_R(x|y) = -\delta(x-y)$ as 
well.

With the Green function at hand, we can now construct a solution
for an initial value problem.
For this procedure, we keep the initial time $x^-_{\rm init}$ to be 
finite
and take the limit $x^-_{\rm init} \to -\infty$ only at the end.
By applying $\int_{x^-_{\rm init}}^\infty dy^-\int_{\bfy} G_R(x|y)$ to
\be
(\partial^2 + m^2 + V''(\barphi))_y \phi_r(y) = J_a(y)
\ee
and performing integrations by parts,
we obtain
\be
\phi_r(x) = \phi_h(x) 
- \int_{x^-_{\rm init}}^\infty dy^- \int_\bfy G_{R}(x|y) J_a(y)
\label{eq:phi_r_sol}
\ee
where
\be
\phi_h(x^-, \bfx)
& = &
2
\int_{\bfy}
\left( \partial_{y^+} G_R(x^-, \bfx| x^-_{\rm init}, \bfy) \right)
\phi_i(\bfy)
\label{eq:LCphih}
\ee
is the homogeneous solution coming from the surface term in carrying out 
the integration by parts.
The boundary condition
\be
\lim_{x^-\to x^-_{\rm init}+0^+}\phi_h(x^-,\bfx) = \phi_i(\bfx)
\ee
is trivially satisfied because of Eq.(\ref{eq:yplusg}).
Here $0^+$ in the limit indicates that the
limit must be approached from above (Recall that we have defined 
$\theta(0) = 0$ when it multiplies a $G_R$.).
Since only the first derivative of $x^-$ appears in $\partial^2$,
only the initial data on the field is required for the initial value
problem.

Putting the solution (\ref{eq:phi_r_sol}) into the 
generating functional, we obtain
\be
\calZ_0[J_a, J_r| \barJ]
= 
\exp\left( -i\int J_r G_R J_a \right)
\int[d\phi_i]\, \rho_{\rm v}[\phi_i]\,
\exp\left(i\int J_r \phi_h \right)
\ee
which clearly shows
\be
{\delta\over \delta iJ_r(x)}
{\delta\over \delta iJ_a(x)} \ln \calZ_0 
=
\ave{\phi_r(x)\phi_a(y)}
=
iG_R(x|y)
\ee

To get the symmetric propagator 
$G_S(x|y) = \ave{\phi_r(x)\phi_r(y)}$ in terms of 
the retarded propagator,
we first require that the 
the classical field $\barphi$ satisfy the null boundary condition
\be
\lim_{x^-\to x^-_{\rm init}} \barphi(x) = 0
\ee
so that the space in the far past is the vacuum.
The spectrum of initial state vacuum fluctuation
is then given by a Gaussian \cite{Heinzl:2007ca}
\be
\rho_{\rm v}[\phi_i]
& = &
\exp\left(
-2
\int_{\bfk}
\phi_i(\bfk)\phi_i(-\bfk)|k^-|
\right)
\label{eq:scalarrhov}
\ee
where $\bfk = (k^-, \bfkperp)$.

Using Eq.(\ref{eq:LCphih}), we have
\be
\ave{\phi_r(x)\phi_r(y)}
& = &
\ave{\phi_h(x)\phi_h(y)}
\non
& = &
4
\int_{\bfu,\bfv} 
\left( \partial_{u^+} G_R(x^-, \bfx| x^-_{\rm init}, \bfu) \right)
\left( \partial_{v^+} G_R(y^-, \bfy| x^-_{\rm init}, \bfv) \right)
\ave{ \phi_i(\bfu) \phi_i(\bfv) }_{\rm v}
\non
\ee
where the average $\ave{\cdots}_{\rm v}$ is carried out with
$\rho_{\rm v}$.
Carrying out the Gaussian integral results in
\be
\ave{\phi_i(\bfu)\phi_i(\bfv)}_{\rm v}
=
\int_{\bfk}
{1\over 4|k^-|}
e^{- ik^- (u^+ - v^+) + i\bfkperp{\cdot}(\bfuperp-\bfvperp)}\,
\ee
Defining the free field propagator
\be
G_S^0(x)
& = &
\int_k
e^{-ik^+ x^- - ik^- x^+ + i\bfkperp{\cdot}\bfxperp}\,
\pi\delta(2k^+ k^- - \bfkperp^2)
\non
& = &
\int_{\bfk}
{1\over 4|k^-|}
e^{-i(\bfkperp^2/2k^-) x^- - ik^- x^+ + i\bfkperp{\cdot}\bfxperp}\,
\label{eq:GS0h}
\ee
we finally get
\be
G_S(x|y)
& = &
\ave{\phi_h(x)\phi_h(y)}
\non
& = &
4\int_{\bfu,\bfv}
[\partial_{u^+}G_R(x|x^-_{\rm init},\bfu)]
G_{S}^0(0,\bfu-\bfv)
[\partial_{v^+}G_A(x^-_{\rm init}, \bfv|y)]
\label{eq:GSfullh}
\ee
where we used the fact that the advanced
Green function can be obtained by the relationship
$G_R(x|y) = G_A(y|x)$.
We can readily check
that $G_R^0$ given above and $G_S^0$ given in Eq.(\ref{eq:GS0h})
is consistent with Eq.(\ref{eq:GSfullh}).

\section{Propagators in SK-QCD -- Light-cone metric}
\label{app:props_details}

We start from Eq.(\ref{eq:resummedGR}):
\be
\lefteqn{
\tildeG_R(x|y)
} && 
\non
& = &
G_R^0(x|y)
\non & & {}
+
2(ig)
\int d^4u_1\, 
(\partial_{u_1^+}G_R^0(x|u_1)
\ul{\tildecalA}_-(u_1^-,\bfuperp{}_1)
G_R^0(u_1|y)
\non & & {}
+
2(ig)^2
\int d^4u_1\, 
(\partial_{u_1^+} G_R^0(x|u_1))
\ul{\tildecalA}_-(u_1^-,\bfuperp{}_1)
\int du_2^-\,
\theta(u_1^- - u_2^-)
\ulV^\dagger(u_1^-, u_2^-; \bfuperp{}_1)
\non & & {} \qquad\qquad\times
\ul{\tildecalA}_-(u_2^-,\bfuperp{}_1)
G_R^0(u_2^-,\bfu_1|y)
\non
& = &
G_R^0(x|y)
\non & & {}
+
2(ig)
\int d^4u_1\, 
(\partial_{u_1^+}G_R^0(x|u_1)
\ul{\tildecalA}_-(u_1^-,\bfuperp{}_1)
G_R^0(u_1|y)
\non & & {}
-
2
\int d^4u_1\, 
(\partial_{u_1^+} G_R^0(x|u_1))
\int du_2^-\,
\theta(u_1^- - u_2^-)
\left(
\partial_{u_1^-}
\partial_{u_2^-}
\ulV^\dagger(u_1^-, u_2^-; \bfuperp{}_1)
\right)
G_R^0(u_2^-,\bfu_1|y)
\non
& = &
G_R^0(x|y)
\non & & {}
-2
\int_{0}^{\epsilon_Y} du_1^-
\int_{0}^{\epsilon_Y} du_2^-
\int_{\bfu_1} 
(\partial_{u_1^+}G_R^0(x|u_1))
\left[
\partial_{u_1^-}
\left(\theta(u^-_1-u^-_2)\partial_{u_2^-}\ulV^\dagger(u_1^-,u_2^-;\bfu_{1,\perp})
\right)
\right]
G_R^0(u_2^-,\bfu_1|y) 
\non
\label{eq:appGRresummed}
\ee
using the facts that the time-ordered exponentials obey
\be
\partial_{u_1^-}\ulV^\dagger(u_1^-,u_2^-;\bfu_{1,\perp})
=
ig\ul{\tildecalA}_-(u_1^-,\bfu_{1,\perp}) 
\ulV^\dagger(u_1^-,u_2^-;\bfu_{1,\perp})
\ee
and
\be
\partial_{u_2^-}\ulV^\dagger(u_1^-,u_2^-;\bfu_{1,\perp})
=
-ig
\ulV^\dagger(u_1^-,u_2^-;\bfu_{1,\perp})
\ul{\tildecalA}_-(u_2^-,\bfu_{1,\perp}) 
\ee
From this expression, it is easy to see that for $y^- > \epsilon_Y$
or $x^- < 0$, $\tildeG_R(x|y) = \tildeG_R^0(x|y)$ since the integral 
part vanishes in this case due to the enforced time ordering.

The expressions for $\tildeG_R$ for the cases where both $x^-$ and $y^-$ are 
not in $(0,\epsilon_Y)$ are already given in Section \ref{sec:qcd_props}.
To find expressions for other time orderings, we integrate by parts
with respect to both $u_1^-$ and $u_2^-$ to get
\be
\lefteqn{\tildeG_R(x|y)}
&&
\non
& = &
G_R^0(x|y)
\non & & {}
-2
\int_{\bfu_1} 
(\partial_{u_1^+}G_R^0(x|\epsilon_Y, \bfu_1))
G_R^0(\epsilon_Y,\bfu_1|y) 
\non && {}
+2
\int_{\bfu_1} 
(\partial_{u_1^+}G_R^0(x|\epsilon_Y, \bfu_1))
\ulV^\dagger(\epsilon_Y,0;\bfu_{1,\perp})
G_R^0(0,\bfu_1|y) 
\non && {}
+2
\int_{0}^{\epsilon_Y} du_2^-
\int_{\bfu_1} 
(\partial_{u_1^+}G_R^0(x|\epsilon_Y, \bfu_1))
\ulV^\dagger(\epsilon_Y,u_2^-;\bfu_{1,\perp})
(\partial_{u_2^-} G_R^0(u_2^-,\bfu_1|y) )
\non & & {}
+2
\int_{0}^{\epsilon_Y} du_1^-
\int_{\bfu_1} 
(\partial_{u_1^-} \partial_{u_1^+}G_R^0(x|u_1))
G_R^0(u_1^-,\bfu_1|y) 
\non & & {}
-2
\int_{0}^{\epsilon_Y} du_1^-
\int_{\bfu_1} 
(\partial_{u_1^-} \partial_{u_1^+}G_R^0(x|u_1))
\ulV^\dagger(u_1^-,0;\bfu_{1,\perp})
G_R^0(0,\bfu_1|y) 
\non & & {}
-2
\int_{0}^{\epsilon_Y} du_1^-
\int_{0}^{u_1^-} du_2^-
\int_{\bfu_1} 
(\partial_{u_1^-} \partial_{u_1^+}G_R^0(x|u_1))
\ulV^\dagger(u_1^-,u_2^-;\bfu_{1,\perp})
(\partial_{u_2^-} G_R^0(u_2^-,\bfu_1|y))
\label{eq:appGRfull}
\non
\ee
For $\epsilon_Y > x^- > 0 > y^-$, the surviving terms are
\be
\lefteqn{\tildeG_R(x|y)}
&&
\non
& = &
G_R^0(x|y)
\non & & {}
+2
\int_{0}^{\epsilon_Y} du_1^-
\int_{\bfu_1} 
(\partial_{u_1^-} \partial_{u_1^+}G_R^0(x|u_1))
G_R^0(u_1^-,\bfu_1|y) 
\non & & {}
-2
\int_{0}^{\epsilon_Y} du_1^-
\int_{\bfu_1} 
(\partial_{u_1^-} \partial_{u_1^+}G_R^0(x|u_1))
\ulV^\dagger(u_1^-,0;\bfu_{1,\perp}) G_R^0(0,\bfu_1|y) 
+ O(\epsilon_Y)
\non
& = &
\ulV^\dagger(x^-,0;\bfxperp) G_R^0(0,\bfx|y) + O(\epsilon_Y)
\label{eq:GR_ordering1}
\ee
The last line in Eq.(\ref{eq:appGRfull}) is $O(\epsilon_Y)$ since
the integrand is at most $O(1/\epsilon_Y)$.
The last line in Eq.(\ref{eq:GR_ordering1}) follows upon using
\be
(\partial^2 + m^2)_xG_R^0(x|y) = -\delta(x-y)
\ee
and the fact that spatial derivatives are regular.

For $x^- > \epsilon_Y > y^- > 0$, the surviving terms are, after 
integrating by parts with respect to $u_1^+$ and 
using Eq.(\ref{eq:GRrel}),
\be
\lefteqn{\tildeG_R(x|y)}
&&
\non
& = &
-2
\int_{0}^{\epsilon_Y} du_2^-
\int_{\bfu_1} 
G_R^0(x|\epsilon_Y, \bfu_1)
\ulV^\dagger(\epsilon_Y,u_2^-;\bfu_{1,\perp})
(\partial_{u_1^+} \partial_{u_2^-} G_R^0(u_2^-,\bfu_1|y) )
\non & & {}
+2
\int_{0}^{\epsilon_Y} du_1^-
\int_{\bfu_1} 
(\partial_{u_1^-} \partial_{u_1^+}G_R^0(x|u_1))
G_R^0(u_1^-,\bfu_1|y) 
\non & & {}
-2
\int_{0}^{\epsilon_Y} du_1^-
\int_{0}^{u_1^-} du_2^-
\int_{\bfu_1} 
(\partial_{u_1^-} \partial_{u_1^+} G_R^0(x|u_1))
\ulV^\dagger(u_1^-,u_2^-;\bfu_{1,\perp})
(
\partial_{u_2^-} G_R^0(u_2^-,\bfu_1|y))
\non
& = &
G_R^0(x|\epsilon_Y, \bfy)
\ulV^\dagger(\epsilon_Y,y;\bfyperp)
+ O(\epsilon_Y)
\ee
The last line follows again upon using 
$(\partial^2 + m^2)_x G_R^0(x|y) = -\delta(x-y)$.

When $\epsilon_Y > x^- > y^- > 0$, the surviving terms are
\be
\lefteqn{\tildeG_R(x|y)}
&&
\non
& = &
G_R^0(x|y)
\non & & {}
+2
\int_{0}^{\epsilon_Y} du_1^-
\int_{\bfu_1} 
(\partial_{u_1^-} \partial_{u_1^+}G_R^0(x|u_1))
G_R^0(u_1^-,\bfu_1|y) 
\non & & {}
-2
\int_{0}^{\epsilon_Y} du_1^-
\int_{0}^{u_1^-} du_2^-
\int_{\bfu_1} 
(\partial_{u_1^-} \partial_{u_1^+}G_R^0(x|u_1))
\ulV^\dagger(u_1^-,u_2^-;\bfu_{1,\perp})
(\partial_{u_2^-} G_R^0(u_2^-,\bfu_1|y))
\non
& = &
\int_{0}^{x^-} du_2^-
\ulV^\dagger(x^-,u_2^-;\bfxperp)
(\partial_{u_2^-} G_R^0(u_2^-,\bfx|y))
+ O(\epsilon_Y)
\non
& = &
\ulV^\dagger(x^-,y^-;\bfxperp)
G_R^0(y^-,\bfx|y))
+ O(\epsilon_Y)
\ee
The last line follows because 
$G_R^0(x|y) = \theta(x^- - y^-)g_R^0(x|y)$
and $\partial_-g_R^0$ is regular.

In summary,
\be
\tildeG_R(x|y)
& = &
\left\{
\begin{array}{ll}
\ulV^\dagger(x^-,0;\bfxperp) G_R^0(0,\bfx|y) 
& \hbox{for $y^- < 0 < x^- \le \epsilon_Y$}
\\
\ulV^\dagger(x^-,y^-;\bfxperp) G_R^0(y^-,\bfx|y) 
& \hbox{for $0 < y^- \le x^- \le \epsilon_Y$} 
\\
2
\int_{\bfu_1}
(\partial_{u_1^+}G_R^0(x|\epsilon_Y,\bfu_1))
\ulV^\dagger(\epsilon_Y, 0; \bfuperp{}_1) G_R^0(0,\bfu_1|y)
& 
\hbox{for $y^- < 0 < \epsilon_Y < x^-$}
\\
G_R^0(x|\epsilon_Y,\bfy) \ulV^\dagger(\epsilon_Y, y^-; \bfyperp)
& 
\hbox{for $0< y^- < \epsilon_Y < x^-$}
\\
G_R^0(x|y) & \hbox{otherwise}
\end{array}
\right.
\non
\ee
Since we have used $\theta$ functions in integration,
the $G_R^0(y^-,\bfx|y)$ in the second line is really $g_R^0(y^-,\bfx|y)$.

The symmetric propagator is given by
\be
\tildeG_S^{ab}(x|y)
& = &
4
\int_{\bfu,\bfv}
\left[\partial_{u^+}\tildeG_R(x|x^-_{\rm init},\bfu) \right]_{ac}
G_S^0(0,\bfu-\bfv)
\left[\partial_{v^+}\tildeG_A(x^-_{\rm init},\bfv|y) \right]_{cb}
\ee
When both $x^- < 0$ and $y^- < 0$, 
this is just $G_S^0(x|y)$.
In the next section, we need
$\tildeG_S(x|y)$ at $x^- = y^- = \epsilon_Y$:
\be
\lefteqn{
\tildeG_S^{ab}(\epsilon_Y, \bfx|\epsilon_Y, \bfy)
} && 
\non
& = &
4
\int_{\bfu,\bfv}
\left[
V^\dagger(\epsilon_Y,\bfxperp)
\partial_{u^+}
G_R^0(0,\bfx|x^-_{\rm init},\bfu)
\right]_{ac}
G_S^0(0,\bfu-\bfv)
\left[
V(\epsilon_Y,\bfyperp)
\partial_{v^+}
G_A^0(x^-_{\rm init}, \bfv|0,\bfy)
\right]_{cb}
\non
& = &
V_{ac}^\dagger(\epsilon_Y,\bfxperp)
G_S^0(0,\bfx|0,\bfy)
V_{cb}(\epsilon_Y,\bfyperp)
\non
& = &
B(x^+ - y^+)\delta(\bfxperp-\bfyperp)\delta_{ab}
\ee
where we used
\be
G_S^0(0,\bfx)
& = &
\int_\bfk {1\over 4|k^-|} 
e^{-ik^-(x^+ - y^+)+i\bfkperp{\cdot}\bfxperp}
\non
& = &
B(x^+ - y^+)\delta(\bfxperp-\bfyperp)
\ee
with a distribution $B(x^+)$ defined by
\be
B(x^+) = \int {dk^-\over 8\pi |k^-|} e^{-ik^-(x^+ - y^+)}
\ee
And for $x^- = \epsilon_Y$ and $y^- < 0$,
\be
\lefteqn{
\tildeG_S^{ab}(\epsilon_Y, \bfx|y)
} && 
\non
& = &
4
\int_{\bfu,\bfv}
\left[
V^\dagger(\epsilon_Y,\bfxperp)
\partial_{u^+}
G_R^0(0,\bfx|x^-_{\rm init},\bfu)
\right]_{ab}
G_S^0(0,\bfu-\bfv)
\partial_{v^+}
G_A^0(x^-_{\rm init}, \bfv|y)
\non
& = &
V_{ab}^\dagger(\epsilon_Y,\bfxperp)
G_S^0(0,\bfx|y)
\ee
Similarly for $x^- < 0$ and $y^- = \epsilon_Y$,
\be
\tildeG_S^{ab}(x|\epsilon_Y,\bfy)
& = &
G_S^0(x|0,\bfy)V_{ab}(\epsilon_Y,\bfyperp)
\ee
For $x^- > \epsilon_Y$ and $y^- > \epsilon_Y$,
\be
\tildeG_S^{ab}(x|y)
& = &
4
\int_{\bfu,\bfv}
\left[\partial_{u^+}\tildeG_R(x|x^-_{\rm init},\bfu) \right]_{ac}
G_S^0(0,\bfu-\bfv)
\left[\partial_{v^+}\tildeG_A(x^-_{\rm init},\bfv|y) \right]_{cb}
\non
& = &
16
\int_{\bfu,\bfv,\bfu_1,\bfv_1}
\left(\partial_{u_1^+}G_R^0(x|\epsilon_Y,\bfu_1)\right)
\ulV_{ac}^\dagger(\epsilon_Y,\bfu_{1,\perp})
G_R^0(0,\bfu_1|x^-_{\rm init},\bfu)
\non & & {} \qquad
G_S^0(0,\bfu-\bfv)
G_A^0(x^-_{\rm init},\bfv|0,\bfv_1) 
\ulV_{cb}(\epsilon_Y,\bfv_{1,\perp})
\left(\partial_{v_1^+}G_A^0(\epsilon_Y,\bfv_1|y)\right)
\non
& = &
4
\int_{\bfu_1,\bfv_1}
\left(\partial_{u_1^+}G_R^0(x|\epsilon_Y,\bfu_1)\right)
\ulV_{ac}^\dagger(\epsilon_Y,\bfu_{1,\perp})
G_S^0(0,\bfu_1-\bfv_1)
\ulV_{cb}(\epsilon_Y,\bfv_{1,\perp})
\left(\partial_{v_1^+}G_A^0(\epsilon_Y,\bfv_1|y)\right)
\non
& = &
\delta_{ab}G_S^0(x|y)
\ee
For $x^- > \epsilon_Y$ and $y^- < 0$,
\be
\tildeG_S^{ab}(x|y)
& =&
4
\int_{\bfu,\bfv}
\left[\partial_{u^+}\tildeG_R(x|x^-_{\rm init},\bfu) \right]_{ac}
G_S^0(0,\bfu-\bfv)
\left[\partial_{v^+}\tildeG_A(x^-_{\rm init},\bfv|y) \right]_{cb}
\non
& =&
8
\int_{\bfu,\bfv,\bfu_1}
(\partial_{u_1^+}G_R^0(x|\epsilon_Y,\bfu_1))
\ulV_{ab}^\dagger(\epsilon_Y,\bfu_{1,\perp})
(\partial_{u^+} G_R^0(0,\bfu_1|x^-_{\rm init},\bfu))
G_S^0(0,\bfu-\bfv)
\partial_{v^+}G_A^0(x^-_{\rm init},\bfv|y) 
\non
& =&
2
\int_{\bfu_1}
(\partial_{u_1^+}G_R^0(x|\epsilon_Y,\bfu_1))
\ulV_{ab}^\dagger(\epsilon_Y,\bfu_{1,\perp})
G_S^0(0,\bfu_1|y) 
\label{eq:GS_x_pos_y_neg}
\ee
For $x^- < 0$ and $y^- > \epsilon_Y$,
\be
\tildeG_S^{ab}(x|y)
& = &
8
\int_{\bfu,\bfv,\bfw}
\left[\partial_{u^+}G_R^0(x|x^-_{\rm init},\bfu) \right]
G_S^0(0,\bfu-\bfv)
\partial_{v^+}
G_A^0(x^-_{\rm init},\bfv|0,\bfw)\ulV_{ab}(\epsilon_Y,\bfw)
\partial_{w^+}G_A^0(\epsilon_Y,\bfw|y)
\non
& = &
2
\int_{\bfu,\bfw}
G_S^0(x|0,\bfw)
\ulV_{ab}(\epsilon_Y,\bfw)
\partial_{w^+}G_A^0(\epsilon_Y,\bfw|y)
\ee
For 
$0 < x^- <\epsilon_Y$ and $0 < y^- <\epsilon_Y$,
\be
\tildeG_S^{ab}(x|y)
& =&
4
\int_{\bfu,\bfv}
\left[\partial_{u^+}\tildeG_R(x|x^-_{\rm init},\bfu) \right]_{ac}
G_S^0(0,\bfu-\bfv)
\left[\partial_{v^+}\tildeG_A(x^-_{\rm init},\bfv|y) \right]_{cb}
\non
& = &
4
\int_{\bfu,\bfv}
\ulV_{ac}^\dagger(x^-,\bfxperp)
\left[\partial_{u^+} G_R^0(0,\bfx|x^-_{\rm init},\bfu) \right]
G_S^0(0,\bfu-\bfv)
\left[\partial_{v^+}G_A^0(x^-_{\rm init},\bfv|0,\bfy) \right]
\ulV_{cb}(y^-,\bfyperp)
\non
& = &
\ulV_{ac}^\dagger(x^-,\bfxperp)
\ulV_{cb}(y^-,\bfyperp)
G_S^0(0,\bfx-\bfy)
\non
& = &
\ulV_{ac}^\dagger(x^-,\bfxperp)
\ulV_{cb}(y^-,\bfxperp)
G_S^0(0,\bfx-\bfy)
\ee
The last line follows because
\be
G_S^0(0,\bfx-\bfy)
= B(x^+ - y^+)\delta(\bfxperp-\bfyperp)
\ee
Lastly, for 
$0 < x^- <\epsilon_Y$ and $y^- < 0$,
\be
\tildeG_S^{ab}(x|y)
& =&
4
\int_{\bfu,\bfv}
\left[\partial_{u^+}\tildeG_R(x|x^-_{\rm init},\bfu) \right]_{ac}
G_S^0(0,\bfu-\bfv)
\left[\partial_{v^+}\tildeG_A(x^-_{\rm init},\bfv|y) \right]_{cb}
\non
& = &
4
\int_{\bfu,\bfv}
\ulV^\dagger(x^-,\bfxperp)
\left[\partial_{u^+}G_R^0(0,\bfx|x^-_{\rm init}, \bfu)
\right]_{ac}
G_S^0(0,\bfu-\bfv)
\left[\partial_{v^+}G_A^0(x^-_{\rm init},\bfv|y) \right]_{cb}
\non
& = &
\ulV_{ab}^\dagger(x^-,\bfxperp)
G_S^0(0,\bfx|y)
\ee

For convenience, we list here a few properties of the
free-field propagators.
Integrating $G_R^0$ over the time yields the transverse
Green function,
\be
\lefteqn{
\int_{\epsilon_Y}^{\infty} dy^-\, 
G_R^0(y^-,x^+,\bfxperp|\epsilon_Y,\bfu)
} && 
\non
& = &
\int_{\epsilon_Y}^{\infty} dy^-\, 
\int {d^3k\over (2\pi)^3}\,
{1\over 2ik^-}
e^{-i{\bfkperp^2\over 2k^-}(y^- - \epsilon_Y)
- ik^- (x^+ - u^+) + i\bfk_\perp(\bfx_\perp-\bfu_\perp)}
\non
& = &
\int {d^3k\over (2\pi)^3}\,
{1\over -\bfkperp^2}
e^{- ik^- (x^+ - u^+) + i\bfk_\perp(\bfx_\perp-\bfu_\perp)}
\non
& = &
-\delta(x^+ - u^+)G_T(\bfx_\perp-\bfu_\perp)
\label{eq:appGRtoGT}
\ee
Integrating $G_S^0$ over both $x^-$ and $x^+$ also yields
the transverse Green function,
\be
\lefteqn{
\int_{-\infty}^{0} dz^-
\int_{-\infty}^{y^+} dz^+
G_S^0(0,x^+,\bfuperp|z^-,z^+,\bfyperp)
} &&
\non
& = &
\int_{-\infty}^{0} dz^-
\int_{-\infty}^{y^+} dz^+
\int {d^4k\over (2\pi)^4}
e^{-ik^+(-z^-)-ik^-(x^+ - z^+) + i\bfkperp{\cdot}(\bfuperp-\bfyperp)}
\pi\delta(2k^+k^- - \bfkperp^2)
\non
& = &
\int {d^4k\over (2\pi)^4}
{1\over -k^+ k^-}
e^{i\bfkperp{\cdot}(\bfuperp-\bfyperp)}
\pi\delta(2k^+k^- - \bfkperp^2)
\non
& = &
\int {dk^-\over (2\pi)}
{e^{-ik^-(x^+ - y^+)}\over 4|k^-|}
\int {d^2k_\perp\over (2\pi)^2}
{2\over -\bfkperp}
e^{i\bfkperp{\cdot}(\bfuperp-\bfyperp)}
\non
& = &
-2B(x^+ - y^+) G_T(\bfuperp-\bfyperp)
\label{eq:appGStoGT}
\ee
The following identity is needed in the 2-point function
calculation
\be
\lefteqn{
\int_{\bfuperp} 
\partial_l^x
G_T(\bfxperp-\bfuperp)
\partial_y^l
G_T(\bfuperp-\bfyperp)
} &&
\non
&=&
\int_{\bfuperp} 
\int_{\bfkperp} 
{(-ik_l)e^{i\bfkperp\cdot(\bfxperp-\bfuperp)}\over \bfkperp^2}
\int_{\bfpperp}
{(ip^l)e^{i\bfpperp\cdot(\bfuperp-\bfyperp)}\over \bfpperp^2}
\non
& = &
\int_{\bfkperp} 
{(k_l k^l)e^{i\bfkperp\cdot(\bfxperp-\bfyperp)}\over \bfkperp^4}
\non
& = & -G_T(\bfxperp-\bfyperp)
\label{eq:appGTtoGT}
\ee

\section{Details of the 2-point function calculation}
\label{app:details_2pts}

In this section, we explain in more detail the calculation of
\be
\bar\chi(\epsilon_Y,\bfx|\epsilon_Y,\bfy)
& = &
\partial_{i}^x\partial_{j}^y \chi^{ij}(\epsilon_Y,\bfx|\epsilon_Y,\bfy)
\ee
where 
\be
\chi^{ij}(x|y)
& = &
\left(
4\ave{\tildea^i(x)\tildea^j(y)}
-2\ave{\tildea^i(x) \partial^j \tildeomega(y)}
-2\ave{\partial^i \tildeomega(x) \tildea^j (y)}
+\ave{\partial^i \tildeomega(x) \partial^j \tildeomega(y)}
\right)
\label{eq:app_chi}
\ee
What enters the JIMWLK equation is
\be
\eta(\bfxperp|\bfyperp) 
&=& 
{1\over 4B(x^+ - y^+)}
\int_{\bfuperp,\bfvperp}
G_T(\bfxperp-\bfuperp)\,
\barchi(\epsilon_Y,x^+,\bfuperp|\epsilon_Y,y^+,\bfvperp)
G_T(\bfvperp-\bfyperp)
\ee
The first term in Eq.(\ref{eq:app_chi}) is simple
\be
\ave{\tildea_a^j(\epsilon_Y,\bfx)\tildea_b^k(\epsilon_Y,\bfy)}
& \approx &
g^{jk}
\ulV_{ac}^\dagger(\epsilon_Y,\bfxperp)
G_{S}^0(0,\bfx|0,\bfy)
\ulV_{cb}(\epsilon_Y,\bfyperp)
\non
& = &
-g^{jk}\delta_{ab}
\delta(\bfxperp-\bfyperp)B(x^+ - y^+)
\ee
using $G_S^0(0,\bfx|0,\bfy) = B(x^+ - y^+)\delta(\bfxperp-\bfyperp)$
where $B(x^+)$ defined in Eq.(\ref{eq:B_def}).

For the second term,
we need Eq.(\ref{eq:bapprox}) and Eq.(\ref{eq:GS_x_neg}).
Using them yields
\be
{\ave{\tildeomega_a(\epsilon_Y,\bfx)
\tildea_b^j(\epsilon_Y,\bfy)}}
& = &
-
\ulV_{ac}^\dagger(\epsilon_Y,\bfxperp)
\int_{-\infty}^{0}dz^-
\int_{-\infty}^{x^+} dz^+\,
\ave{ \partial_i \tildea_c^i(z^-,z^+,\bfxperp) 
\tildea_b^j(\epsilon_Y,\bfy)}
\non
& = &
-
\ulV_{ac}^\dagger(\epsilon_Y,\bfxperp)
\int_{-\infty}^{0}dz^-
\int_{-\infty}^{x^+} dz^+\,
(-g^{ij})
\partial_i^x 
G_S^0(z^-,z^+,\bfxperp|0,\bfy)
\ulV_{cb}(\epsilon_Y,\bfyperp)
\non
& = &
\ulV_{ac}^\dagger(\epsilon_Y,\bfxperp)
\ulV_{cb}(\epsilon_Y,\bfyperp)
\partial^j_x 
\int_{-\infty}^{0}dz^-
\int_{-\infty}^{x^+} dz^+\,
G_S^0(z^-,z^+,\bfxperp|0,\bfy)
\non
& = &
-2B(x^+ - y^+)\ulV_{ac}^\dagger(\epsilon_Y,\bfxperp)
\ulV_{cb}(\epsilon_Y,\bfyperp)
\partial^j_x 
G_T(\bfxperp - \bfyperp)
\label{eq:app_omega_aj}
\ee
where we used Eq.(\ref{eq:appGStoGT}).
Similarly,
\be
{\ave{\tildea_a^i(\epsilon_Y,\bfx) \tildeomega_b(\epsilon_Y,\bfy)}}
& = &
-
\ulV_{bc}^\dagger(\epsilon_Y,\bfyperp)
\int_{-\infty}^{0}dz^-
\int_{-\infty}^{y^+} dz^+\,
\ave{\tildea_a^i(\epsilon_Y,\bfx)
\partial_j \tildea_c^j(z^-,z^+,\bfyperp)} 
\non
& = &
\ulV^\dagger_{bc}(\epsilon_Y,\bfyperp)
\ulV_{ac}^\dagger(\epsilon_Y,\bfxperp)
\partial^i_y 
\int_{-\infty}^{0}dz^-
\int_{-\infty}^{y^+} dz^+\,
G_S^0(0,\bfx|z^-,z^+,\bfyperp) 
\non
& = &
-
2B(x^+ - y^+)
\ulV^\dagger_{bc}(\epsilon_Y,\bfyperp)
\ulV_{ac}^\dagger(\epsilon_Y,\bfxperp)
\partial^i_y G_T(\bfxperp-\bfyperp)
\ee
For $\ave{\tildeomega\tildeomega}$, we get
\be
{\ave{\tildeomega_a(\epsilon_Y,\bfx)\tildeomega_b(\epsilon_Y,\bfy)}}
& \approx &
\ulV_{ac}^\dagger(\epsilon_Y,\bfxperp)
\ulV_{bd}^\dagger(\epsilon_Y,\bfyperp)
\int_{-\infty}^{0}dz^-
\int_{-\infty}^{x^+} dz^+\,
\int_{-\infty}^{0}dz'^-
\int_{-\infty}^{y^+} dz'^+\,
\non & & {}
\ave{
\partial_i \tildea_c^i(z^-,z^+,\bfxperp) 
\partial_j \tildea_d^j(z'^-,z'^+,\bfyperp) 
}
\non
& = &
-
\ulV_{ac}^\dagger(\epsilon_Y,\bfxperp)
\ulV_{bc}^\dagger(\epsilon_Y,\bfyperp)
\non & & {}
\partial_i^x \partial^i_y
\int_{-\infty}^{0}dz^-
\int_{-\infty}^{x^+} dz^+\,
\int_{-\infty}^{0}dz'^-
\int_{-\infty}^{y^+} dz'^+\,
G_S^0(z^-,z^+,\bfxperp|z'^-,z'^+,\bfyperp) 
\non
& = &
4B(x^+ - y^+)
\ulV_{ac}^\dagger(\epsilon_Y,\bfxperp)
\ulV_{bc}^\dagger(\epsilon_Y,\bfyperp)
G_T(\bfxperp-\bfyperp)
\ee
where we used
\be
\partial_i^x \partial_y^i
\int_{-\infty}^{0}dz^-
\int_{-\infty}^{x^+} dz^+\,
\int_{-\infty}^{0}dz'^-
\int_{-\infty}^{y^+} dz'^+\,
G_S^0(z^-,z^+,\bfxperp|z'^-,z'^+,\bfyperp) 
= -4B(x^+-y^+)G_T(\bfxperp-\bfyperp)
\non
\ee

The kernel in the JIMWLK equation is now
\be
\eta(\bfxperp|\bfyperp) 
&=& 
{1\over 4B(x^+ - y^+)}
\int_{\bfuperp,\bfvperp}
G_T(\bfxperp-\bfuperp)\,
\barchi(\epsilon_Y,x^+,\bfuperp|\epsilon_Y,y^+,\bfvperp)
G_T(\bfvperp-\bfyperp)
\non
& = &
\int_{\bfuperp,\bfvperp}
G_T(\bfxperp-\bfuperp)\,
\partial_i^u \partial_j^v
\left(
-g^{ij}\delta_{ab}
\delta(\bfuperp-\bfvperp)
\right) G_T(\bfvperp-\bfyperp)
\non & & {}
+
\int_{\bfuperp,\bfvperp} G_T(\bfxperp-\bfuperp)
\partial_j^v \partial_i^u \partial_u^i
\left(
\ulV_{ac}^\dagger(\epsilon_Y,\bfuperp)
\ulV_{cb}(\epsilon_Y,\bfvperp)
\partial^j_u 
G_T(\bfuperp - \bfvperp)
\right) G_T(\bfvperp-\bfyperp)
\non & & {}
+
\int_{\bfuperp,\bfvperp} G_T(\bfxperp-\bfuperp)
\partial_i^u \partial_j^v \partial_v^j
\left(
\ulV^\dagger_{bc}(\epsilon_Y,\bfvperp)
\ulV_{ac}^\dagger(\epsilon_Y,\bfuperp)
\partial^i_v G_T(\bfuperp-\bfvperp)
\right) G_T(\bfvperp-\bfyperp)
\non & & {}
+
\int_{\bfuperp,\bfvperp} G_T(\bfxperp-\bfuperp) 
\partial_j^v \partial_v^j \partial_i^u \partial_u^i
\left(
\ulV_{ac}^\dagger(\epsilon_Y,\bfuperp)
\ulV_{bc}^\dagger(\epsilon_Y,\bfvperp)
G_T(\bfuperp-\bfvperp)
\right) G_T(\bfvperp-\bfyperp)
\non
\ee
Integrating by parts and using
\be
\partial_i\partial^i G_T(\bfxperp-\bfyperp)
& = & \delta(\bfxperp-\bfyperp)
\ee
we finally get
\be
\eta(\bfxperp|\bfyperp) 
& = &
-
\delta_{ab}
\int_{\bfuperp}
\partial_i^x 
G_T(\bfxperp-\bfuperp)\,
\partial_y^i
G_T(\bfuperp-\bfyperp)
\non & & {}
+
\int_{\bfvperp} 
\left(
\ulV_{ac}^\dagger(\epsilon_Y,\bfxperp)
\ulV_{cb}(\epsilon_Y,\bfvperp)
\partial^j_x 
G_T(\bfxperp - \bfvperp)
\right) 
( \partial_j^y G_T(\bfvperp-\bfyperp) )
\non & & {}
+
\int_{\bfuperp} 
(\partial_i^x G_T(\bfxperp-\bfuperp))
\left(
\ulV_{ac}^\dagger(\epsilon_Y,\bfuperp)
\ulV_{cb}(\epsilon_Y,\bfyperp)
\partial^i_y G_T(\bfuperp-\bfvperp)
\right) 
\non & & {}
-
\ulV_{ac}^\dagger(\epsilon_Y,\bfxperp)
\ulV_{cb}(\epsilon_Y,\bfyperp)
\int_{\bfuperp}
(\partial_i^x G_T(\bfxperp-\bfuperp))
(\partial^i_y G_T(\bfuperp-\bfyperp))
\non
\ee
where we used Eq.(\ref{eq:appGTtoGT}).

\section{Details of the 1-point average calculation}
\label{app:Lines1to6}

We start from Eq.(\ref{eq:xi}) 
\be
\ave{ \sigma_1(\epsilon_Y, \bfx) }
& = &
ig 
\int_{\epsilon_Y}^{\infty} dy^-\, 
\ave{
[\acutea_i(y^-,x^+,\bfxperp), \partial_- \acutea^i(y^-,x^+,\bfxperp)]
}
\non & & {}
+ig
\int_{\epsilon_Y}^{\infty} dy^-\, 
\ave{
[[\calD_i,\omega(y^-,x^+,\bfxperp)], 
[\calD^i, \acutea^+(y^-,x^+,\bfxperp)]]
}
\non & & {}
-2ig
\int_{\epsilon_Y}^{\infty} dy^-\, 
\ave{
[\acutea_i(y^-,x^+,\bfxperp), 
[\calD_i, \acutea^+(y^-,x^+,\bfxperp)]]
}
\non & & {}
+ig
\ave{
[[\calD_i, \omega(\epsilon_Y,x^+,\bfxperp)], \acutea^i(\epsilon_Y,x^+,\bfxperp)]
}
\label{eq:appxi}
\ee
For the first and the third lines, the time coordinates 
are both above or on $\epsilon_Y$.
In this case, the resulting $G_S^{ab}(x|y) \propto \delta_{ab}$ as shown
in the last section.
Hence after factoring out $\ulV$ the averages contain
\be
if_{abc}\ave{\tildea_i^b(y^-,\bfx)\tildea_c^i(y^-,\bfx)} 
\propto f_{abb} = 0
\ee
The 4-th line is
\be
L_4
& = &
ig
\ave{
[[\calD_i, \omega(\epsilon_Y,x^+,\bfxperp)], \acutea^i(\epsilon_Y,x^+,\bfxperp)]
}
\non
& = &
ig
\ave{
[[\calD_i,
V(\epsilon_Y,\bfxperp)\tildeomega(\epsilon_Y,x^+,\bfxperp)
V^\dagger(\epsilon_Y,\bfxperp)
], V(\epsilon_Y,\bfxperp)\tildea^i(\epsilon_Y,x^+,\bfxperp)
V^\dagger(\epsilon_Y,\bfxperp)
]
}
\non
& = &
ig
\ave{
[
V(\epsilon_Y,\bfxperp)
\partial_i\tildeomega(\epsilon_Y,x^+,\bfxperp)
V^\dagger(\epsilon_Y,\bfxperp),
V(\epsilon_Y,\bfxperp)
\tildea^i(\epsilon_Y,x^+,\bfxperp)
V^\dagger(\epsilon_Y,\bfxperp)
]
}
\non
& = &
-g
t_a f_{abc}
\ulV_{bd}(\epsilon_Y,\bfxperp)
\ulV_{ce}(\epsilon_Y,\bfxperp)
\lim_{\bfyperp\to\bfxperp}
\partial_i^y
\ave{
\tildeomega_d(\epsilon_Y,x^+,\bfyperp)
\tildea_e^i(\epsilon_Y,x^+,\bfxperp)
}
\non
& \approx &
2 g B(0) 
t_a f_{abc}
\ulV_{bd}(\epsilon_Y,\bfxperp)
\ulV_{ce}(\epsilon_Y,\bfxperp)
\lim_{\bfyperp\to\bfxperp}
\partial_i^y
\left(
\ulV_{dg}^\dagger(\epsilon_Y,\bfyperp)
\ulV_{ge}(\epsilon_Y,\bfxperp)
\partial^j_y 
G_T(\bfyperp - \bfxperp)
\right)
\non
& = 0
\ee
where we used Eq.(\ref{eq:app_omega_aj}).  This vanishes because the result of
$\partial_i^y$ operation  yields either $f_{abc}\delta_{bc}$ 
or $\partial^j_yG(0) = 0$

The only remaining term is the second line in Eq.(\ref{eq:appxi}).
In the limit $x^- \to \epsilon_Y + 0^+$,  $\xi(x)$ becomes
\be
\lefteqn{
\ave{\sigma_1(\epsilon_Y,\bfx)} 
} &&
\non
& = &
ig 
\int_{\epsilon_Y}^{\infty} dy^-\, 
V(\epsilon_Y,\bfxperp)
\ave{[
\partial_i \tildeomega(y^-,x^+,\bfxperp), \partial^i \tildea^+(y^-,x^+,\bfxperp)
]}
V^\dagger(\epsilon_Y,\bfxperp)
\non
& = &
ig 
\int_{\epsilon_Y}^{\infty} dy^-
\int_{-\infty}^{y^-} dz^-
\int_{-\infty}^{x^+} dz^+
\int_{-\infty}^{x^+} dw^+
\non & & {}
V(\epsilon_Y,\bfxperp)
\ave{ [
\partial_i^x\left( 
V^\dagger(\epsilon_Y, z^-; \bfxperp)
\partial_k \tildea^k(z^-, z^+, \bfxperp)
V(\epsilon_Y, z^-; \bfxperp)
\right),
\partial^i 
\partial_l \tildea^l(y^-,w^+,\bfxperp)
] }
V^\dagger(\epsilon_Y,\bfxperp)
\non
\ee
using $\tildeomega = V^\dagger(\int dy^- \acutea^+) V$ and 
$\tildea^+ = -\int dy^+\partial_l \tildea^l$.
The $z^-$ integrals can be divided into 3 parts. 
In the region $\epsilon_Y < z^- < y^-$,
$V^\dagger(\epsilon_Y,z^-;\bfxperp) = 1$. The average then contains
$f_{abc}\delta_{bc} = 0$.
The contribution from the region $0 < z^- < \epsilon_Y$ is
$O(\epsilon_Y)$ since the integrand is regular.
Therefore,
\be
\lefteqn{
\ave{\sigma_1(\epsilon_Y,\bfx)} }
&&
\non
& \approx &
ig 
\int_{\epsilon_Y}^{\infty} dy^-
\int_{-\infty}^{0} dz^-
\int_{-\infty}^{x^+} dz^+
\int_{-\infty}^{x^+} dw^+
\non & & {}
V(\epsilon_Y,\bfxperp)
[
\partial_i^x 
\left(
V^\dagger(\epsilon_Y, \bfxperp)
\partial_k \tildea^k(z^-, z^+, \bfxperp)
V(\epsilon_Y, \bfxperp)
\right),
\partial_x^i 
\partial^x_l \tildea^l(y^-,w^+,\bfxperp)]
V^\dagger(\epsilon_Y,\bfxperp)
\non
& = &
-
ig 
\int_{\epsilon_Y}^{\infty} dy^-\, 
\int_{-\infty}^{x^+} dw^+\,
\int_{-\infty}^{0} dz^-
\int_{-\infty}^{x^+} dz^+
\non & &{}
V(\epsilon_Y,\bfxperp)
[
\partial_x^i 
\partial^x_l \tildea^l(y^-,w^+,\bfxperp) ,
\partial^x_i 
\left(V^\dagger(\epsilon_Y, \bfxperp)
(\partial^x_k \tildea^k(z^-, z^+, \bfxperp))V(\epsilon_Y, \bfxperp)\right)
]
V^\dagger(\epsilon_Y,\bfxperp)
\non & & {}
\ee
In the color vector notation, this is
\be
\lefteqn{
\ave{\sigma_1^a(\epsilon_Y,\bfx)}} &&
\non
& = &
-\lim_{\bfyperp\to\bfxperp}
ig 
\int_{\epsilon_Y}^{\infty} dy^-\, 
\int_{-\infty}^{x^+} dw^+\,
\int_{-\infty}^{0} dz^-
\int_{-\infty}^{x^+} dz^+
\non & &{}
\ulV_{ab}(\epsilon_Y,\bfxperp)
(if_{bcd})
\ave{
\left(
\partial_x^i 
\partial^x_l \tildea_c^l(y^-,w^+,\bfxperp) 
\right)
\partial^y_i 
\left(\ulV^\dagger_{de}(\epsilon_Y, \bfyperp)
(\partial^y_k \tildea_e^k(z^-, z^+, \bfyperp))
\right)
}
\ee
Using Eq.(\ref{eq:GS_x_pos_y_neg}), we get
\be
\lefteqn{
\ave{\sigma_1^a(\epsilon_Y,\bfx)}} &&
\non
& \approx &
-2g 
\lim_{\bfyperp\to\bfxperp}
\int_{\epsilon_Y}^{\infty} dy^-\, 
\int_{-\infty}^{x^+} dw^+\,
\int_{-\infty}^{0} dz^-
\int_{-\infty}^{x^+} dz^+
\non & &{}
\ulV_{ab}(\epsilon_Y,\bfxperp)
f_{bcd}
\partial_x^i
\partial_i^y 
\non & & {}
\left(
\ulV_{de}^\dagger(\epsilon_Y, \bfyperp)
\int_{\bfu} \,
\partial_l^x
(\partial_{u^+}G_R^0(y^-,w^+,\bfxperp|\epsilon_Y,\bfu))
\ulV^\dagger_{ce}(\epsilon_Y, \bfuperp) 
(\partial^l_y G_S^0(0,\bfu|z^-,z^+,\bfyperp))
\right)
\non
\ee
Since $G_S^0(x|y)$ only depends on the difference $x-y$, we can
change $\partial_{u^+}$ to $-\partial_{w^+}$ in the above expression
and carry out the $w^+$ integral to get
\be
\lefteqn{
\ave{\sigma_1^a(\epsilon_Y,\bfx)}} &&
\non
& = & 
2g 
\lim_{\bfyperp\to\bfxperp}
\non & &{}
\ulV_{ab}(\epsilon_Y,\bfxperp)
f_{bcd}
\partial_x^i
\partial_i^y 
\Bigg(
\ulV_{de}^\dagger(\epsilon_Y, \bfyperp)
\int_{\bfu} \,
\int_{\epsilon_Y}^{\infty} dy^-\, 
(\partial_l^x G_R^0(y^-,x^+,\bfxperp|\epsilon_Y,\bfu))
\non & & {}
\qquad \qquad \qquad \qquad
\ulV^\dagger_{ce}(\epsilon_Y, \bfuperp) 
\int_{-\infty}^{0} dz^-
\int_{-\infty}^{x^+} dz^+
(\partial^l_y G_S^0(0,\bfu|z^-,z^+,\bfyperp))
\Bigg)
\non & & {}
\ee
Using Eq.(\ref{eq:appGRtoGT}) and Eq.(\ref{eq:appGStoGT}) from the last section,
we arrive at
\be
\ave{\sigma_1^a(\epsilon_Y,\bfx)}
& = & 
4B(0)g 
\ulV_{ab}(\epsilon_Y,\bfxperp)
f_{bcd}
\non & & {}
\lim_{\bfyperp\to\bfxperp}
\partial_i^y 
\Bigg(
\ulV_{de}^\dagger(\epsilon_Y, \bfyperp)
\int_{\bfuperp} \,
(\partial_x^i \partial_l^x G_T(\bfxperp-\bfuperp))
\ulV_{ec}(\epsilon_Y, \bfuperp) 
(\partial^l_y G_T(\bfuperp-\bfyperp))
\Bigg)
\non
& = & 
4B(0)g 
\ulV_{ab}(\epsilon_Y,\bfxperp)
f_{bcd}
\non & & {}
\partial_i^x 
\Bigg(
\ulV_{de}^\dagger(\epsilon_Y, \bfxperp)
\int_{\bfuperp} \,
(\partial_x^i \partial_l^x G_T(\bfxperp-\bfuperp))
\ulV_{ec}(\epsilon_Y, \bfuperp) 
(\partial^l_x G_T(\bfuperp-\bfxperp))
\Bigg)
\ee
where we used $\ulV_{ce}^\dagger = \ulV_{ec}$.
The last line follows because the difference contains
$(\partial_i^x \partial_x^i \partial_l^x G_T(\bfxperp-\bfuperp))
= \partial_l^x \delta(\bfxperp-\bfuperp)$.
Hence, integrating over $\bfuperp$ results in either 
$f_{bcd}\delta_{dc} = 0$ or $\partial_l^xG_T(0) = 0$.
Using the fact that $G_T(\bfxperp-\bfuperp) = G_T(\bfuperp-\bfxperp)$,
we finally get
\be
\lefteqn{\ave{\sigma_1^a(\epsilon_Y,\bfxperp)}} &&
\non
& = &
2B(0)g 
\ulV_{ab}(\epsilon_Y,\bfxperp)
f_{bcd}
\non & & {}
\partial_i^x 
\left(
\ulV_{de}^\dagger(\epsilon_Y, \bfxperp)
\partial_x^i
\int_{\bfuperp} \,
(\partial_l^x G_T(\bfxperp-\bfuperp))
\ulV_{ec}(\epsilon_Y, \bfuperp) 
(\partial^l_x G_T(\bfuperp-\bfxperp))
\right)
\ee

To calculate $\ave{\zeta}$, we also need
\be
R_a(\bfx) = 
{1\over 2} 
gf_{abc}
\int_{\bfzperp} 
\ave{ \tildenu_{b}(x^+,\bfzperp) \tildenu_{c}(x^+, \bfxperp) }
G_T(\bfzperp - \bfxperp)
\ee
From the results obtained in Section \ref{sec:2pt_corr},
we get
\be
\lefteqn{
{1\over 4B(0)}
\int_{\bfzperp} 
\barchi_{bc}(\bfz,\bfx) G_T(\bfzperp-\bfxperp)
}&&
\non
&=& 
{1\over 4B(0)}
\int_{\bfxperp}
\barchi_{bc}(\epsilon_Y,x^+,\bfzperp|\epsilon_Y,y^+,\bfxperp)
G_T(\bfzperp-\bfxperp)
\non
& = &
\int_{\bfzperp}
\partial_i^z \partial_j^x
\left(
-g^{ij}\delta_{bc}
\delta(\bfzperp-\bfxperp)
\right)
G_T(\bfzperp-\bfxperp)
\non & & {}
+
\int_{\bfzperp} 
\partial_j^x \partial_i^z \partial_z^i
\left(
\ulV_{bd}^\dagger(\epsilon_Y,\bfzperp)
\ulV_{dc}(\epsilon_Y,\bfxperp)
\partial^j_z 
G_T(\bfzperp - \bfxperp)
\right)
G_T(\bfzperp-\bfxperp)
\non & & {}
+
\int_{\bfzperp} 
\partial_i^z \partial_j^x \partial_x^j
\left(
\ulV_{bd}^\dagger(\epsilon_Y,\bfzperp)
\ulV_{dc}(\epsilon_Y,\bfxperp)
\partial^i_x G_T(\bfzperp-\bfxperp)
\right)
G_T(\bfzperp-\bfxperp)
\non & & {}
+
\int_{\bfzperp}
\partial_j^x \partial_x^j \partial_i^z \partial_z^i
\left(
\ulV_{bd}^\dagger(\epsilon_Y,\bfzperp)
\ulV_{dc}(\epsilon_Y,\bfxperp)
G_T(\bfzperp-\bfxperp)
\right)
G_T(\bfzperp-\bfxperp)
\ee
After much algebra and integrations by parts,
we get
\be
\lefteqn{
{1\over 4B(0)}
\int_{\bfzperp} 
\barchi_{bc}(\bfz,\bfx) G_T(\bfzperp-\bfxperp)
}&&
\non
& = &
\partial_j^x
\int_{\bfzperp} 
\left(
\ulV_{bd}^\dagger(\epsilon_Y,\bfzperp)
(\partial_x^j \ulV_{dc}(\epsilon_Y,\bfxperp))
\partial^i_x G_T(\bfzperp-\bfxperp)
\right)
\partial_i^x 
G_T(\bfzperp-\bfxperp)
\non & & {}
+
\left(
\ulV_{bd}^\dagger(\epsilon_Y,\bfxperp)
(\partial_j^x \partial_x^j \ulV_{dc}(\epsilon_Y,\bfxperp))
G_T(0)
\right)
\ee

The 1-point average $\ave{\zeta_a}$ is given by
Eq.(\ref{eq:matching_cond_temp2}) reproduced here
\be
\ave{\tildezeta_a(\bfx)}
& = &
-\ave{\tildexi_a(x)}
-
{1\over 2}
f_{abc}
\int_{\bfzperp} 
\ave{\tildezeta_{b}(x^+, \bfzperp)\tildezeta_{c}(x^+,\bfxperp)}
G_T(\bfxperp - \bfzperp)
\ee
Combining our results, we get
\be
\ave{\zeta_a(\bfx)}
& = &
-\ave{\tildexi_a(x)}
-
{1\over 2}
f_{abc}
\int_{\bfzperp} 
\ave{\tildezeta_{b}(x^+, \bfzperp)\tildezeta_{c}(x^+,\bfxperp)}
G_T(\bfxperp - \bfzperp)
\non
& = &
-2B(0)g 
f_{abc}
\partial_i^x 
\left(
\ulV_{cd}^\dagger(\epsilon_Y, \bfxperp)
\partial_x^i
\int_{\bfuperp} \,
(\partial_l^x G_T(\bfxperp-\bfuperp))
\ulV_{db}(\epsilon_Y, \bfuperp) 
(\partial^l_x G_T(\bfuperp-\bfxperp))
\right)
\non & & {}
-2B(0)
gf_{abc}
\partial_i^x
\int_{\bfzperp} 
\left(
\ulV_{bd}^\dagger(\epsilon_Y,\bfzperp)
(\partial_x^l \ulV_{dc}(\epsilon_Y,\bfxperp))
\partial^l_x G_T(\bfzperp-\bfxperp)
\right)
\partial_l^x 
G_T(\bfzperp-\bfxperp)
\non & & {}
-2B(0) gf_{abc}
\left(
\ulV_{bd}^\dagger(\epsilon_Y,\bfxperp)
(\partial_j^x \partial_x^j \ulV_{dc}(\epsilon_Y,\bfxperp))
G_T(0)
\right)
\non
& = &
-2B(0)g f_{abc}
\partial_i^x 
\partial_x^i
\left(
\ulV_{cd}^\dagger(\epsilon_Y, \bfxperp)
\int_{\bfzperp} \,
(\partial_l^x G_T(\bfxperp-\bfzperp))
\ulV_{db}(\epsilon_Y, \bfzperp) 
(\partial^l_x G_T(\bfzperp-\bfxperp))
\right)
\non & & {}
-2B(0)
gf_{abc}
\left(
(\partial_j^x \partial_x^j \ulV_{cd}^\dagger(\epsilon_Y,\bfxperp))
\ulV_{db}(\epsilon_Y,\bfxperp)
G_T(0)
\right)
\ee

To see that the second term indeed cancels the divergence in the second
derivative,
consider the $\nabla_\perp^2 V^\dagger$ part of the above expression divided
by $4B(0)$
It is
\be
V_2 & = &
-{g \over 2}
f_{abc}
\left(
\partial_i^x \partial_x^i
\ulV_{bd}^\dagger(\epsilon_Y, \bfxperp)
\right)
\left(
\int_{\bfuperp} \,
\left(\partial_l^x G_T(\bfxperp-\bfuperp)\right)
\ulV_{dc}(\epsilon_Y, \bfuperp) 
(\partial^l_x G_T(\bfuperp-\bfxperp))
\right)
\non
& & {}
-
{1\over 2}gf_{abc}
\left( \partial_i^x \partial_x^i
\ulV_{bd}^\dagger(\epsilon_Y,\bfxperp)\right)
\ulV_{dc}(\epsilon_Y,\bfxperp)
G_T(0)
\ee
Integrating by parts and upon using $\partial_l\partial^lG_T = \delta$,
we get
\be
V_2
& = &
{g \over 2}
f_{abc}
\left(
\partial_i^x \partial_x^i
\ulV_{bd}^\dagger(\epsilon_Y, \bfxperp)
\right)
\left(
\int_{\bfuperp} \,
\left(\partial_l^u G_T(\bfxperp-\bfuperp)\right)
(\partial^l_u \ulV_{dc}(\epsilon_Y, \bfuperp)) 
G_T(\bfuperp-\bfxperp)
\right)
\non
\ee
In the momentum space,
\be
\lefteqn{
\int_{\bfuperp} \,
\left(\partial_l^u G_T(\bfxperp-\bfuperp)\right)
(\partial^l_u \ulV_{dc}(\epsilon_Y, \bfuperp)) 
G_T(\bfuperp-\bfxperp)
} && 
\non
& = &
\int_{\bfuperp} \,
\int_{\bfpperp,\bfkperp,\bfqperp} 
{ip_le^{i\bfpperp{\cdot}(\bfxperp-\bfuperp)}\over \bfpperp^2}
(-iq^l) e^{i\bfqperp\cdot\bfuperp} \ulV^\dagger(\bfqperp) 
{e^{i\bfkperp{\cdot}(\bfuperp-\bfxperp)}\over \bfkperp^2}
\non
& = &
-
\int_{\bfpperp,\bfqperp} 
(\bfpperp\cdot\bfqperp)
{e^{i\bfqperp{\cdot}\bfxperp}
\over \bfpperp^2(\bfpperp-\bfqperp)^2}
\ulV^\dagger(\bfqperp) 
\ee
The integration over $\bfqperp$ is cut-off by $e^{i\bfqperp\cdot\bfxperp}$
as well as $ \ulV^\dagger(\bfqperp)  $.
The integration over $\bfpperp$ converges in UV.
Hence, $V_2$ is UV-finite. 

To complete the JIMWLK derivation we need the relationship between
$\nu$ and $\eta$.
In the color vector space, we have
\be
\lefteqn{
\eta_{ab}(\bfxperp|\bfyperp)
} && 
\non
& = &
-
\int d^2u_\perp\,
\left(
\delta_{ab}
-
\ulV_{ac}^\dagger(\epsilon_Y,\bfuperp)
\ulV_{cb}(\epsilon_Y,\bfyperp)
-
\ulV_{ac}^\dagger(\epsilon_Y,\bfxperp) 
\ulV_{cb}(\epsilon_Y,\bfuperp)
+
\ulV_{ac}^\dagger(\epsilon_Y,\bfxperp)
\ulV_{cb}(\epsilon_Y,\bfyperp)
\right)
\non & & {}
\qquad
\partial_x^i G_T(\bfxperp-\bfuperp)
\partial_i^y G_T(\bfuperp-\bfyperp)
\ee
Taking a functional derivative with respect to 
$\tildecalA^-_b(z^-,\bfzperp)$ yields
\be
\lefteqn{
\lim_{z^-\to \epsilon_Y\atop \bfzperp\to\bfyperp}
{\delta\over \delta\tildecalA_b^-(z^-,\bfzperp)}
\eta_{ab}(\bfxperp|\bfyperp)
} && 
\non
& = &
\int d^2u_\perp\,
\left(
ig(T^b)_{ad}\delta(\bfyperp-\bfuperp)
\ulV_{dc}^\dagger(\epsilon_Y,\bfuperp) \ulV_{cb}(\epsilon_Y,\bfyperp)
-
\ulV_{ac}^\dagger(\epsilon_Y,\bfuperp) \ulV_{cd}(\epsilon_Y,\bfyperp)
ig(T^b)_{db}\delta(\bfyperp-\bfyperp)
\right)
\non & & {}
\qquad
\partial_x^i G_T(\bfxperp-\bfuperp)
\partial_i^y G_T(\bfuperp-\bfyperp)
\non & & {}
+
\int d^2u_\perp\,
\left(
ig(T^b)_{ad}\delta(\bfyperp-\bfxperp)
\ulV_{dc}^\dagger(\epsilon_Y,\bfxperp) \ulV_{cb}(\epsilon_Y,\bfuperp)
-
\ulV_{ac}^\dagger(\epsilon_Y,\bfxperp) \ulV_{cd}(\epsilon_Y,\bfuperp)
ig(T^b)_{db}\delta(\bfyperp-\bfuperp)
\right)
\non & & {}
\qquad
\partial_x^i G_T(\bfxperp-\bfuperp)
\partial_i^y G_T(\bfuperp-\bfyperp)
\non & & {}
-
\int d^2u_\perp\,
\left(
ig(T^b)_{ad}\delta(\bfyperp-\bfxperp)
\ulV_{dc}^\dagger(\epsilon_Y,\bfxperp) \ulV_{cb}(\epsilon_Y,\bfyperp)
-
\ulV_{ac}^\dagger(\epsilon_Y,\bfxperp) \ulV_{cd}(\epsilon_Y,\bfyperp)
ig(T^b)_{db}\delta(\bfyperp-\bfyperp)
\right)
\non & & {}
\qquad
\partial_x^i G_T(\bfxperp-\bfuperp)
\partial_i^y G_T(\bfuperp-\bfyperp)
\ee
Since $(T^b)_{ac} = -if_{bac}$, all the terms in the above
expression vanish except 
\be
\lefteqn{
\int_{\bfyperp}
\lim_{z^-\to \epsilon_Y\atop \bfzperp\to\bfyperp}
{\delta\over \delta\tildecalA_b(z^-,\bfzperp)}
\eta_{ab}(\bfxperp|\bfyperp)
} && 
\non
& = &
\int_{\bfyperp, \bfuperp}
\left(
ig(T^b)_{ad}\delta(\bfyperp-\bfxperp)
\ulV_{dc}^\dagger(\epsilon_Y,\bfxperp) \ulV_{cb}(\epsilon_Y,\bfuperp)
\right)
\partial_x^i G_T(\bfxperp-\bfuperp)
\partial_i^y G_T(\bfuperp-\bfyperp)
\non
& = &
\int_{\bfuperp}
\left(
ig(T^b)_{ad}
\ulV_{dc}^\dagger(\epsilon_Y,\bfxperp) \ulV_{cb}(\epsilon_Y,\bfuperp)
\right)
\partial_x^i G_T(\bfxperp-\bfuperp)
\partial_i^x G_T(\bfuperp-\bfxperp)
\non
& = &
g
\int_{\bfuperp}
\left(
f_{bad}
\ulV_{dc}^\dagger(\epsilon_Y,\bfxperp) \ulV_{cb}(\epsilon_Y,\bfuperp)
\right)
\partial_x^i G_T(\bfxperp-\bfuperp)
\partial_i^x G_T(\bfuperp-\bfxperp)
\non
& = &
g
\int_{\bfuperp}
\left(
f_{adb}
\ulV_{dc}^\dagger(\epsilon_Y,\bfxperp) \ulV_{cb}(\epsilon_Y,\bfuperp)
\right)
\partial_x^i G_T(\bfxperp-\bfuperp)
\partial_i^x G_T(\bfuperp-\bfxperp)
\non
& = &
-g
\int_{\bfuperp}
\left(
f_{abd}
\ulV_{dc}^\dagger(\epsilon_Y,\bfxperp) \ulV_{cb}(\epsilon_Y,\bfuperp)
\right)
\partial_x^i G_T(\bfxperp-\bfuperp)
\partial_i^x G_T(\bfuperp-\bfxperp)
\non
& = &
2\nu_a(\bfxperp)
\ee

\end{document}